\documentclass[a4paper,12pt,oneside]{book}

\usepackage[T1]{fontenc}
\usepackage[ansinew]{inputenc}
\usepackage{amsmath}
\usepackage{amssymb}
\usepackage{graphicx}
\usepackage{color}
\usepackage{euscript}
\usepackage[noautoscale]{youngtab}
\newlength{\defbaselineskip}
\newcommand{\setlinespacing}[1]%
           {\setlength{\baselineskip}{#1 \defbaselineskip}}

\newcommand{\dslash}{\partial \hspace{-0.5em}/\hspace{0.1em}}
\newcommand{\Dslash}{D \hspace{-0.65em}/\hspace{0.2em}}
\setlinespacing{1.66}
\begin{document}
\thispagestyle{empty}
 \setlinespacing{1.66}
 \begin{center}
\vskip 2cm \Huge Particles Under Extreme Conditions\\
\vskip 0.4cm
 \Large Part I: Quantum Modified
Null Trajectories in Schwarzschild Spacetime
\\

Part II: Superfluid Behaviour of the 2+1d NJL Model at High Density\\
\large \vskip 1cm Avtar Singh Sehra
\\

\vskip 2cm
\begin{center}
\begin{minipage}{5cm}
\begin{center}
        \includegraphics[width=3cm]{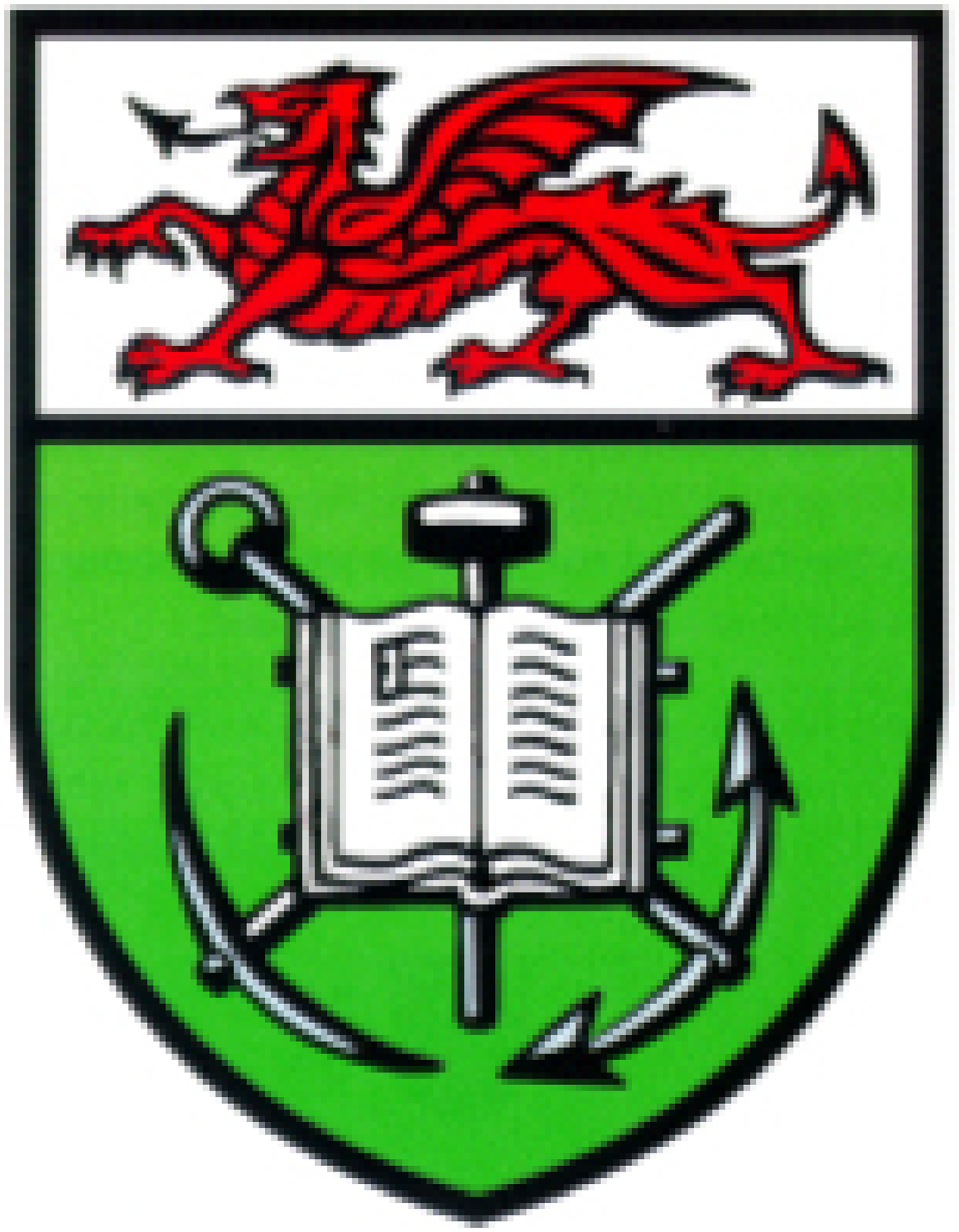}
\end{center}
\end{minipage}
\end{center}
\vskip 2cm
\small SUBMITTED TO THE UNIVERSITY OF WALES\\
IN FULFILMENT OF THE REQUIREMENTS OF\\
DOCTOR OF PHILOSOPHY\\
AT\\
UNIVERSITY OF WALES SWANSEA\\
SINGLETON PARK SWANSEA\\
SA2 8PP\\
\end{center}

\frontmatter \large
\chapter{Abstract}
 \normalsize \setlinespacing{1.66} In part I we study quantum modified photon trajectories in a Schwarzschild blackhole
spacetime. The photon vacuum polarization effect in curved spacetime
leads to birefringence, i.e. the photon velocity becomes $c\pm
\delta c$ depending on its polarization. This velocity shift then
results in modified photon trajectories.

We find that photon trajectories are shifted by equal and opposite amounts for the two photon polarizations, as expected by
the sum rule \cite{graham2}.  Therefore, the critical circular orbit at $u=1/3M$ in Schwarzschild spacetime, is split
depending on polarization as $u=1/3M\pm A\delta(M)$ (to first order in $A$), where $A$ is a constant found to be $\sim
10^{-32}$ for a solar mass blackhole. Then using general quantum modified trajectory equations we find that photons
projected into the blackhole for a critical impact parameter tend to the critical orbit associated with that polarization.
We then use an impact parameter that is lower than the critical one.  In this case the photons tend to the event horizon in
coordinate time, and according to the affine parameter the photons fall into the singularity.  This means even with the
quantum corrections the event horizon behaves in the classic way, as expected from the horizon theorem \cite{graham2}.

We also construct a quantum modified Schwarzschild metric, which encompasses the quantum polarization corrections.  This is
then used to derive the photons general quantum modified equations of motion, as before.  We also show that when this
modified metric is used with wave vectors for radially projected photons we obtain the classic equations of motion, as
expected, because radial velocities are not modified by the quantum polarization correction.
\\
\\
In Part II we use the 2+1d Nambu--Jona-Lasino (NJL) model to study
the superfluid behaviour of two-dimensional quark matter. In
previous work, \cite{hands1}, it was suggested that the high density
phase of the 2+1d NJL model could be a relativistic gapless thin
film BCS superfluid. In this work we find that as we raise the
baryon chemical potential ($\mu$) the baryon supercurrent jumps from
a non-superfluid (zero) phase to a superfluid (non-zero) phase. This
sharp transition is seen to occur in the region $0.65<\mu_c<0.68$,
which was shown in \cite{hands1} to be the region of chiral symmetry
restoration. In this analysis we prove that at high density the
$2+1d$ NJL model is in a superfluid phase.

We then go on to study the dynamics of the superfluid phase,
represented by the helicity modulus ($\Upsilon$ ), which is the
constant of proportionality between the supercurrent and the
gradient of the diquark state function. We find that below the
temperature associated with lattice size $L_t=4$, the system is in a
non-superfluid phase, and above $L_t=24$ the system is in a
superfluid phase. We also find a possible 2nd order transition at
$L_t\approx6$, which corresponds to the critical point as described
by Kosterlitz and Thouless' theory of $2D$ critical systems with
$U(1)$ global symmetry - such as the $XY$ model.

\thispagestyle{empty}

\vskip 3cm \hskip 10cm \large \textit{For my family}

\normalsize

\thispagestyle{empty}
\begin{center}
\Large UNIVERSITY OF WALES SWANSEA
\end{center}
\vskip 1.2cm Author: \hskip 1.95cm \textbf{Avtar Singh Sehra}
\\
Title: \hskip 2.4cm \textbf{Particles Under Extreme Conditions}
\\
Department: \hskip 1.1cm \textbf{Department of Physics}
\\
Degree: \hskip 2.05cm \textbf{Ph.D.}
\\
Year: \hskip 2.45cm \textbf{2005}
\\
\\
\\
\\
\normalsize \setlinespacing{1} This work has not previously been accepted in substance for any degree and is not being
concurrently submitted in candidature for any degree.

\begin{flushright}\includegraphics[width=6cm]{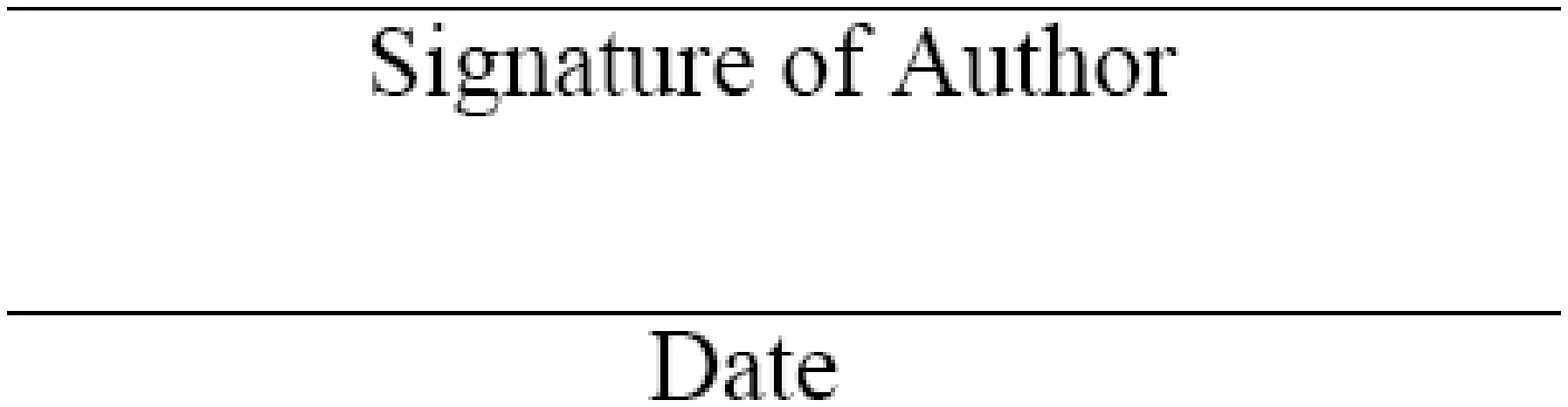}\end{flushright}
\vskip 0.8cm
This thesis is the result of my own investigations, except where otherwise stated.  Other sources are
acknowledged by explicit references.  A bibliography is appended.

\begin{flushright}\includegraphics[width=6cm]{sign}\end{flushright}
\vskip 0.8cm
I hereby give consent for my thesis, if accepted, to be available for photocopying and for inter-library loan,
and for the title and summary to be made available to outside organisations.

\begin{flushright}\includegraphics[width=6cm]{sign}\end{flushright}
\vskip 0.8cm \normalsize \setlinespacing{1.66}

\chapter{Acknowledgments}

There are many people to thank for their support and encouragement,
without whom this thesis would not have been possible. Firstly my
supervisors, Prof. Simon Hands and Prof. Graham Shore, for their
guidance, both academic and personal, throughout the course of my
postgraduate studies. Thanks also to my colleagues in the Swansea
particle theory group, Andrew Buxton, Steve Bidder and Iorwerth
Thomas. Also Aurora Trivini, with whom I have had many enlightening
discussions - physics related and otherwise. I also want to thank
all the friends I have made while living in Swansea, particularly
Robin Price, Antonio Di-Caprio, Gareth Coltman, Javier Silla, Jane
K\"{a}ehler, and Matthew Elwick.

I would like to thank both Swansea Physics Department and PPARC for
providing resources and funding for my research throughout this PhD.
\\
\\
Finally my family, to whom this thesis is dedicated, I cannot show
enough gratitude for the endless support and encouragement they have
given me.

\chapter{Preface}
In this work we will be dealing with quantum particles under extreme
conditions.  This will include the study of quantum modified photon
trajectories in a highly curved background spacetime (Schwarzschild
blackhole spacetime) and high density quark matter in two
dimensional films.

\subsubsection{Part 1: Quantum Modified Null
Trajectories in Schwarzschild Spacetime} In this part we will be
studying the quantum modifications to null geodesics in
Schwarzschild spacetime, in particular the modifications to the
critical stable orbit.  In Chapter~\ref{introduction} we will give
an introduction to the the vacuum polarization effect that leads to
the modification of photon trajectories in general relativity. In
Chapter~\ref{null dynamics} we will give an introduction to the
derivation of the orbit equation in Schwarzschild spacetime,
including the techniques involved in solving it and the
interpretation of results, i.e. for critical stable orbits and
trajectories to the singularity. In Chapter~\ref{optics} we will
then provide an introduction to the techniques of geometric optics,
i.e. the derivation of the light-cone and geodesic equation from
Maxwell's equations of motion in curved spacetime (with and without
the quantum corrections).  We will also give a brief overview of
previous theories associated with quantum modified null geodesics,
such as the horizon theorem and the polarization sum rule. In
Chapter~\ref{quant-mod-trajectories-chapter} we will go onto using
theses techniques in order to determine the quantum modified
critical circular orbits, and the changes in the associated critical
impact parameters. These finding will then be extended, by deriving
a general quantum modified equation of motion.  Using this, with the
modified impact parameters, we will study the modifications to
general null geodesics as they tend to the critical stable orbits.
Also, using the quantum modified general equation of motion, we will
show that a general geodesic aimed into the black hole behaves in
the classical way around the event horizon, as described by the
horizon theorem. In Chapter~\ref{Quantum Modified Schwarzschild
Metric} we will encompass these results into a quantum modified
Schwarzschild metric.

 \subsubsection{Part 2: Study of the 2+1d NJL Gapless Superfluid}
 In this part we will be studying high density quark matter in two
 dimensional films using the 2+1d Nambu-Jona-Lasino (NJL) model, in
 particular to isolate a critical gapless BCS superfluid phase, which
 was suspected to exist in \cite{hands1}.
In Chapter~\ref{chap:introduction} we will set the scene for this
work by giving an overview of quantum chromodynamics, its symmetries
and aspects of its phase diagram.  In Chapter~\ref{nabu-model} we
will give a brief introduction to the NJL model; then we will go on
to discuss the techniques and results of the previous simulations of
the NJL model, given in 2+1d \cite{hands1}, i.e. simulation of the
diquark condensate through the introduction of diquark sources, and
the evidence of chiral symmetry restoration and the non-zero baryon
density. In Chapter~\ref{forced baryon current flow} we will extend
the baryon density to a baryon three current by the use of Ward
identities, these are then implemented into the simulation.  We also
introduce a spatially varying diquark source, referred to as a
twisted source. In this way a gradient in the diquark pair wave
function is introduced, which forces a flow of the baryon current,
which will then be measured. We will then explore the behaviour of
the 3-current (using the helicity modules, $\Upsilon$,
Sec.~\ref{helicity}) with variations in spatial volume, temperature,
and variations in the diquark source, in order to isolate the
superfluid (and non-superfluid) state of the 2+1d NJL model.  In
Chapter~\ref{thin film} we will study the variation of the helicity
modulus with temperature; which will be done in order to determine
the critical point of the system, i.e. the point where vortex and
antivortex pairs come together to form the superfluid phase (as
predicted in the condensed matter study of 2 dimensional systems,
e.g. the XY model).

%%%%%%%%%%%%%%%%%%%%%%%%%%%%%%%%%%%%%%%%%%%%%%%%%%%%%%%%%%%%%%%%%%%%%%%%%%%%%%%%%%%%
%%%%%%%%%%%%%%%%%%%%%%%%%%%%%%%%%%%%%%%%%%%%%%%%%%%%%%%%%%%%%%%%%%%%%%%%%%%%%%%%%%%%%
%%%%%%%%%%%%%%%%%%%%%%%%%%%%%%%%%%%%%%%%%%%%%%%%%%%%%%%%%%%%%%%%%%%%%%%%%%%%%%%%%%%%%
%%%%%%%%%%%%%%%%%%%%%%%%%%%%%%%%%%%%%%%%%%%%%%%%%%%%%%%%%%%%%%%%%%%%%%%%%%%%%%%%%%%%%
%%%%%%%%%%%%%%%%%%%%%%%%%%%%%%%%%%%%%%%%%%%%%%%%%%%%%%%%%%%%%%%%%%%%%%%%%%%%%%%%%%%%%
\setlinespacing{1.66}

 \tableofcontents

%%%%%%%%%%%%%%%%%%%%%%%%%%%%%%%%%%%%%%%%%%%%%%%%%%%%%%%%%%%%%%%%%%%%%%%%%%%%%%%%%%%%%
%%%%%%%%%%%%%%%%%%%%%%%%%%%%%%%%%%%%%%%%%%%%%%%%%%%%%%%%%%%%%%%%%%%%%%%%%%%%%%%%%%%%%
%%%%%%%%%%%%%%%%%%%%%%%%%%%%%%%%%%%%%%%%%%%%%%%%%%%%%%%%%%%%%%%%%%%%%%%%%%%%%%%%%%%%%
%%%%%%%%%%%%%%%%%%%%%%%%%%%%%%%%%%%%%%%%%%%%%%%%%%%%%%%%%%%%%%%%%%%%%%%%%%%%%%%%%%%%%
%%%%%%%%%%%%%%%%%%%%%%%%%%%%%%%%%%%%%%%%%%%%%%%%%%%%%%%%%%%%%%%%%%%%%%%%%%%%%%%%%%%%%
\mainmatter
\part{Quantum Modified Null Trajectories in Schwarzschild Spacetime}

{\typeout{Introduction}
\chapter{Introduction}
\label{introduction}
\section{General Relativity}
Since the birth of special relativity, in 1905, the nature of space
and time has been demoted to a relative entity, known as spacetime,
which is stretched and contracted depending on an observer's frame
of reference, while the speed of light, $c$, has taken the pedestal
of an absolute and universal speed limit, unaffected by any
transformation of reference frame. From this emerged a generalized
theory of relativity, which portrayed the gravitational field in a
new and revolutionary way: where it didn't depend on a propagating
field but on the nature of spacetime itself. In this view matter (or
energy) is said to curve and modify the surrounding spacetime, this
then results in photons and particles tracing out shortest paths
between two points, known as geodesics. Therefore, gravitational
forces become a manifestation of the curved spacetime due to the
presence of matter \cite{kenyon,weinberg}. In this general
relativistic framework spacetime is described by the metric
$g_{\mu\nu}$ and the motion of particles are described by the
interval equation:
\begin{equation}
k^2=g_{\mu\nu}k^{\mu}k^{\nu} \qquad \Rightarrow \qquad
                              \begin{array}{c}
                                >0 \qquad \textmd{Time-like ($c<1$)} \\
                                =0 \qquad \textmd{Light-like ($c=0$)} \\
                                <0 \qquad \textmd{Space-like ($c>1$)} \\
                              \end{array}
\end{equation}
where $k^{\mu}=\frac{dx^{\mu}}{d\tau}$\footnote{For photons this
becomes $k^{\mu}=\frac{dx^{\mu}}{d\lambda}$, written in terms of the
affine parameter $\lambda$}.  For flat spacetime or a local inertial
frame (LIF), where $g_{\mu\nu}$ is replaced by the diagonal
Minkowski metric $\eta_{\mu\nu}=(1,-1,-1,-1)$, the interval equation
becomes:
\begin{equation}
k^2=\eta_{\mu\nu}k^{\mu} k^{\nu}=\frac{dt}{d\tau}^2-\frac{d
x}{d\tau}^2 -\frac{d y}{d\tau}^2-\frac{d z}{d\tau}^2
\end{equation}
for $c=1$.  Apart from resolving the problems associated with
Newtonian mechanics, such as describing the perihelion advance of
Mercury, the strongest aspect of general relativity was its
predictive power. One of its most radical claims, and the building
blocks of the theory itself, was that gravitational fields affect
radiation, which was then confirmed through the observation of
starlight deflection by the sun.  From this emerged some profound
and fantastic possibilities such as black holes and gravitational
lensing. The gravitational effect on light rays also leads to the
possibility that photons could follow stable orbits around stars
(discussed in chapter~\ref{null dynamics}), and it's this
possibility in which we will be interested.
\section{QED in a Curved Spacetime}
\label{qed curved space} Even though photon trajectories are
modified in a curved spacetime, and the resulting curved paths are
described by general relativity, this bending of light was, for a
long time, considered to have no effect on the velocity of the
photon. This view shifted slightly when, in 1980, Drummond and
Hathrell\cite{drummond} proposed that a photon propagating in a
curved spacetime may, depending on its direction and polarization,
travel with a velocity that exceeds the normal speed of light $c$.
This change in velocity would then result in trajectories other then
the ones described by "classical" general relativity. This effect is
simply described as a modification of the light cone in a LIF:
    \begin{equation}
        k^2=\eta_{\mu\nu}k^{\mu}k^{\nu}=0\quad\rightarrow\quad(\eta_{\mu\nu}+\alpha\sigma_{\mu\nu}(R))k^{\mu}k^{\nu}
    \end{equation}
Where $\alpha$ is the fine structure constant and
$\sigma_{\mu\nu}(R)$ is a modification to the metric that depends on
the Riemann curvature at the origin of the LIF.  This correction is
seen to arise from photon vacuum polarizations in a curved space
time, Fig.~\ref{vacuum}.
\begin{figure}
    \begin{center}
         \includegraphics[width=0.6\textwidth]{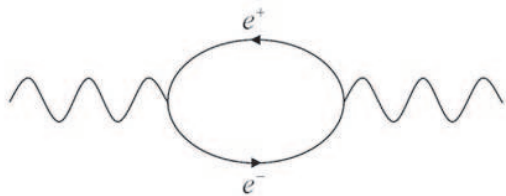}\\
        \caption{First order, $\alpha$, vacuum polarization Feynman diagram contributing to the photon propagator}
        \label{vacuum}
    \end{center}
\end{figure}
Qualitatively it can be thought of as a photon splitting into a
virtual $e^+e^-$ pair, so at the quantum level it is characterized
by the Compton wavelength $\lambda_c$; then, when this quantum cloud
of size $\mathcal{O}(\lambda_c)$ passes through a curved spacetime
its motion would be affected differently to that described by
general relativity, possibly in a polarization-dependent way
\cite{graham2}.

This effect of vacuum polarization is considered through the
effective action:
\begin{equation}
S=S_1+S_2 \label{total action}
 \end{equation}
where $S_1$ is the Maxwell electromagnetic action in curved space
time:
\begin{eqnarray}
S_1&=&-\frac{1}{4}\int d^4 x
\sqrt{-g}F_{\mu\nu}F^{\mu\nu}\nonumber\\
F_{\mu\nu}&=&\partial_{\mu}A_{\nu}-\partial_{\nu}A_{\mu}
\label{classic curved action}
\end{eqnarray}
and $S_2$ is the part of the action that incorporates the effects of
virtual electron loops in a background gravitational field.  As we
are only concerned with the propagation of individual photons it
must be quadratic in $A_{\mu}$, and the constraint of gauge
invariance then implies that it must depend on $F_{\mu\nu}$ rather
than $A_{\mu}$.  Also, as the virtual loops give the photon a size
of $\lambda_c$, $S_2$ can be expanded in powers of
$\lambda_c^2=m^{-2}$, thus the lowest term in the expansion would be
of order $m^{-2}$, which is the term corresponding to one electron
loop.  With these constraints there are only four independent gauge
invariant terms, which can be chosen to be:
\begin{eqnarray}
S_2&=&\frac{1}{m^2}\int d^4 x
\sqrt{-g}(aRF_{\mu\nu}F^{\mu\nu}+bR_{\mu\nu}F^{\mu\sigma}F^{\nu}_{\sigma}\nonumber\\
&+&cR_{\mu\nu\sigma\tau}F^{\mu\nu}F^{\sigma\tau}+d
D_{\mu}F^{\mu\nu}D_{\sigma}F^{\sigma}_{\nu}) \label{quantum curved
action}
\end{eqnarray}
The first three terms represent a direct coupling of the
electromagnetic field to the curvature, and they vanish in
flat-spacetime.  The fourth, however, is also applicable in the case
of flat-spacetime, and represents off-mass-shell effects in the
vacuum polarization. In \cite{drummond} the values for $a$, $b$,
$c$, and $d$ have been determined to $\mathcal{O}(e^2)$.  The
constant $d$ is obtained by comparing the coefficient of the
renormalized flat-spacetime photon propagator associated with the
Feynman diagram in Fig.~\ref{vacuum}\footnote{The photon propogator
with the vacuum polarization in flat-spacetime is given (in the
Feynman gauge) by:
$\frac{\eta_{\mu\nu}}{q^2}\rightarrow\frac{\eta_{\mu\nu}}{q^2}+\frac{1}{q^4}I_{\mu\nu}$,
where
$I_{\mu\nu}=(\eta_{\mu\nu}q^2-q_{\mu}q_{\nu})(1-\frac{e^2}{60\pi^2}\frac{q^2}{m_e^2}+\ldots)$
} to the result of the same order given by the effective action $S$;
and $a$, $b$, and $c$ are obtained by comparing the coefficients of
the coupling of a graviton to two photons\footnote{Deduced from the
matrix element $<\gamma(q_2,\beta)T^{\mu\nu}\gamma(q_1,\alpha)>$,
where $T^{\mu\nu}$ is the energy momentum tensor, \cite{drummond}}
to the same result obtained from $S_2$.  In this way the constants
are given as:
\begin{equation}
a=-\frac{5}{720}\frac{\alpha}{\pi} \qquad
b=\frac{26}{720}\frac{\alpha}{\pi} \qquad
c=-\frac{2}{720}\frac{\alpha}{\pi} \qquad
d=-\frac{24}{720}\frac{\alpha}{\pi}
\end{equation}
Then, as the equations of motion for the electromagnetic field are
given by:
\begin{equation}
\frac{\delta S}{\delta A_{\mu}(x)}=0
\end{equation}
using the modified action (\ref{total action}) we find:
\begin{equation}
D_{\mu}F^{\mu\nu}+\frac{\delta S_2}{\delta A_{\mu}}=0
\end{equation}
From this we can see that $D_{\mu}F^{\mu\nu}$ is of
$\mathcal{O}(e^2)$, therefore, the term with coefficient $d$ in
Eqn.~(\ref{quantum curved action}) will be of $\mathcal{O}(e^4)$,
hence we can omit it from the final equation of motion.  In this way
we find \cite{graham}:
\begin{equation}
D_{\mu}F^{\mu\nu}-\frac{1}{m_e^2}[2bR_{\mu\lambda}D^{\mu}F^{\lambda\nu}
+4cg^{\nu\tau}R_{\mu\tau\lambda\rho}D^{\mu}F^{\lambda\rho}]=0
\label{quantum maxwell}
\end{equation}
which is the Maxwell equation in curved spacetime, incorporating the
coupling of curvature with vacuum polarization effects. Using this
modified Maxwell equation and the methods of geometric optics
(described in Chapter~\ref{optics}) it is possible to derive the
quantum modified light cone and geodesic equations.  Then, using
these, we are able to determine the quantum modified trajectories in
curved spacetime.
\section{The Equivalence Principle and Causality}
The equivalence principle exists in two forms: weak and strong. The
weak equivalence principle states that at each point in spacetime
there exists a local Minkowski frame, which is a fundamental
requirement of general relativity\footnote{This implies that general
relativity is formulated on a Riemannian manifold}. The strong
equivalence principle (SEP), on the other hand, states that the laws
of physics are the same in all LIFs at different points in
spacetime, and at the origin of each LIF they take the special
relativistic form. Then the coupling of curvature to the
electromagnetic field in the effective action, Eqn.~(\ref{total
action}), is a violation of the SEP.  Due too this violation of the
SEP, QED in curved spacetime remains a causal theory - despite the
modification to the physical light cone.

This can be seen more clearly by considering Global Lorentz
invariance\footnote{Global Lorentz invariance states that the laws
of physics are the same in all inertial frames.} (GLI), which is the
special relativistic equivalent of the SEP.  In special relativity
GLI states that faster than light signals automatically imply the
possibility of unacceptable closed-time-loops, Fig.~\ref{time-loop}.
That is, if you can send a signal backwards in time in one frame,
then it should be possible in any frame.  However, if you break this
GLI, a signal backwards in time in one frame does not automatically
imply you can send a signal backwards in time from any frame.
Therefore, in the case of QED in curved spacetime, due to the
breakdown of the SEP we retain the fundamental property of
causality: faster than light signals can be seen to go backwards in
time in a certain frame, but this no longer implies that you can
send a signal backwards in time from any frame. Thus, in this case,
spacelike motion does not necessarily imply a causality
violation\footnote{Further discussion of causality is given in
\cite{graham3}}.
\begin{figure}
    \begin{center}
         \includegraphics[width=0.5\textwidth]{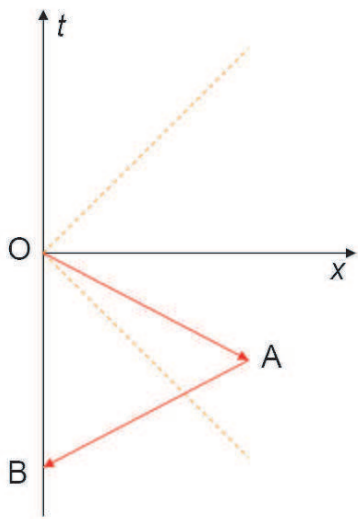}\\
        \caption{An unacceptable closed-time-loop.}
        \label{time-loop}
    \end{center}
\end{figure}

\section{Critical Stable Orbits} One of the simplest curved
spacetimes is the (Ricci-flat) Schwarzschild spacetime, which
describes the spherically symmetric geometry outside a star. The
geometry of such a spacetime structure is given by the line
interval:
\begin{equation}
  ds^{2}=(1-\frac{2M}{r})dt^{2}-(1-\frac{2M}{r})^{-1}dr^{2}-r^{2}d\theta^{2}-(r^{2}\sin^{2}\theta)d\phi^{2}
  \label{schwarzschild line interval}
\end{equation}
In this spacetime the equations of motion, derived for $k^2=0$, have
a special solution for null geodesics.  This solution describes a
light ray in a circular orbit with a radius $r=3M$ and an associated
critical impact parameter.  Therefore, a photon projected from
$r\rightarrow\infty$ with the critical parameter tends to the stable
circular orbit by spiralling around it.  But, in the context of
vacuum polarization effects, we may derive the null equation of
motion using the quantum modified Maxwell equation.  In this way the
orbit equation should give us stable circular orbits that are
dependent on the photon polarization, i.e. splitting the circular
orbit depending on polarization. Therefore, a photon coming in from
$r\rightarrow\infty$, with a particular polarization and impact
parameter, should tend to its associated critical stable orbit by
spiralling around it in the classic way.
}

{\typeout{Null Dynamics in Schwarzschild Spacetime }
\chapter{Null Dynamics in Schwarzschild Spacetime }
\label{null dynamics}  In this chapter we will give a brief overview
of null dynamics in Schwarzschild spacetime.  Starting with the
geodesic equations and the line interval for photons we will derive
the orbit equation, which will then be used to determine the
critical stable orbit for photons and the associated impact
parameter.  We will then go on to show that decreasing the impact
parameter, from the critical value, causes the photon trajectory,
given as a function of $\phi$, to spiral into the singularity, but
as a function of $t$ it tends to the event horizon.  The main aim of
this chapter is to familiarise ourselves with the classical
solutions of photon trajectories in Schwarzschild spacetime, so we
can then compare the equivalent results for the quantum modified
case.
\section{Equations of Motion}
When we consider motion in a plane, where $\theta=\frac{\pi}{2}$ and
$\dot{\theta}=\ddot{\theta}=0$, the geodesic equations for
Schwarzschild spacetime, (\ref{equation of motion t})-(\ref{equation
of motion
 phi}), can written as:
    \begin{equation}
        (1-\frac{2M}{r})\ddot{t}+\frac{2M}{r^2}\dot{r}\dot{t}=0
         \label{t with theta=0}
    \end{equation}
    \begin{equation}
        \frac{M}{r^2}\dot{t}^2-r \dot{\phi}^{2} +(1-\frac{2M}{r})^{-1}\ddot{r}
        -\frac{M}{r^{2}}(1-\frac{2M}{r})^{-2}\dot{r}^{2}=0
         \label{r with theta=0}
    \end{equation}
    \begin{equation}
        2r\dot{r} \dot{\phi}+r^2 \ddot{\phi}=0
         \label{phi with theta=0}
    \end{equation}
And the spacetime line interval becomes:
    \begin{equation}
        (1-\frac{2M}{r})\dot{t}^{2}-(1-\frac{2M}{r})^{-1}\dot{r}^{2}-r^{2}\dot{\phi}^{2}=K
         \label{interval with theta=0}
    \end{equation}
where $K=0$ for photons and $\pm1$ for space-like or time-like
motion. Now, to derive the orbit equation $\frac{du}{d\phi}$ we use
Eqns.~(\ref{t with theta=0}), (\ref{phi with theta=0}) and the
interval equation (\ref{interval with theta=0}), in the simplified
form:
 \begin{eqnarray}
\label{equation of motion t theta=0}
        0&=&\ddot{t}+\frac{2M}{r^2}(1-\frac{2M}{r})^{-1}\dot{r}\dot{t}\\
         \label{equation of motion phi theta=0}
        0&=&\ddot{\phi}+\frac{2\dot{r}}{r} \dot{\phi}\\
 \label{interval equation theta=0}
        K(1-\frac{2M}{r})&=&-\dot{r}^{2}+(1-\frac{2M}{r})^{2}\dot{t}^{2}-r^{2}(1-\frac{2M}{r})\dot{\phi}^{2}
         \end{eqnarray}
In order to specify photon trajectories we can set $K=0$ at anytime,
however, then $\tau$ would be interpreted as an affine parameter and
not proper time. To proceed we divide~(\ref{equation of motion t
theta=0}) by $\frac{dt}{d\tau}$ and~(\ref{equation of motion phi
theta=0}) by $\frac{d\phi}{d\tau}$:
    \begin{eqnarray}
        0&=&\frac{\ddot{t}}{\dot{t}}+\frac{2M}{r^2}(1-\frac{2M}{r})^{-1}\dot{r}\\
        0&=&\frac{ \ddot{\phi}}{\dot{\phi}}+\frac{2\dot{r}}{r}
    \end{eqnarray}
These can written as:
    \begin{eqnarray}
        0&=&\frac{d}{d\tau}(\ln{\dot{t}}+\ln(1-\frac{2M}{r}))\\
        0&=&\frac{d}{d\tau}(\ln{\dot{\phi}}+\frac{2\dot{r}}{r})
    \end{eqnarray}
Then, solving these we have:
    \begin{eqnarray}
 \label{dot-t}
        \dot{t}&=&(1-\frac{2M}{r})^{-1}E\\
   \label{dot-phi}
        \dot{\phi}&=&\frac{J}{r^2}
    \end{eqnarray}
Where $E$ and $J$ are constants of integration denoting total energy
and angular momentum about an axis normal to the plane
$\theta=\pi/2$. Now, using Eqns.~(\ref{dot-t}) and (\ref{dot-phi})
to substitute for $\dot{t}$ and $\dot{\phi}$ in Eqn~(\ref{interval
equation theta=0}) and rearranging, we have:
    \begin{equation}
        \dot{r}^2+r^2(1-\frac{2M}{r})\frac{J^2}{r^4}+K(1-\frac{2M}{r})=(1-\frac{2M}{r})^{2}(1-\frac{2M}{r})^{-1}E^2
         \label{interval with dot-t and dot-phi replaced}
    \end{equation}
Rearranging and simplifying, this becomes:
    \begin{eqnarray}
  \label{dr/dtau-1}
        (\frac{dr}{d\tau})^2+(1-\frac{2M}{r})(K+\frac{J^2}{r^2})=E^2\\
        (\frac{dr}{d\tau})=E[1-(1-\frac{2M}{r})(\frac{K}{E^2}+\frac{J^2}{E^{2}r^2})]^{\frac{1}{2}}
    \end{eqnarray}
We now have the components of the four momentum $P^{\alpha}$ in
spherical polar coordinates
    \begin{eqnarray}
        P^{\alpha}&=&(\dot{t},\dot{r},\dot{\theta},\dot{\phi})\nonumber\\
        &=&E(F^{-1},[1-F(\frac{K}{E^2}+\frac{D^2}{r^2})]^{\frac{1}{2}},0,\frac{D}{r^2})
         \label{four momentum}
    \end{eqnarray}
where $D=\frac{J}{E}$ is the impact parameter, $F=(1-\frac{2M}{r})$
and $K=0$ for photons and $K=+1$ for particles.
\subsection{Orbit Equation}
In order to derive orbit equations that are physically
understandable we need to represent them as $r(\phi)$, $r(t)$ and
$\phi(t)$. In this form we can analyse the orbit paths as a function
of rotational angle $\phi$ and the time taken to reach a certain
point along the angle. Therefore, by using
    \begin{equation}
        \frac{dr}{d\tau}=\frac{dr}{d\phi}\cdot\frac{d\phi}{d\tau}=\frac{dr}{d\phi}\frac{J}{r^2}
         \label{dr/dtau=diff}
    \end{equation}
    we can rewrite~(\ref{dr/dtau-1}) as
    \begin{eqnarray}
        E^2&=&(\frac{dr}{d\phi})^2\frac{J^2}{r^4}+(1-\frac{2M}{r})(K+\frac{J^2}{r^2})\nonumber\\
     \Rightarrow \qquad  (\frac{dr}{d\phi})^2&=&(E^2-K)\frac{r^4}{J^2}+2MK\frac{r^3}{J^2}-r^2+2Mr
         \label{dr/dphi}
    \end{eqnarray}
Now, transforming $u=r^{-1}$, so at $r=\infty$ we have $u=0$,
Eqn.~(\ref{dr/dphi}) becomes:
    \begin{eqnarray}
 \label{du/dr with u=1/r}
        (\frac{du}{d\phi})^2r^4&=&(E^2-K)\frac{r^4}{J^2}+2MK\frac{r^3}{J^2}-r^2+2Mr\nonumber\\
  \label{du/dphi}
        (\frac{du}{d\phi})^2&=&\frac{(E^2-K)}{J^2}+2MK\frac{u}{J^2}-u^2+2Mu^3
   \end{eqnarray}
Doing similar manipulation for $\phi(t)$ we have
    \begin{equation}
        \frac{d\phi}{dt}=\frac{d\phi}{d\tau}\cdot\frac{d\tau}{dt}
         \label{dphi/dt=diff}
    \end{equation}
    and using Eqns.~(\ref{dot-t}) and (\ref{dot-phi}) in this, and transforming $u=\frac{1}{r}$, we
    have:
    \begin{equation}
        \frac{d\phi}{dt}=\frac{D}{r^2}(1-\frac{2M}{r})=Du^2(1-2Mu)
         \label{dphi/dt}
    \end{equation}
Finally, we can determine $u(t)$ by inserting:
    \begin{equation}
        \frac{du}{d\phi}=\frac{du}{dt}\cdot\frac{dt}{d\phi}=\frac{du}{dt}[D u^2(1-2M
        u)]^{-1}
         \label{du/dphi=diff}
    \end{equation}
into Eqn.~(\ref{du/dphi}), which gives
    \begin{equation}
        (\frac{du}{dt})^2=[Du^2(1-2Mu)]^2[\frac{(E^2-K)}{J^2}+2MK\frac{u}{J^2}-u^2+2Mu^3]
         \label{du/dt}
    \end{equation}
Now, the Eqns.~(\ref{du/dphi}), (\ref{dphi/dt}) and (\ref{du/dt})
are the general equations that determine photon and particle
trajectories in the plane $\theta=\pi/2$. If we take $K=0$ we then
have the required photon trajectory equations
    \begin{eqnarray}
\label{du/dphi=f(u)}
        (\frac{du}{d\phi})^2&=&\frac{1}{D^2}-u^2+2Mu^3=f(u)\\
 \label{du/dt=f(u)}
        (\frac{du}{dt})^2&=&[Du^2(1-2Mu)]^2[\frac{1}{D^2}-u^2+2Mu^3]\nonumber\\
        &=&[Du^2(1-2Mu)]^2f(u)
    \end{eqnarray}
We can also write another equation, specifically for null radial
geodesics.  By using the fact that $\dot{\phi}=0$,
Eqn.~(\ref{dot-phi}) implies $J=0$, then Eqns~(\ref{dot-t}) and
(\ref{dr/dtau-1}) become
    \begin{eqnarray}
        (1-\frac{2M}{r})\frac{dt}{d\tau}=E\\
        \frac{dr}{d\tau}=\pm E
         \label{dr/dtau J=0}
    \end{eqnarray}
    Combining these we have:
    \begin{equation}
        \frac{dr}{dt}=\pm (1-\frac{2M}{r})
         \label{dr/dt J=0}
    \end{equation}
\section{Orbits in Schwarzschild Spacetime}
We will now discuss the solutions of Eqns.~(\ref{du/dphi=f(u)}),
(\ref{du/dt=f(u)}) and (\ref{dr/dt J=0}).   We will start off with
the simple case of the radial geodesic and then go onto the case of
the general orbits. For the general case we will solve
(\ref{du/dphi=f(u)}) and (\ref{du/dt=f(u)}) to determine the
critical stable orbits and the associated impact parameters.  We
will then go on to study the trajectories of photons as they fall
into the singularity
\subsection{Null Radial Geodesic}
Eqn.~(\ref{dr/dt J=0}) can be solved by rewriting it as:
    \begin{eqnarray}
  \label{rearranged dr/dt J=0}
        \frac{dt}{dr}&=&\pm (1-\frac{2M}{r})^{-1}\\
    \label{coordinate time solution}
        \Rightarrow \qquad t&=&\pm(r+2M\log(\frac{r}{2M}-1))+ constant_{\pm}
    \end{eqnarray}
This solution (specifically the $-$ one) gives the trajectory of a
photon coming from infinity and into the black hole.  It shows that
the photon takes an infinite coordinate time to reach the horizon,
which can be seen in Fig.~\ref{coordinate-radial}. However, if we
solve Eqn.~(\ref{dr/dtau J=0}), i.e. in terms of the affine
parameter, we can show that the photon reaches and crosses the event
horizon without ever noticing it,
    \begin{equation}
        r=\pm E \tau + constant_{\pm}
        \label{proper time solution}
    \end{equation}
this can be seen in Fig.~\ref{proper-radial}
\begin{figure}
    \begin{center}
        \includegraphics[width=0.8\textwidth]{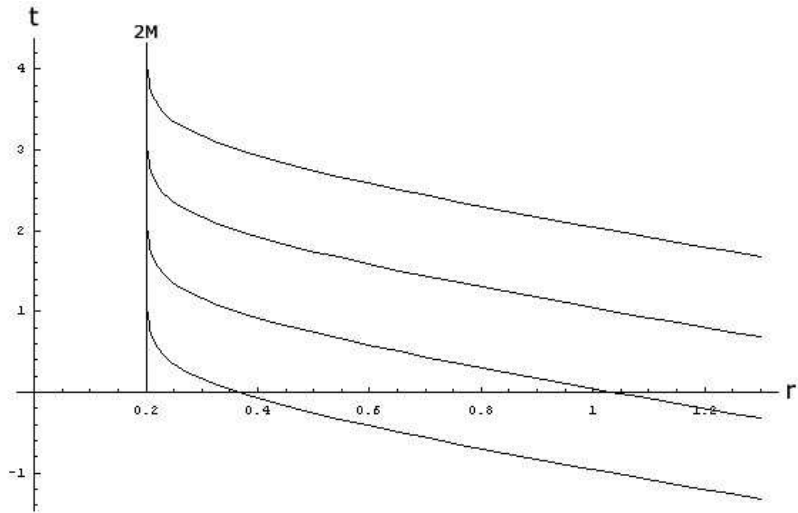}\\
        \caption{Radial Null geodesics take an infinite coordinate
         time to reach the event horizon at 2M}\label{coordinate-radial}
    \end{center}
\end{figure}
\begin{figure}
    \begin{center}
        \includegraphics[width=0.8\textwidth]{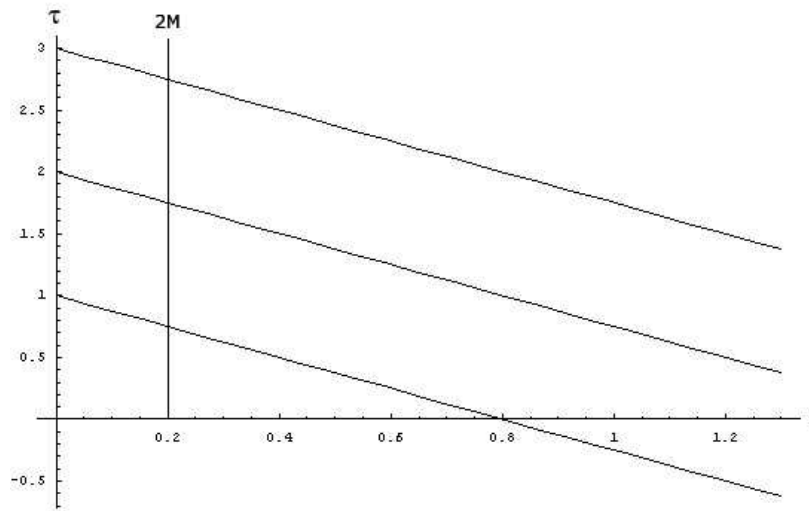}\\
        \caption{Radial Null geodesics reach and cross the event horizon
        at 2M without noticing it when defined with the affine parameter}\label{proper-radial}
    \end{center}
\end{figure}
\subsection{General Null Geodesics and Critical Orbits}
We will first solve Eqn.~(\ref{du/dphi=f(u)}) in order to determine
the the critical stable photon orbits.  In order to do this we first
consider the point of equilibrium and the associated impact
parameter.  This equilibrium point occurs when
    \begin{equation}
        \frac{du}{d\phi}=0
        \label{du/dphi=0}
    \end{equation}
which means as the photon is orbiting the black hole $u$ does not
change; so if the radial distance does not change it implies a
circular orbit. Therefore we must solve
    \begin{equation}
         f(u)=\frac{1}{D^2}-u^2+2Mu^3=0
         \label{f(u)=0}
    \end{equation}
The sum and product of the roots $u_1$, $u_2$, and $u_3$ of this
equation are given by\footnote{for $au^3+bu^2+c=0$ we have
$u_1+u_2+u_3=-\frac{b}{a}$ and $u_1u_2u_3=-\frac{c}{a}$}
    \begin{equation}
        u_1+u_2+u_3=\frac{1}{2M} \qquad \textrm{and} \qquad u_1u_2u_3=-\frac{1}{2MD^2}
        \label{root-condition}
    \end{equation}
This shows that $f(u)=0$ must have a real negative root, and the two
remaining roots can be real (distinct or coincident) or be a complex
conjugate pair; however, the occurrence of coincident positive real
roots implies the existence of a circular orbit. Thus, if a
coincident root occurs it should be at the point given by the
derivative of $f(u)$
    \begin{equation}
         f'(u)=6Mu^2-2u=0
         \label{f'(u)=0}
    \end{equation}
Which then has the solution $u_1=u_2=(3M)^{-1}$.  For this solution
the impact parameter of equation~(\ref{f(u)=0}) is $D=(3\sqrt{3})M$.
From the product condition, equation~(\ref{root-condition}), we find
that the roots of $f(u)=0$ are
    \begin{equation}
        u_1=u_2=\frac{1}{3M} \qquad \textrm{and} \qquad
        u_3=-\frac{1}{6M} \qquad \textrm{and} \qquad D=(3\sqrt{3})M
        \label{classical roots of orbits}
    \end{equation}
Therefore, when the impact parameter is $D=(3\sqrt{3})M$ then
$\frac{du}{d\phi}$ vanishes for $u=(3M)^{-1}$, which implies a
circular orbit of radius $3M$ is an allowed null geodesic
\cite{chandrasekhar}.

Now we can consider a photon at $u=0$ with an impact parameter
$D=(3\sqrt{3})M$.  This, then, gives a trajectory of a photon
spiralling in and tending to the critical orbit at $u=(3M)^{-1}$.
The general differential equation for this impact parameter is given
by rearranging and substituting for $D$ in Eqn.~(\ref{du/dphi=f(u)})
    \begin{equation}
        (\frac{du}{d\phi})^2=2M(u+\frac{1}{6M})(u-\frac{1}{3M})^2
        \label{du/dphi for D=critical}
    \end{equation}
From \cite{chandrasekhar} we have the solution to this as
    \begin{equation}
        u=-\frac{1}{6M}+\frac{1}{2M}\tanh^2{\frac{1}{2}(\phi-\phi_0)}
        \label{u=critical sol}
    \end{equation}
where $\phi_0$ is a constant of integration, given by:
    \begin{equation}
        \tanh^2{(-\frac{1}{2}\phi_0)}=\frac{1}{3},
    \end{equation}
which gives: $u=0$ ($r\rightarrow\infty$) when $\phi=0$, and
$u=\frac{1}{3M}$ when $\phi\rightarrow\infty$. Therefore a null
geodesic arriving from infinity with an impact parameter
$D=(3\sqrt{3})M$ approaches the circle of radius $3M$,
asymptotically, by spiralling around it, as can be seen in
Fig.~\ref{u vs phi with D=critical}\footnote{This figure was plotted
using Eqn~(\ref{u=critical sol}).  We also obtained the same plot by
numerically solving Eqn~(\ref{du/dphi for D=critical}) in
Mathematica.}. Also, numerically solving Eqn~(\ref{du/dt=f(u)}) we
can show that as time increases $u$ tends to $\frac{1}{3M}$, which
can be seen in Fig.~\ref{u vs t with D=critical}.
\begin{figure}
    \begin{center}
        \includegraphics[width=0.9\textwidth]{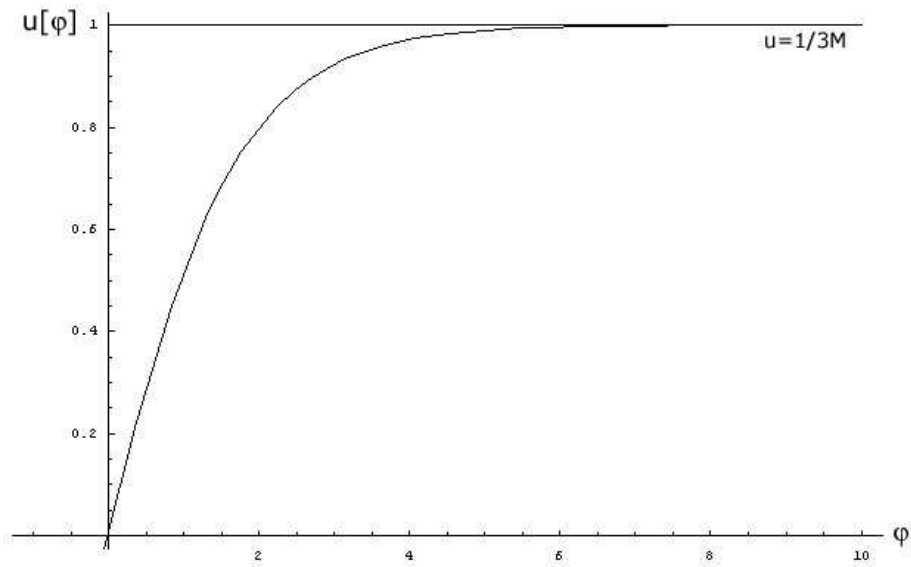}\\
        \caption{Null geodesic, with impact parameter $D=(3\sqrt{3})M$,
        arriving from infinity and approaching $u=\frac{1}{3M}$ asymptotically (M=1/3)}
        \label{u vs phi with D=critical}
    \end{center}
\end{figure}
\begin{figure}
    \begin{center}
        \includegraphics[width=0.9\textwidth]{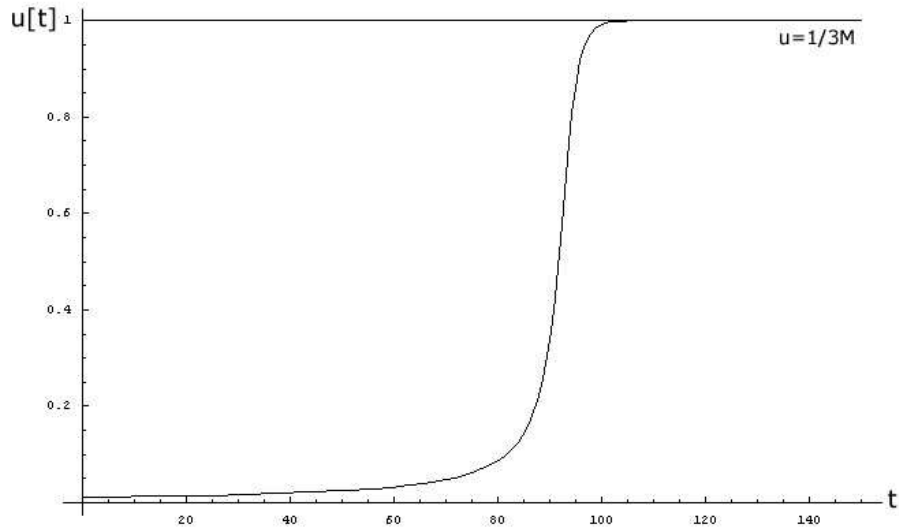}\\
        \caption{For impact parameter $D=(3\sqrt{3})M$ a null trajectory tends to
        $u=\frac{1}{3M}$ asymptotically with time.}
        \label{u vs t with D=critical}
    \end{center}
\end{figure}

Finally we can show that when an impact parameter other than
$D=(3\sqrt{3})M$ is used, for example if we set
$D=(3\sqrt{3})M-0.1$, the solution of Eqn.~(\ref{du/dphi=f(u)})
shows that the photon falls past the critical orbit, through the
event horizon, and into the singularity, Fig.~\ref{u vs phi with
D=critical-0.1}. Using this new impact parameter in
Eqn.~(\ref{du/dt=f(u)}) and, again solving numerically, we see that
the photon comes in from infinity and tends to the event horizon,
$u=\frac{1}{2M}$, asymptotically in time $t$, Fig.~\ref{u vs t with
D=critical-0.1}.
\begin{figure}
    \begin{center}
        \includegraphics[width=0.9\textwidth]{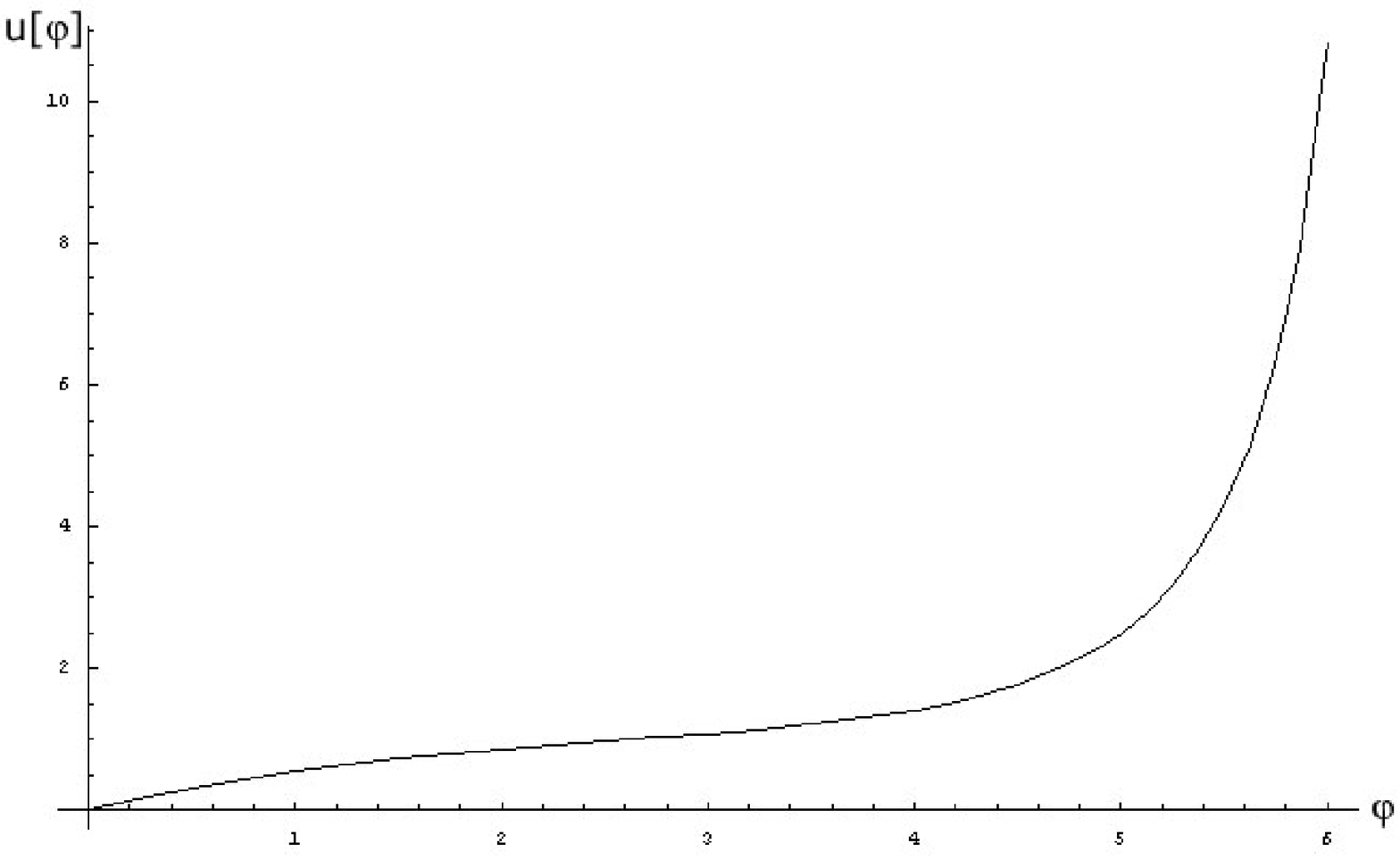}\\
        \caption{Null geodesic, with impact perimeter
        $D=(3\sqrt{3})M-0.1$, arrives from infinity and spirals into the singularity.}
        \label{u vs phi with D=critical-0.1}
    \end{center}
\end{figure}
\begin{figure}
    \begin{center}
        \includegraphics[width=0.9\textwidth]{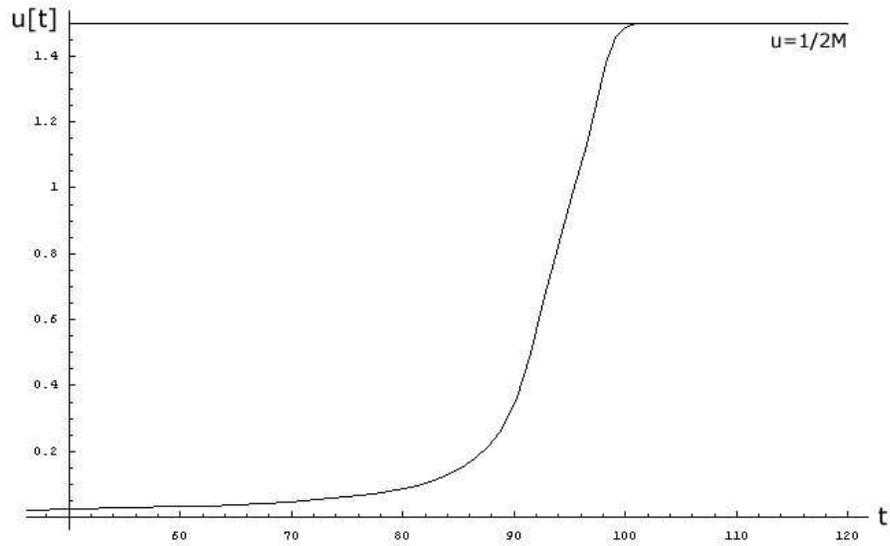}\\
        \caption{Null geodesic, with impact perimeter
        $D=(3\sqrt{3})M-0.1$, arrives from infinity and tends to $u=\frac{1}{2M}$ asymptotically in time $t$.}
        \label{u vs t with D=critical-0.1}
    \end{center}
\end{figure}
}

{\typeout{Quantum Gravitational Optics}
\chapter{Quantum Gravitational Optics}
\label{optics}
\section{Photon Propagation in Curved Spacetime}
Maxwell's equations, in curved spacetime,
\begin{equation}
0=D_{\mu}F^{\mu\nu} \label{max1}
\end{equation}
\begin{eqnarray}
\label{max2}
0&=&D_{\mu}F_{\nu\lambda}+D_{\nu}F_{\lambda\mu}+D_{\lambda}F_{\mu\nu}\\
\label{elect-vec} &\Rightarrow&
F_{\mu\nu}=\partial_{\mu}A_{\nu}-\partial_{\nu}A_{\mu},
\end{eqnarray}
cannot be solved explicitly, and even in cases of extreme symmetry
explicit solutions are difficult; this is due to the fact that as
curved space acts as a dispersive material (i.e. bending light rays)
plane wave solutions do not exist.
\subsection{Geometric Optics in Curved Spacetime}
In dispersive materials, where light rays are bent, we can consider
the solution of Maxwell's equations to be a simple perturbation of
the plane wave solution. For example in curved space, relative to an
observer, the electromagnetic waves can appear to be plane and
monochromatic on a scale that is much larger compared to the typical
wavelength, but very small compared with the typical radius of
curvature of space time. Such "locally plane" waves can be
represented, in geometric optics, by approximate solutions of
Maxwell's equations of the form\cite{schneider}:
\begin{equation}
\mathcal{F}^{\mu\nu}=Re(F_{1}^{\mu\nu}+i\varepsilon
F_{2}^{\mu\nu}+\ldots)e^{i\frac{\theta}{\varepsilon}} \label{faraday
tensor}
\end{equation}
and the electromagnetic field vector, defined by
Eqn.~(\ref{elect-vec}), takes the form:
\begin{equation}
\mathcal{A}^{\mu}=Re(A_1^{\mu}+i\varepsilon
A_2^{\mu}+\ldots)e^{i\frac{\theta}{\varepsilon}}
\end{equation}
where the electromagnetic field is written as a slowly-varying
amplitude and a rapidly-varying phase. The parameter $\varepsilon$
is introduced in order to keep track of the relative order of
magnitude of terms, so in curved space Maxwell's equations can be
solved order-by-order in $\varepsilon$.  In this formulation the
wave vector is then defined as the gradient of the phase of the
field,
$k_{\mu}=D_{\mu}\frac{i}{\varepsilon}\theta=\partial_{\mu}\frac{i}{\varepsilon}\theta$,
which in terms of the quantum interpretation is identified as the
photon momentum. We can also write $A^{\mu}=Aa^{\mu}$, where $A$
represents the amplitude and $a_{\mu}$ (normalized as
$a_{\mu}a^{\mu}=-1$) specifies the wave polarization.  These vectors
then satisfy the condition $k_{\mu}a^{\mu}=0$.
\subsubsection{Geometric Optics and Null Dynamics}
In this notation Eqn.~(\ref{max1}) can be written, to leading order
$\mathcal{O}(\frac{1}{\varepsilon})$, as:
\begin{eqnarray}
\partial_{\mu}(F_1^{\mu\nu}e^{i\frac{\theta}{\varepsilon}})&=&
F_1^{\mu\nu}e^{i\frac{\theta}{\varepsilon}}(\frac{i}{\varepsilon}\partial_{\mu}\theta)\nonumber\\
\Rightarrow \qquad k_{\mu}F_1^{\mu\nu}&=&0
\end{eqnarray}
and Eqn.~(\ref{elect-vec}) becomes:
\begin{eqnarray}
F_{1\mu\nu} e^{i\frac{\theta}{\varepsilon}}&=&(\partial_{\mu}A_{1\nu}-\partial_{\nu}A_{1\mu})e^{i\frac{\theta}{\varepsilon}}\nonumber\\
&=&\frac{i}{\varepsilon}(\partial_{\mu}\theta
A_{1\nu}-\partial_{\nu}\theta
A_{1\mu})e^{i\frac{\theta}{\varepsilon}}\nonumber\\
\Rightarrow \qquad F_{1\mu\nu}&=&k_{\mu}A_{1\nu}-k_{\nu}A_{1\mu}
\label{elect-vect-geometric}
\end{eqnarray}
Now, by combining these, we have:
\begin{eqnarray}
k_{\mu}F_1^{\mu\nu}&=&k_{\mu}(k^{\mu}A_1^{\nu}-k^{\nu}A_1^{\mu})=0\nonumber\\
&=&(k_{\mu}k^{\mu})A_1^{\nu}-k^{\nu} (k_{\mu}A_1^{\mu})\nonumber\\
 \Rightarrow \qquad k^2a^{\nu}&=&0
 \end{eqnarray}
and from this we can deduce that $k^2=0$, i.e. $k^{\mu}$ is a null
vector.  Also, it follows from the definition of $k^{\mu}$ as a
gradient that $D_{\mu}k_{\nu}=D_{\nu}k_{\mu}$, so
\begin{equation}
k^{\mu}D_{\mu}k^{\nu}=k^{\mu}D^{\nu}k_{\mu}=\frac{1}{2}D^{\nu}k^2=0
\label{condition for geodesic}
\end{equation}
Using this, and the fact that light rays are defined as the curves
given by $x_{\mu}(s)$ where $\frac{dx_{\mu}}{ds}=k_{\mu}$, we can
derive the geodesic equation as follows\cite{graham}:
\begin{eqnarray}
0&=&k^{\mu}D_{\mu}k^{\nu}\nonumber\\
&=&\frac{d^2x^{\nu}}{ds^2}+\Gamma^{\nu}_{\mu\lambda}\frac{dx^{\mu}}{ds}\frac{dx^{\lambda}}{ds}
\end{eqnarray}
\section{Quantum Modified Null Dynamics}
As was seen in Sec.~\ref{qed curved space}, using the effective
action, Eqn.~(\ref{total action}), the equation of motion becomes:
\begin{equation}
0=D_{\mu}F^{\mu\nu}-\frac{1}{m^2}[2bR_{\mu\lambda}D^{\mu}F^{\lambda\nu}+4cg^{\nu\tau}R_{\mu\tau\lambda\rho}D^{\mu}F^{\lambda\rho}]
\label{quantum modified maxwell equation}
\end{equation}
and the Bianchi identity, Eqn.~(\ref{max2}), remains unchanged.
Now, as before we can determine the quantum modified light cone and
geodesic equations.
\subsubsection{Quantum Modified Light Cone}
Substituting Eqn.~(\ref{faraday tensor}) in (\ref{quantum modified
maxwell equation}) we find, again to
$\mathcal{O}(\frac{1}{\varepsilon})$:
\begin{eqnarray}
0=\frac{i}{\varepsilon}(D_{\mu}\theta)e^{i\frac{\theta}{\varepsilon}}
F_1^{\mu\nu}&-&\frac{1}{m^2}[2bR_{\mu\lambda}\frac{i}{\varepsilon}(D^{\mu}\theta)e^{i\frac{\theta}{\varepsilon}}
F_1^{\lambda\nu}\nonumber\\
&+&4cg^{\nu\tau}R_{\mu\tau\lambda\rho}\frac{i}{\varepsilon}(D^{\mu}\theta)e^{i\frac{\theta}{\varepsilon}}
F_1^{\lambda\rho}]\nonumber
\end{eqnarray}
\begin{eqnarray}
\Rightarrow \qquad
0=k_{\mu}F_1^{\mu\nu}-\frac{1}{m_e^2}[2bR_{\mu\lambda}k^{\mu}F_1^{\lambda\nu}+4cg^{\nu\tau}R_{\mu\tau\lambda\rho}k^{\mu}F_1^{\lambda\rho}]
\end{eqnarray}
and using Eqn.~(\ref{elect-vect-geometric}) we have:
\begin{eqnarray}
 0=k_{\mu}( k^{\mu}A_1^{\nu}-k^{\nu}A_1^{\mu})
&-&\frac{1}{m_e^2}[2bR_{\mu\lambda}k^{\mu}
(k^{\lambda}A_1^{\nu}-k^{\nu}A_1^{\lambda})\nonumber\\
&+&4cg^{\nu\tau}R_{\mu\tau\lambda\rho}k^{\mu}
(k^{\lambda}A_1^{\rho}-k^{\rho}A_1^{\lambda}) ]\nonumber
\end{eqnarray}
using $k_{\mu}A^{\mu}=0$ this becomes:
\begin{eqnarray}
0=k_{\mu}k^{\mu}A_1^{\nu}&-& \frac{1}{m_e^2}[2bR_{\mu\lambda}k^{\mu}
(k^{\lambda}A_1^{\nu}-k^{\nu}A_1^{\lambda})\nonumber\\
&+&4cg^{\nu\tau}R_{\mu\tau\lambda\rho}k^{\mu}
(2k^{\lambda}A_1^{\rho}) ]
\end{eqnarray}
where $A_1^{\nu}=A_1a^{\nu}$, and the last line is simplified by
relabeling of indices. Now, Contracting with $Aa_{\nu}$ and
eliminating $A$ we have:
\begin{eqnarray} 0=-k_{\mu}k^{\mu}+ \frac{2b}{m_e^2}R_{\mu\lambda}k^{\mu}
k^{\lambda}-\frac{8c}{m_e^2}R_{\mu\tau\lambda\rho}k^{\mu}
k^{\lambda}a_1^{\tau}a_1^{\rho}\nonumber
\end{eqnarray}
This then gives the quantum modified light cone:
\begin{eqnarray}
k^2-\frac{2b}{m^2}R_{\mu\lambda}k^{\mu} k^{\lambda}
-\frac{8c_{\alpha}}{m_{e}^2}R_{\mu\tau\lambda\rho}k^{\mu}
k^{\lambda}a^{\tau}a^{\rho}=k^2-\delta k(a)=0 \label{quantum
modified light cone equation}
    \end{eqnarray}
where we have replaced the constant $c=-\frac{\alpha}{360\pi}$ with
$c_{\alpha}=\frac{\alpha}{360\pi}$ for convenience of
interpretation, i.e. the sign of the light cone is immediately
obvious from $R_{\mu\tau\lambda\rho}k^{\mu}
k^{\lambda}a^{\tau}a^{\rho}$. As we will be working in the
Schwarzschild spacetime the quantum modified light cone for the
Ricci flat case ($R_{\mu\lambda}=0$) is given by:
\begin{equation}
k^2=\frac{8c_{\alpha}}{m_{e}^2}R_{\mu\tau\lambda\rho}k^{\mu}
k^{\lambda}a^{\tau}a^{\rho}=\delta k (a) \label{quantum modified
light cone correction}
\end{equation}
Here the sign of the light cone depends on polarization and the
photon trajectory; if the correction is positive we have space-like
motion, and if it's negative we have time-like motion.
\subsubsection{Quantum Modified Geodesic Equation}
The photon trajectories corresponding to the quantum modified
equation of motion, (\ref{quantum modified maxwell equation}), can
be represented by a generalised version of Eqn.~(\ref{condition for
geodesic}):
\begin{eqnarray}
0&=&\frac{1}{2}D_{\nu}[ k^2-\frac{2b}{m^2}R_{\mu\lambda}k^{\mu}
k^{\lambda}
-\frac{8c_{\alpha}}{m_{e}^2}R_{\mu\tau\lambda\rho}k^{\mu}
k^{\lambda}a^{\tau}a^{\rho}]\nonumber\\
0&=&\frac{1}{2}D_{\nu}k^2-\frac{1}{m^2}D_{\nu}[bR_{\mu\lambda}k^{\mu}
k^{\lambda} +4c_{\alpha}R_{\mu\tau\lambda\rho}k^{\mu}
k^{\lambda}a^{\tau}a^{\rho}]\nonumber\\
\Rightarrow \qquad
0&=&\frac{d^2x^{\nu}}{ds^2}+\Gamma_{\mu\lambda}^{\nu}\frac{dx^\mu}{ds}\frac{dx^\lambda}{ds}\nonumber\\
&-&\frac{1}{m^2}\partial_{\nu}[(bR_{\beta\gamma}+4c_{\alpha}R_{\beta\sigma\gamma\tau}a^\sigma
a^\tau)\frac{dx^\beta}{ds}\frac{dx^\gamma}{ds}]
    \label{quantum modified geodesic equation}
\end{eqnarray}
where we have used $k^{\mu}=\frac{dx^{\mu}}{ds}$, and covariant
derivative in the second term is replaced by a partial derivative as
it's acting on a scalar.  In Ricci spacetime this equation becomes:
\begin{equation}
0=\frac{d^2x^{\nu}}{ds^2}+\Gamma_{\mu\lambda}^{\nu}\frac{dx^\mu}{ds}\frac{dx^\lambda}{ds}-\frac{4c_{\alpha}}{m^2}\partial_{\nu}[R_{\beta\sigma\gamma\tau}a^\sigma
a^\tau\frac{dx^\beta}{ds}\frac{dx^\gamma}{ds}]
    \label{quantum modified geodesic equation ricci}
\end{equation}
\section{Horizon Theorem and Polarization Rule}
There are two general features associated with quantum modified
photon propagation\cite{graham}.  First, it is a general result that
the velocity of radially directed photons remains equal to $c$ at
the event horizon. Second, for Ricci flat spacetimes (such as
Schwarzschild\cite{drummond} and Kerr\cite{daniels} spacetimes), the
velocity shifts for the two transverse polarizations are always
equal and opposite.  However, this is no longer true for non-Ricci
flat cases (such as Robertson-Walker spacetime\cite{drummond}).  In
these cases, the polarization averaged velocity shift is
proportional to the matter energy-momentum tensor. These features
can be easily shown by using the Newman-Penrose formalism: this
characterises spacetimes using a set of complex scalars, which are
found by contracting the Weyl tensor with elements of a null
tetrad\cite{chandrasekhar}.
\subsubsection{Newman-Penrose Formalism}
We choose the basis vectors of the null tetrad as\cite{graham}:
$l^{\mu}=k^{\mu}$, the photon momentum.  Then, we denote the two
spacelike, normalized, transverse polarization vectors by $a^{\mu}$
and $b^{\mu}$ and construct the null vectors
$m^{\mu}=\frac{1}{\sqrt{2}}(a^{\mu}+ib^{\mu})$ and
$\bar{m}^{\mu}=\frac{1}{\sqrt{2}}(a^{\mu}-ib^{\mu})$.  We complete
the tetrad with a further null vector $n^{\mu}$, which is orthogonal
to $m^{\mu}$ and $\bar{m}^{\mu}$.  We then have the conditions:
\begin{equation}
l\cdot m=l \cdot \bar{m}=n\cdot m=n\cdot \bar{m}=0
\end{equation}
from orthogonality, and:
\begin{equation}
l\cdot l=n \cdot n=m\cdot m=\bar{m}\cdot \bar{m}=0
\end{equation}
since the basis vectors are null.  Finally, we impose:
\begin{equation}
l\cdot n = -m\cdot \bar{m}=1
\end{equation}
The Weyl tensor, given in terms of the Riemann and Ricci tensors,
is:
\begin{eqnarray}
C_{\mu\nu\gamma\delta}=R_{\mu\nu\gamma\delta}&-&\frac{1}{2}(\eta_{\mu
\gamma}R_{\nu\delta}-\eta_{\nu\gamma}R_{\mu\delta}-
\eta_{\mu\delta}R_{\nu\gamma}+\eta_{\nu\delta}R_{\mu\gamma})\nonumber\\
&+&\frac{1}{6}(\eta_{\mu\gamma}\eta_{\nu\delta}-\eta_{\mu\delta}\eta_{\nu\gamma})R
\label{weyl Tensor}
\end{eqnarray}
where $R_{\mu\gamma}=\eta^{\nu\delta}R_{\mu\nu\gamma\delta}$ and
$R=\eta^{\mu\nu}R_{\mu\nu}$; and the Weyl tensor satisfies the
trace-free condiation:
\begin{equation}
\eta^{\mu\delta}C_{\mu\nu\gamma\delta}=0
\end{equation}
and the cyclicity property:
\begin{equation}
C_{1234}+C_{1342}+C_{1423}=0
\end{equation}
Now, using the null tetrad, we can denote the ten independent
components of the Weyl tensor by the five complex Newman-Penrose
scalars:
\begin{eqnarray}
\Psi_0&=&-C_{\mu\nu\gamma\delta}l^{\mu}m^{\nu}l^{\gamma}m^{\delta}\nonumber\\
\Psi_1&=&-C_{\mu\nu\gamma\delta}l^{\mu}n^{\nu}l^{\gamma}m^{\delta}\nonumber\\
\Psi_2&=&-C_{\mu\nu\gamma\delta}l^{\mu}m^{\nu}\bar{m}^{\gamma}n^{\delta}\nonumber\\
\Psi_3&=&-C_{\mu\nu\gamma\delta}l^{\mu}n^{\nu}\bar{m}^{\gamma}n^{\delta}\nonumber\\
\Psi_4&=&-C_{\mu\nu\gamma\delta}n^{\mu}\bar{m}^{\nu}n^{\gamma}\bar{m}^{\delta}
\label{penrose scalars}
\end{eqnarray}
\subsection{Polarization Sum Rule}
\subsubsection{Ricci Flat Spacetime} In Ricci flat spacetime, summing
the quantum correction over the two polarizations leads to the
following polarization sum rule:
\begin{equation}
\sum_{a}\delta k(a)=0
\end{equation}
This can be proven by suming the quantum correction in
Eqn.~(\ref{quantum modified light cone correction}) over the two
polarizations,
\begin{equation}
\sum_{a}\delta
k(a)=\frac{8c_{\alpha}}{m_e^2}\sum_{a}R_{\mu\nu\gamma\delta}k^{\mu}k^{\gamma}a^{\nu}a^{\delta}
\end{equation}
In the Newman-Penrose basis, using $k=l$,
$a=\frac{\sqrt{2}}{2}(m+\bar{m})$,
$b=-\frac{i\sqrt{2}}{2}(m-\bar{m})$, and the fact that
$C_{\mu\nu\gamma\delta}=R_{\mu\nu\gamma\delta}$ for the Ricci flat
case, we have:
\begin{eqnarray}
\sum_{a}R_{\mu\nu\gamma\delta}k^{\mu}k^{\gamma}a^{\nu}a^{\delta}&=&\frac{1}{2}C_{\mu\nu\gamma\delta}l^{\mu}l^{\gamma}(m^{\nu}
+\bar{m}^{\nu})(m^{\delta}+\bar{m}^{\delta})\nonumber\\
&-&\frac{1}{2}C_{\mu\nu\gamma\delta}l^{\mu}l^{\gamma}(m^{\nu}-\bar{m}^{\nu})(m^{\delta}-\bar{m}^{\delta})\nonumber\\
&=&C_{\mu\nu\gamma\delta}l^{\mu} l^{\gamma} (m^{\nu}\bar{m}^{\delta}
+\bar{m}^{\nu}m^{\delta})
\end{eqnarray}
This particular contraction is equal to zero as it's not part of the
complex scalars in Eqns.~(\ref{penrose scalars}); hence the sum of
the two quantum corrections is zero. This implies the trajectory
(and velocity) shifts are equal and opposite.
\subsubsection{Non-Ricci Flat Spacetime} For the non-Ricci flat
spacetimes the polarization sum rule is given as:
\begin{equation}
\sum_{a}\delta
k(a)=-\frac{8\pi}{m_e^2}(2b-8c_{\alpha})T_{\mu\nu}k^{\mu}k^{\nu}
\label{polar-sum rule-non-ricci flat}
\end{equation}
where $T_{\mu\nu}$ is the energy-momentum tensor. This can be shown
by proceeding as before, but now we include the Ricci tensor and
scalar, as in Eqn.~(\ref{weyl Tensor}):
\begin{equation}
\sum_{a}R_{\mu\nu\gamma\delta}k^{\mu}k^{\gamma}a^{\nu}a^{\delta}=C_{\mu\nu\gamma\delta}l^{\mu}l^{\gamma}(m^{\nu}\bar{m}^{\delta}
+\bar{m}^{\nu}m^{\delta})-R_{\mu\gamma}l^{\mu}l^{\gamma}
\end{equation}
As before the first term on the RHS is zero, and the second term is
only dependent on the photon momentum. Then, combining this with the
Ricci term in Eqn.~(\ref{quantum modified light cone equation}) we
have:
\begin{equation}
\sum_{a}\delta
k(a)=\frac{1}{m_e^2}(2b-8c_{\alpha})R_{\mu\nu}k^{\mu}k^{\nu}
\end{equation}
Finally replacing the Ricci tensor with the energy-momentum tensor,
by using the Einstein equation
\begin{equation}
R_{\mu\nu}=-8\pi T_{\mu\nu}+\frac{1}{2}Rg_{\mu\nu}
 \label{einstein}
\end{equation}
we obtain Eqn.~(\ref{polar-sum rule-non-ricci flat}).
\subsection{Horizon Theorem}
At the event horizon, photons with momentum directed normal to the
horizon have velocity equal to $c$, i.e. the light cone remains
$k^2=0$, independent of polarization\cite{graham}.
\\
\\
This can be easily proven for the Ricci flat spacetime, using the
orthonormal vectors $k^a=(E_k,E_k,0,0)$, $a^b=(0,0,1,0)$ and
$a^b=(0,0,0,1)$.  Therefore, using these vectors in
Eqn.~(\ref{quantum modified light cone correction}), we have:
\begin{equation}
k^2=\frac{8c_{\alpha}}{m_{e}^2}R_{abcd}k^ak^ca^ba^d=0
\end{equation}
So, in Ricci flat spacetime all radially projected photon
trajectories remain unchanged.

It is also possible to prove the horizon theorem for the general
case (for Ricci and non-Ricci flat spacetimes) that the light cone
at the event horizon is unchanged.  This can be seen in the null
tetrad, so that the physical, space-like, polarization vectors
$a^{\mu}$ and $b^{\mu}$ lie parallel to the event horizon 2-surface,
while $k^{\mu}$ is the null vector normal to the surface. Then, from
Eqn.~(\ref{quantum modified light cone equation}) we have for the
two polarizations:
\begin{eqnarray}
k^2&=&\frac{2b}{m_e^2}R_{\mu\gamma}k^{\mu}k^{\gamma}+\frac{8c_{\alpha}}{m_{e}^2}R_{\mu\nu\gamma\delta}k^{\mu}k^{\gamma}a^{\nu}a^{\delta}\nonumber\\
&=&\frac{2b}{m_e^2}R_{\mu\gamma}l^{\mu}l^{\gamma}+\frac{8c_{\alpha}}{m_{e}^2}[-\frac{1}{2}R_{\mu\gamma}l^{\mu}l^{\gamma}\pm C_{\mu\nu\gamma\delta}l^{\mu}l^{\gamma}\frac{1}{2}(m^{\nu}\pm\bar{m}^{\nu})(m^{\delta}\pm\bar{m}^{\delta})]\nonumber\\
&=&\frac{1}{m_e^2}(2b-4c_{\alpha})R_{\mu\gamma}l^{\mu}l^{\gamma}\pm\frac{4c_{\alpha}}{m_e^2}C_{\mu\nu\gamma\delta}l^{\mu}l^{\gamma}\frac{1}{2}(m^{\nu}\pm\bar{m}^{\nu})(m^{\delta}\pm\bar{m}^{\delta})
\end{eqnarray}
Using Eqn.~(\ref{einstein}) and the fact that
$C_{\mu\nu\gamma\delta}l^{\mu}l^{\gamma}(m^{\nu}\bar{m}^{\delta}+\bar{m}^{\nu}m^{\delta})=0$,
we can write this as:
\begin{eqnarray}
k^2&=&-\frac{8\pi}{m_e^2}(2b-4c_{\alpha})T_{\mu\gamma}l^{\mu}l^{\gamma}\pm\frac{4c_{\alpha}}{m_e^2}C_{\mu\nu\gamma\delta}l^{\mu}l^{\gamma}(m^{\nu}m^{\delta}+\bar{m}^{\nu}\bar{m}^{\delta})
\end{eqnarray}
and in terms of the Newman-Penrose scalars this can be written as:
\begin{eqnarray}
k^2&=&-\frac{8\pi}{m_e^2}(2b-4c_{\alpha})T_{\mu\gamma}l^{\mu}l^{\gamma}\pm\frac{8c_{\alpha}}{m_e^2}\Psi_0,
\label{equation}
\end{eqnarray}
where the simplification in the last term on the RHS is possible as
$\Psi_0$ is real for Schwarzschild spacetime. In general
Eqn.~(\ref{equation}) is non zero, however, at the event horizon the
terms: $T_{\mu\gamma}l^{\mu}l^{\gamma}$ and $\Psi_0$ are zero for
stationary spacetimes\cite{graham,hawking}\footnote{Stationary
spacetimes are independent of time and may or may not be symmetric
under time reversal}.

Physically the Ricci term represents the flow of matter across the
horizon and the Weyl term represents the flow of gravitational
radiation\cite{graham}, and both are zero in classic general
relativity; and as, even with the quantum modification, the light
cone remains unchanged at the event horizon, means the event horizon
is fixed and light cannot escape from inside the black hole.
}

{\typeout{Quantum Modified Trajectories}
\chapter{Quantum Modified Trajectories}
\label{quant-mod-trajectories-chapter} In this chapter we will
analyse how the classical null trajectories, described in
Chapter~\ref{null dynamics}, are modified in Schwarzschild
spacetime, due to quantum modifications of the equations of motion
of general relativity. Using Eqn.~(\ref{quantum modified light cone
equation}) we will calculate the quantum corrected version of
Eqn.~(\ref{du/dphi=f(u)}), which will then describe the quantum
modified motion of a null trajectory in a Schwarzschild spacetime.
Using this, and following the methods of Chapter~\ref{null
dynamics}, we will begin by studying simple critical circular orbits
at the stationary point, $u=1/3M$.  As stated in the polarization
rule, the critical orbit should be shifted up and down by equal
amounts, depending on the polarization of each photon. Also, we will
show that these modifications are only valid if the "classic" impact
parameter corresponding to the stationary orbit is adjusted,
depending on the photon's polarization, in order to compensate for
the orbit shift. This information, of the modified orbits and the
corresponding impact parameter for each polarization, will then be
used to determine the general trajectory of a (vertically or
horizontally polarized) photon coming in from infinity and tending
to one of the two shifted critical orbits. We will then go on to
show that these quantum modifications have no effect on the event
horizon, that is, when the impact parameter is accordingly adjusted
and a photon falls into the singularity the event horizon remains
fixed at $u=1/2M$.
\section{Quantum Modified Circular Orbits}
In Schwarzschild spacetime, using $k^{\nu}=dx^{\nu}/d\tau$,
Eqn.~(\ref{quantum modified light cone correction}) can be written
as:
    \begin{eqnarray}
            0&=&\dot{r}^2-(1-\frac{2M}{r})^2\dot{t}^2+r^2(1-\frac{2M}{r})\dot{\theta}^2\nonumber\\
            &+&(1-\frac{2M}{r})r^2\dot{\phi}^2+\frac{8c_{\alpha}}{m_{e}^2}(1-\frac{2M}{r})R_{\mu\nu\gamma\delta}
            k^{\mu} k^{\gamma} a^{\nu} a^{\delta}
            \label{quantum light cone}
    \end{eqnarray}
and Eqn.~(\ref{quantum modified geodesic equation ricci}) as:
    \begin{eqnarray}
            0&=&\ddot{t}+\frac{2M}{r^2}(1-\frac{2M}{r})^{-1}\dot{r}\dot{t}\nonumber\\
            &-&\frac{4c_{\alpha}}{m_e^2}\partial_t
            (R_{\mu\nu\gamma\delta}a^{\nu}
            a^{\delta}\frac{dx^{\mu}}{d\tau}\frac{dx^{\gamma}}{d\tau})
            \label{quantum trajectory t}
    \end{eqnarray}
    \begin{eqnarray}
            0&=&\ddot{r}-\frac{M}{r^2}(1-\frac{2M}{r})^{-1}\dot{r}^2-r(1-\frac{2M}{r})\dot{\phi}^2\nonumber\\
            &+&\frac{M}{r^2}(1-\frac{2M}{r})\dot{t}^2
            -\frac{4c_{\alpha}}{m_e^2}\partial_r(R_{\mu\nu\gamma\delta}a^{\nu}
            a^{\delta}\frac{dx^{\mu}}{d\tau}\frac{dx^{\gamma}}{d\tau})
             \label{quantum trajectory r}
    \end{eqnarray}
    \begin{eqnarray}
            0&=&\ddot{\phi}+\frac{2\dot{r}\dot{\phi}}{r}-\frac{4c_{\alpha}}{m_e^2}\partial_\phi
            (R_{\mu\nu\gamma\delta}a^{\nu}
            a^{\delta}\frac{dx^{\mu}}{d\tau}\frac{dx^{\gamma}}{d\tau})
             \label{quantum trajectory phi}
    \end{eqnarray}
   \begin{eqnarray}
          0&=&-\frac{4c_{\alpha}}{m_e^2}\partial_\theta
            (R_{\mu\nu\gamma\delta}a^{\nu}
            a^{\delta}\frac{dx^{\mu}}{d\tau}\frac{dx^{\gamma}}{d\tau})
             \label{quantum trajectory theta}
    \end{eqnarray}
   where $c_{\alpha}=\frac{\alpha}{360\pi}$.  Now, in order to consider quantum modified circular orbits we
require three things: (i) the Riemann tensor components, (ii) the
photon wave vectors, $k^\mu=\frac{dx^\mu}{d\tau}$, for circular
orbits, and finally (iii) the photon polarization vectors, $a^\mu$.
Due to the circular nature of the orbit the simplest basis to work
in is the orthonormal basis. In this frame the required polarization
and wave vectors, for circular orbits, are simply given as:
    \begin{equation}
        a^\mu=(0,1,0,0) \qquad \textrm{Planar Polarized}
        \label{plane polarized a}
    \end{equation}
    \begin{equation}
        a^\mu=(0,0,1,0)  \qquad \textrm{Vertically Polarized}
         \label{vertcal polarized a}
    \end{equation}
    \begin{equation}
        k^\mu=\frac{dx^\mu}{d\tau}=(E_k,0,0,E_k) \qquad \textrm{$4$-momentum
        along $\phi$}
        \label{k photon wave vector}
    \end{equation}
Now using these photon vectors, the quantum modifications in
Eqns.~(\ref{quantum light cone})-(\ref{quantum trajectory theta}),
can be expanded to give:
\begin{itemize}
    \item For the planar polarized case we have:
    \begin{eqnarray}
        R_{abcd}k^a k^c a^b a^d&=&R_{arcr} k^a k^c=E_k^2(R_{arcr} \hat{k}^a
        \hat{k}^c)\nonumber\\
        &=&E_k^2(R_{trtr}+R_{\phi r \phi r}+2R_{tr\phi r})
        \label{planar modification}
    \end{eqnarray}
    \item For the vertically polarized case we have:
    \begin{eqnarray}
        R_{abcd}k^a k^c a^b a^d&=&R_{a\theta c \theta} k^a k^c=E_k^2(R_{a \theta c \theta} \hat{k}^a
        \hat{k}^c)\nonumber\\
        &=&E_k^2(R_{t \theta t \theta
        }+R_{\phi\theta\phi\theta})
        \label{vertical modification}
    \end{eqnarray}
\end{itemize}
Using the six independent components of the Riemann tensor in the
orthonormal basis\footnote{Which we have calculated in Appendix A},
we find:
    \begin{equation}
        R_{trrt}=-R_{trtr}=-\frac{2M}{r^3} \Rightarrow
        R_{trtr}=\frac{2M}{r^3}
    \end{equation}
    \begin{equation}
        R_{\phi rr \phi}=-R_{\phi r \phi r}=-\frac{M}{r^3} \Rightarrow
        R_{\phi r \phi r}=\frac{M}{r^3}
    \end{equation}
    \begin{equation}
        R_{\theta t \theta t}=R_{t\theta t \theta}=-\frac{M}{r^3} \Rightarrow
        R_{t\theta t\theta}=-\frac{M}{r^3}
    \end{equation}
    \begin{equation}
        R_{\phi \theta \theta \phi}=-R_{\phi \theta \phi \theta}=\frac{2M}{r^3} \Rightarrow
        R_{\phi \theta \phi \theta}=-\frac{2M}{r^3}
    \end{equation}
    \begin{equation}
        R_{tr\phi r}=0
    \end{equation}
Then Eqns.~(\ref{planar modification}) and (\ref{vertical
modification}) become:
\begin{itemize}
    \item For the planar polarized case:
    \begin{equation}
        R_{abcd}k^a k^c a^b a^d=E_k^2(R_{trtr}+R_{\phi r \phi
        r}+2R_{tr\phi r})=\frac{3E_k^2M}{r^3}
        \label{planar R term}
    \end{equation}
with the relevant derivatives:
    \begin{eqnarray}
        \partial_t\frac{3E_k^2M}{r^3}&=&0 \qquad
        \partial_r\frac{3E_k^2M}{r^3}=-\frac{9E_k^2M}{r^4}\nonumber\\
        \partial_\theta\frac{3E_k^2M}{r^3}&=&0 \qquad
        \partial_\phi\frac{3E_k^2M}{r^3}=0
        \label{planar R derivative term}
    \end{eqnarray}
    \item For the vertically polarized case:
    \begin{equation}
       R_{abcd}k^a k^c a^b a^d=E_k^2(R_{t \theta t \theta
        }+R_{\phi\theta\phi\theta})=-\frac{3E_k^2M}{r^3}
        \label{vertical R term}
    \end{equation}
and the relevant derivatives:
    \begin{eqnarray}
        -\partial_t\frac{3E_k^2M}{r^3}&=&0 \qquad
        -\partial_r\frac{3E_k^2M}{r^3}=\frac{9E_k^2M}{r^4}\nonumber\\
        -\partial_\theta\frac{3E_k^2M}{r^3}&=&0 \qquad
        -\partial_\phi\frac{3E_k^2M}{r^3}=0
        \label{vertical R derivative term}
    \end{eqnarray}
\end{itemize}
From Eqns.~(\ref{quantum modified light cone correction}),
(\ref{planar R term}) and (\ref{vertical R term}) we can see that:
\begin{eqnarray}
k^2&\sim&\frac{3E_k^2M}{r^3} \qquad \textrm{For Planar
Polarization}\\
 k^2&\sim&-\frac{3E_k^2M}{r^3} \qquad \textrm{For
Vertical Polarization}
\end{eqnarray}
This implies that, as the light cone for planar polarization is
positive, it represents a photon trajectory with a speed $<c$, and
as the light cone for vertical polarization is negative, it
represents a photon trajectory with a speed $>c$.
\subsubsection{Planar Polarization}
Considering the planar polarized case first, we can use
Eqns.~(\ref{planar R term}) and (\ref{planar R derivative term}) to
rewrite Eqns.~(\ref{quantum light cone})-(\ref{quantum trajectory
phi}) as:
    \begin{eqnarray}
            0&=&\dot{r}^2-(1-\frac{2M}{r})^2\dot{t}^2+r^2(1-\frac{2M}{r})\dot{\theta}^2\nonumber\\
            &+&(1-\frac{2M}{r})r^2\dot{\phi}^2+\frac{8c_{\alpha}}{m_{e}^2}(1-\frac{2M}{r})\frac{3E_k^2M}{r^3}
            \label{quantum light cone with planar term}
    \end{eqnarray}
    \begin{eqnarray}
            0&=&\ddot{t}+\frac{2M}{r^2}(1-\frac{2M}{r})^{-1}\dot{r}\dot{t}
            \label{quantum trajectory t with planar term}
    \end{eqnarray}
    \begin{eqnarray}
            0&=&\ddot{r}-\frac{M}{r^2}(1-\frac{2M}{r})^{-1}\dot{r}^2-r(1-\frac{2M}{r})\dot{\phi}^2\nonumber\\
            &+&\frac{M}{r^2}(1-\frac{2M}{r})\dot{t}^2+\frac{4c_{\alpha}}{m_e^2}\frac{9E_k^2M}{r^4}
             \label{quantum trajectory r with planar term}
    \end{eqnarray}
    \begin{eqnarray}
            0&=&\ddot{\phi}+\frac{2\dot{r}\dot{\phi}}{r}
             \label{quantum trajectory phi with planar term}
    \end{eqnarray}
where Eqn.~(\ref{quantum trajectory theta}) becomes zero.  Now,
using the solutions of (\ref{quantum trajectory t with planar term})
and (\ref{quantum trajectory phi with planar term}), as given by
(\ref{dot-t}) and (\ref{dot-phi}), we can rewrite Eqn.~(\ref{quantum
light cone with planar term}) as a simple quantum modified
trajectory equation for circular orbits with radius $r$ and in a
plane $\theta=\frac{\pi}{2}$:\footnote{We must note that $E$ and
$E_k$ are not equal, $E$ is the energy from the classic relativistic
orbit equations, while $E_k$ is the quantum energy of the photon}
    \begin{eqnarray}
        0&=&\dot{r}^2-E^2+(1-\frac{2M}{r})\frac{J^2}{r^2}+\frac{8c_{\alpha}}{m_{e}^2}(1-\frac{2M}{r})\frac{3E_k^2M}{r^3}\nonumber\\
        \Rightarrow \qquad E^2&=&(\frac{dr}{d\tau})^2+(1-\frac{2M}{r})(\frac{24cE_k^2M}{m_e^2r^3}+\frac{J^2}{r^2})
        \label{plane equation preliminary}
    \end{eqnarray}
To make this equation more meaningful and easier to solve we make
the following transformations\footnote{Using
$J=r^2\frac{d\phi}{d\tau}$}:
    \begin{equation}
        \tau\rightarrow\phi \qquad
        \frac{dr}{d\phi}\cdot\frac{d\phi}{d\tau}=\frac{dr}{d\phi}\cdot\frac{J}{r^2}
    \end{equation}
    \begin{equation}
        r\rightarrow u=\frac{1}{r} \qquad
        \frac{du}{d\phi}\cdot\frac{
        dr}{du}=\frac{du}{d\phi}\cdot(-\frac{1}{u^2})
    \end{equation}
Therefore, Eqn.~(\ref{plane equation preliminary}) becomes:
    \begin{eqnarray}
        E^2&=&(\frac{dr}{d\phi})^2\cdot(\frac{J}{r^2})^2+(1-\frac{2M}{r})(\frac{24c_{\alpha}E_k^2M}{m_e^2r^3}+\frac{J^2}{r^2})\nonumber\\
        &=&(\frac{du}{d\phi})^2\cdot(-\frac{1}{u^2})^2\cdot(J
        u^2)^2+(1-2Mu)(\frac{24c_{\alpha}E_k^2Mu^3}{m_e^2}+J^2u^2)\nonumber\\
        &=&(\frac{du}{d\phi})^2\cdot J^2+(1-2Mu)(\frac{24c_{\alpha}E_k^2Mu^3}{m_e^2}+J^2u^2)
        \end{eqnarray}
Which can be written as:
    \begin{eqnarray}
        (\frac{du}{d\phi})^2&=&2M^2Au^4+(2M-MA)u^3-u^2+\frac{E^2}{J^2}
            \label{planar r polarization trajectory equation}
    \end{eqnarray}
where we have defined the dimensionless constant:
    \begin{equation}
        A=\frac{24c_{\alpha}E_k^2}{J^2m_e^2}=\frac{72 c_{\alpha}}{m_e^2
        D^2},
        \label{constant A}
    \end{equation}
In the last form we have used (without proof) the relation
$E_k=\sqrt{3}E$, which will be proven in Sec.~\ref{gen-vect},
Eqn.~(\ref{norm}).\footnote{$c_{\alpha}$ is dimensionless, while $D$
and $m_e$ have dimensions of length and inverse-length respectively}
\subsubsection{Vertical Polarization}
Similarly, for the vertically polarized case, using
Eqns.~(\ref{vertical R term}) and (\ref{vertical R derivative term})
in Eqns.~(\ref{quantum light cone})-(\ref{quantum trajectory theta})
to give:
    \begin{eqnarray}
            0&=&\dot{r}^2-(1-\frac{2M}{r})^2\dot{t}^2+r^2(1-\frac{2M}{r})\dot{\theta}^2\nonumber\\
            &+&(1-\frac{2M}{r})r^2\dot{\phi}^2-\frac{8c_{\alpha}}{m_{e}^2}(1-\frac{2M}{r})\frac{3E_k^2M}{r^3}
            \label{quantum light cone with vertical term}
    \end{eqnarray}
    \begin{eqnarray}
            0&=&\ddot{t}+\frac{2M}{r^2}(1-\frac{2M}{r})^{-1}\dot{r}\dot{t}
            \label{quantum trajectory t with vertical term}
    \end{eqnarray}
    \begin{eqnarray}
            0&=&\ddot{r}-\frac{M}{r^2}(1-\frac{2M}{r})^{-1}\dot{r}^2-r(1-\frac{2M}{r})\dot{\phi}^2\nonumber\\
            &+&\frac{M}{r^2}(1-\frac{2M}{r})\dot{t}^2-\frac{4c_{\alpha}}{m_e^2}\frac{9E_k^2M}{r^4}
             \label{quantum trajectory r with vertical term}
    \end{eqnarray}
    \begin{eqnarray}
            0&=&\ddot{\phi}+\frac{2\dot{r}\dot{\phi}}{r}
             \label{quantum trajectory phi with vertical term}
    \end{eqnarray}
which, in a similar way as before, gives us the equation of motion
for the vertically polarized photon:
    \begin{equation}
        (\frac{du}{d\phi})^2=-2M^2Au^4+(2M+MA)u^3-u^2+\frac{E^2}{J^2}
        \label{vertical theta polarisation trajectory equation}
    \end{equation}
\subsection{Quantum Modified Critical Orbits} Now the
general equation for the quantum modified circular orbits is:
    \begin{equation}
        (\frac{du}{d\phi})^2=\pm
        AM(2Mu-1)u^3+2Mu^3-u^2+\frac{1}{D^2}=f(u)
        \label{polarization trajectory equation}
    \end{equation}
where $+$ is for planar polarization in the $r$ direction, $-$ is
for vertical polarization in the $\theta$ direction and $D$ is the
impact parameter.  As $A\rightarrow0$ Eqn.~(\ref{polarization
trajectory equation}) tends to the classic orbit equation in general
relativity, Eqns.~(\ref{du/dphi=f(u)}).  Therefore, in order to
determine the magnitude of the quantum correction we can calculate
the order of $A$ using typical values for $m_e$, $D$ and
$c_{\alpha}$, in Eqn.~(\ref{constant A}). Using\footnote{As we were
working with $G=c=\hbar=1$, we must reintroduce these constants to
obtain the correct order of $A$} $m_e c/h\sim 10^{11}$ for the
electron mass (as it is given as inverse length),
$c_{\alpha}=\alpha/360\pi\sim 10^{-6}$, and the mass of the sun
 inserted into the critical impact
parameter: $D=3\sqrt{3} GM_{\bigodot}/c^2\sim 10^{2}$ (given in
terms of length), this then gives us\footnote{This result is also
shown in \cite{graham2}}:
\begin{eqnarray}
A=\frac{72 c_{\alpha}}{m_e^2
        D^2}\sim\frac{10^{-6}}{(10^{11} 10^{2})^2}\sim 10^{-32}
\label{order of A}
\end{eqnarray}
With the order of $A$ being so small the correction in
Eqn.~(\ref{polarization trajectory equation}) will be tiny compared
to the size of the orbit ($r=3GM/c^2\sim10^2$); therefore the
modified orbits will not differ from the classic critical orbit,
given by Eqn.~(\ref{du/dphi=f(u)}), by very much.

We will now determine the quantum modified critical circular orbits.
In order to solve Eqn.~(\ref{polarization trajectory equation}), we
can use the fact, from the polarization rule, that the modified
orbits should be shifted above and below $u=\frac{1}{3M}$ by equal
amounts depending on polarization.  So we expect a solution of the
type $u=\frac{1}{3M}\pm\delta u$; which means we can try a simple
modified solution of the form:
    \begin{equation}
        u=u_0+k u_1
        \label{trial solution}
    \end{equation}
where $u_0=\frac{1}{3M}$, $u_1$ is the quantum modification, and $k$
is a small constant depending on the quantum modification $A$.
\subsubsection{Planar Polarized Critical Orbit}
Working with Planar polarization, we can substitute the solution
(\ref{trial solution}) into the derivative $df/du$ of
Eqn.~(\ref{polarization trajectory equation}):
    \begin{equation}
        \frac{df}{du}=2AM^2u^3+3A M(2Mu-1)u^2+6Mu^2-2u=0
    \end{equation}
and as $A\ll1$ and $k\sim A$, then the only terms of relevance are
the ones first order in $k$ and $A$, everything else can be assumed
to be $\approx0$.  Therefore, we have\footnote{$u=u_0+ku_1 \qquad
u^2=u_0^2+2ku_1u_0 \qquad u^3=u_0^3+3ku_1u_0^2$}:
    \begin{displaymath}
        2AM^2u_0^3+6AM^2u_0^3-3AMu_0^2+6Mu_0^2+12Mku_1u_0-2u_0-2ku_1=0
    \end{displaymath}
    \begin{equation}
        \Rightarrow u_1=\frac{A M}{6k}u_0^2
        \label{trial solution modification term}
    \end{equation}
Now, substituting this into the trial solution (\ref{trial
solution}), we have:
    \begin{equation}
        u=u_0(1+\frac{A M}{6}u_0)
        \label{modified planar circular solution}
    \end{equation}
where $u_0=\frac{1}{3M}$ is the classic solution.  Therefore the
classic orbit $u_0$ is modified by $\frac{M}{6}u_0$ to first order
in $A$. Also, with this orbit modification we require an associated,
modified, impact parameter, which should take the form:
    \begin{equation}
        \frac{1}{D^2}=\frac{1}{D_0^2}+\beta
        \label{trial impact solution}
    \end{equation}
where $1/D_0^2=1/27M^2$.  The modified impact parameter can be found
by substituting (\ref{modified planar circular solution}) and
(\ref{trial impact solution}) into (\ref{polarization trajectory
equation}) and solving for $\beta$. Doing so, we find\footnote{we,
again, work to first order in A: $u=u_0+A\frac{M}{6}u_0^2$,
$u^2=u_0^2+A\frac{M}{3}u_0^3$, $u^3=u_0^3+A\frac{M}{2}u_0^4$ and
$u^4=u_0^4+A\frac{2M}{3}u_0^5$}:
    \begin{eqnarray}
        (\frac{du}{d\phi})^2&=&2M^2Au^4-AMu^3+2Mu^3-u^2+\frac{1}{D_0^2}+\beta=0\nonumber\\
        &=&2M^2Au_0^4-AMu_0^3+2Mu_0^3+M^2Au_0^4-u_0^2\nonumber\\
        &-&\frac{1}{3}AMu_0^3+\frac{1}{D_0^2}+\beta
    \end{eqnarray}
We have $2Mu_0^3-u_0^2+\frac{E^2}{J^2}=0$, as this forms the classic
equation of motion for circular orbits.  Therefore,
    \begin{displaymath}
        2M^2Au_0^2-AMu_0^3+M^2Au_0^4-\frac{1}{3}AMu_0^3+\beta=0
    \end{displaymath}
Now, as $u_0=\frac{1}{3M}$, we have:
    \begin{equation}
        \beta=\frac{A M}{3}u_0^3
    \end{equation}
Therefore the modified impact parameter, for planar polarization,
is:
    \begin{equation}
        \frac{1}{D^2}=\frac{1}{D_0^2}+\frac{A M}{3}u_0^3
    \end{equation}
Then, substituting for $D_0$, we have:
    \begin{eqnarray}
        \frac{1}{D^2}&=&\frac{1}{3(3M)^2}+\frac{AM}{3}u_0^3=\frac{u_0^2}{3}+\frac{AM}{3}u_0^3\nonumber\\
        &=&\frac{u_0^2}{3}(1+AMu_0)
        \label{modified planar impact perimeter}
    \end{eqnarray}
\subsubsection{Vertical Polarized Critical Orbit}
Doing the same for the vertically polarized photon, i.e. by using:
    \begin{equation}
       \frac{df}{du}=-2AM^2u^3-3AM(2Mu-1)u^2+6Mu^2-2u=0
    \end{equation}
and substituting the trial solution (\ref{trial solution}) we find:
    \begin{displaymath}
        -2AM^2u_0^3-6AM^2u_0^3+3AMu_0^2+6Mu_0^2+12Mku_1u_0-2u_0-2ku_1=0
    \end{displaymath}
    \begin{equation}
        \Rightarrow u_1=-\frac{AM}{6k}u_0^2
    \end{equation}
    \begin{equation}
        u=u_0(1-\frac{A M}{6}u_0)
        \label{modified vertical circular solution}
    \end{equation}
which is equal, but opposite in sign, to (\ref{trial solution
modification term}), as is expected from the polarization rule.
Also, as before, the relevant impact parameter is given by
substituting (\ref{modified vertical circular solution}) and
(\ref{trial impact solution}) into the negative equation of
(\ref{polarization trajectory equation}) and working to first order
in $A$.
    \begin{eqnarray}
        \frac{du}{d\phi}&=&-2M^2Au^4+AMu^3+2Mu^3-u^2+\frac{1}{D_0^2}+\beta=0\nonumber\\
        &=&-2M^2Au_0^4+AMu_0^3+2Mu_0^3-MAu_0^4-u_0^2\nonumber\\
        &+&\frac{1}{3}AMu_0^3+\frac{1}{D_0^2}+\beta
    \end{eqnarray}
Eliminating terms and rearranging, as before, we find:
    \begin{equation}
        \beta=-\frac{AM}{3}u_0^3
    \end{equation}
Therefore the modified impact perimeter, for vertical polarization,
is:
    \begin{eqnarray}
        \frac{1}{D^2}&=&\frac{1}{3(3M)^2}-\frac{AM}{3}u_0^3=\frac{u_0^2}{3}-\frac{AM}{3}u_0^3\nonumber\\
        &=&\frac{u_0^2}{3}(1-AMu_0)
            \label{modified vertical impact perimeter}
    \end{eqnarray}
\subsubsection{The Modified Orbits}
We now have the circular orbit solutions for Eqn.~(\ref{polarization
trajectory equation}) and the relevant impact parameters:
\begin{itemize}
    \item Planar $(r)$ polarization solution (c<1)
    \begin{equation}
        u=u_0+A(\frac{M}{6}u_0^2) \qquad \frac{1}{D^2}=\frac{u_0^2}{3}(1+A Mu_0)
        \label{modified planar solution}
    \end{equation}
    \item Vertical($\theta$) polarization solution (c>1)
    \begin{equation}
        u=u_0-A(\frac{M}{6}u_0^2) \qquad \frac{1}{D^2}=\frac{u_0^2}{3}(1-A Mu_0)
            \label{modified vertical solution}
    \end{equation}
\end{itemize}
which are displayed in Fig.~\ref{orbs} We can note that as the
constant $A$ is of the order $10^{-32}$ these modifications are
extremely small.
\begin{figure}
    \begin{center}
        \includegraphics[width=0.8\textwidth]{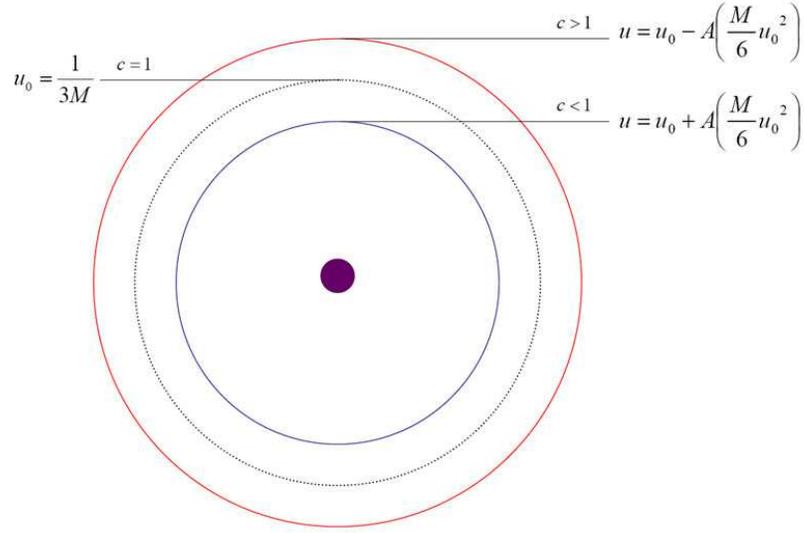}\\
        \caption{Orbit shifts about the classic critical orbit $u_0=1/3M$, with corresponding shifts in the speed of light (Not to scale)}.
        \label{orbs}
        \end{center}
\end{figure}

\section{Quantum Modified General Geodesics}
Using the solutions for the critical circular orbits and the
associated impact parameters we can now study how the general photon
trajectories are modified due to quantum corrections. In the classic
case when a photon comes in from infinity, with the critical impact
parameter, it tends to the critical orbit, as was shown in
Fig.~\ref{u vs phi with D=critical}; and if we slightly decrease the
impact parameter the photon spirals into the singularity,
Fig.~\ref{u vs phi with D=critical-0.1}.  We will now construct a
general quantum modified equation of motion and determine how the
geodesics change, depending on polarization, as they tend to the
critical orbits.
\subsection{General Vectors}
\label{gen-vect} From the quantum modification term in
Eqn.~(\ref{quantum light cone}) we can see that in order to
construct a general quantum modified geodesic equation we require
general photon polarization and wave vectors.  The wave vector,
$k^{\mu}$, can no longer be represented by a simple constant vector
pointing in the $\phi$ direction, as given in the orthonormal frame.
And, even though the vertical polarization will remain a constant,
as before, the planar polarization will now be a more general
vector, constantly changing as the photon moves through the plane.
\subsubsection{General Wave Vectors}
Our general wave vector will be of the form:
    \begin{equation}
        k^\mu=(\frac{dt}{d\tau},\frac{dr}{d\tau},\frac{d\theta}{d\tau},\frac{d\phi}{d\tau})
        \label{general wave}
    \end{equation}
which respects the light cone condition:
    \begin{equation}
        0=g_{\mu\nu}k^{\nu} k^{\mu}=F\dot{t}^2-F^{-1}\dot{r}^2-r^2\dot{\phi}^2
        \label{light cone for quantum general solution}
    \end{equation}
where $g_{\mu\nu}$ is the Schwarzschild metric.  Using previous
results, of Eqns~(\ref{dot-t}) and (\ref{dot-phi}), and  the fact
that we are working in the plane, $\theta=\frac{\pi}{2}$, we can
represent three of the wave vector components as:
    \begin{equation}
        \frac{dt}{d\tau}=EF^{-1} \qquad
        \frac{d\phi}{d\tau}=\frac{J}{r^2} \qquad
        \frac{d\theta}{d\tau}=0
        \label{wave vector components}
    \end{equation}
where we have defined:
    \begin{equation}
        F=(1-\frac{2M}{r})
        \label{F=1-2m/r}
    \end{equation}
Now using Eqn.~(\ref{light cone for quantum general solution}) we
can write the final component as:
        \begin{eqnarray}
            (\frac{dr}{d\tau})^2&=&F^2(\frac{dt}{d\tau})^2-r^2F(\frac{d\phi}{d\tau})^2\nonumber\\
            \frac{dr}{d\tau}&=&\sqrt{E^2-\frac{J^2F}{r^2}}=E(1-\frac{D^2F}{r^2})^{\frac{1}{2}}
            \label{dr/dtau}
        \end{eqnarray}
where we have used $D^2=\frac{J^2}{E^2}$.  Now the general wave
vector can be written as:
    \begin{equation}
        k^\mu=(EF^{-1},E(1-\frac{D^2F}{r^2})^{\frac{1}{2}},0,\frac{J}{r^2})
    \end{equation}
Also:
    \begin{eqnarray}
        k_{\mu}&=&g_{\mu\nu}k^\nu\nonumber\\
        k_{\mu}&=&(E,-F^{-1}E(1-\frac{D^2F}{r^2})^{\frac{1}{2}},0,-J)
    \end{eqnarray}
We now have:
    \begin{displaymath}
        k_{\mu}
        k^{\mu}=E^2F^{-1}-F^{-1}E^2(1-\frac{D^2F}{r^2})-\frac{J^2}{r^2}=0
    \end{displaymath}
as required. Therefore, eliminating $E$, we have the general wave
vectors:
    \begin{eqnarray}
           k^{\mu}&=&\frac{J}{D}(F^{-1},(1-\frac{D^2F}{r^2})^{\frac{1}{2}},0,\frac{D}{r^2})\\
            \label{k^mu general vector}
           k_{\mu}&=&\frac{J}{D}(1,-F^{-1}(1-\frac{D^2F}{r^2})^{\frac{1}{2}},0,-D)
            \label{k_mu general vector}
    \end{eqnarray}
We now need to normalize these vectors so, as they come in from
infinity and tend to the critical circular orbit, they correspond to
the critical orbit vector (\ref{k photon wave vector}). However, as
(\ref{k photon wave vector}) is given in the orthonormal basis, and
we are now working in the coordinate frame, we must use the tetrad
(\ref{inverse tetrad}), given in Appendix B, to transform (\ref{k
photon wave vector}) to its coordinate frame equivalent. Therefore,
(\ref{k photon wave vector}) in the coordinate basis is given as:
    \begin{equation}
         k^{\mu}=(e^{-1})^{\mu}_{a}k^a=(e^{-1})^{\mu}_{a}\left(%
        \begin{array}{c}
         E_k \\
         0 \\
         0 \\
         E_k\\
        \end{array}%
        \right)=E_k\left(%
        \begin{array}{c}
        F^{-\frac{1}{2}} \\
        0 \\
        0 \\
        \frac{1}{r} \\
        \end{array}%
        \right)
        \label{circular orbit transformation}
    \end{equation}
This now corresponds to a wave vector for a circular orbit in the
coordinate frame.  To represent the critical orbit we simply
substitute for $r=3M$, in which case
$F=1-\frac{2M}{r}\rightarrow\frac{1}{3}$, thus (\ref{circular orbit
transformation}) becomes:
    \begin{equation}
        k^{\mu}=E_k(\sqrt{3},0,0,\frac{1}{3M})
        \label{critical orbit in coordinate frame}
    \end{equation}
Now, if we evaluate (\ref{k^mu general vector}) with $r=3M$ and
$D=3\sqrt{3}M$ we have:
    \begin{equation}
        k^\mu=\frac{J}{D}\sqrt{3}(\sqrt{3},0,0,\frac{1}{3M})
        \label{general wavevector at critical point}
    \end{equation}
Therefore, (\ref{general wavevector at critical point}) is similar
to (\ref{critical orbit in coordinate frame}) up to a constant of
normalization, given as:
\begin{equation}
E_k=\frac{J}{D}\sqrt{3}=E\sqrt{3} \label{norm}
\end{equation}
Then, the normalized form of (\ref{k^mu general vector}) is given by
using (\ref{norm}):
    \begin{equation}
        k^{\mu}=\frac{E_k}{\sqrt{3}}(F^{-1},(1-\frac{D^2F}{r^2})^{\frac{1}{2}},0,\frac{D}{r^2})
        \label{normalized general wavevector}
    \end{equation}
\subsubsection{General Polarization Vectors}
\label{polarization vectors} Now we need to construct the, planar
and vertical, polarization vectors, $a^{\mu}$, of the photon; which
must be spacelike normalized as:
    \begin{equation}
        a_{\mu}a^{\mu}=-1 \qquad a_{\mu} k^{\mu}=0
        \label{polarization wave vector conditions}
    \end{equation}
As before, as we are working in a plane, the vertical polarization,
$ a^{\mu}_1 $, will be a constant, and can simply be written as:
    \begin{equation}
        a_{1}^{\mu}=(0,0,1,0)
        \label{vertical polarization general vector}
    \end{equation}
To normalize this we do as follows:
    \begin{equation}
        a_{1\mu}=g_{\mu\nu}a^{\nu}_1=-r^2(0,0,1,0) \qquad
        \Rightarrow \qquad a_{1}^{\mu} a_{1\mu}=-r^2
    \end{equation}
Therefore, the normalized vertical polarization vector is given as:
    \begin{eqnarray}
        a_1^{\mu}&=&(0,0,\frac{1}{r},0)\nonumber\\
        a_{1\mu}&=&-r^2(0,0,\frac{1}{r},0)
        \label{vertical polarization normailized general vector}
    \end{eqnarray}
These now satisfy both the conditions in (\ref{polarization wave
vector conditions}).  The planar polarized vector, $ a^{\mu}_2 $, is
given in the plane of $r$ and $\phi$:
    \begin{equation}
        a^{\mu}_2=(0,A,0,B)
        \label{planar vector}
    \end{equation}
Now, using the two conditions in (\ref{polarization wave vector
conditions}) we can determine $A$ and $B$.  From:
    \begin{equation}
        a_{2\mu}=g_{\mu\nu}a_2^{\nu}=(0,-AF^{-1},0,-r^2B)
        \label{a_mu plane polarization}
    \end{equation}
and the vector (\ref{normalized general wavevector}) we have:
    \begin{equation}
        k^{\mu}a_{2\mu}=(1-\frac{D^2F}{r^2})^{\frac{1}{2}}(AF^{-1})+(\frac{D}{r^2})(r^2B)=0
        \label{wave and polarization}
    \end{equation}
    \begin{equation}
        a^{2\mu}a_{2\mu}=A^2F^{-1}+B^2r^2=-1
    \end{equation}
Solving these for $A$ and $B$:
    \begin{equation}
        B=-\frac{F^{-1}}{D}(1-\frac{D^2F}{r^2})^{\frac{1}{2}}A
    \end{equation}
    \begin{equation}
        A^2=\frac{F}{(1+\frac{r^2F^{-1}}{D^2}(1-\frac{D^2F}{r^2}))}
    \end{equation}
    Therefore, we have:
    \begin{equation}
        A=\frac{DF}{r} \qquad
        B=-\frac{1}{r}(1-\frac{D^2F}{r^2})^{\frac{1}{2}}
    \end{equation}
and the planar polarization vector becomes:
    \begin{equation}
        a_{2}^{\mu}=(0,\frac{DF}{r},0,-\frac{1}{r}(1-\frac{D^2F}{r^2})^{\frac{1}{2}})
    \end{equation}
We now have the required polarization vectors:
    \begin{eqnarray}
        a_1^{\mu}&=&\frac{1}{r}(0,0,1,0)
        \label{final theta polarization}\\
        a_{1\mu}&=&-r^2(0,0,\frac{1}{r},0)
        \label{final theta polarization lower}\\
        a_2^{\mu}&=&\frac{1}{r}(0,DF,0,-(1-\frac{D^2F}{r^2})^{\frac{1}{2}})
        \label{final r-phi polarization}\\
        a_{2\mu}&=&(0,-\frac{D}{r},0,(1-\frac{D^2F}{r^2})^{\frac{1}{2}})
        \label{final r-phi polarization lower}
    \end{eqnarray}
where subscript 1 and 2 are vertical and planar polarizations
respectively. These now satisfy the conditions in (\ref{polarization
wave vector conditions}) with the wave vector (\ref{normalized
general wavevector}).
\newpage
\subsection{Quantum Modification}
\label{quantum mod section} Having derived the general polarization
and wave vectors, we can see from Eqn.~(\ref{quantum modified light
cone correction}) that the quantum modification given by:
    \begin{equation}
        \delta k(a)=\frac{8c_{\alpha}}{m_e^2}R_{abcd}k^ak^ca^ba^d
        \label{quantum correction}
    \end{equation}
also requires the Riemann tensor components in the coordinate frame.
In Appendix A we have calculated the required components as:
    \begin{eqnarray}
        R'_{trrt}&=&-\frac{2M}{r^3} \qquad  R'_{\theta r r
        \theta}=-\frac{MF^{-1}}{r} \qquad R'_{\phi rr
        \phi}=-\frac{MF^{-1}}{r}\nonumber\\
         R'_{\phi \theta \theta
        \phi}&=&2Mr \qquad  R'_{\theta t \theta t}=-\frac{M F}{r}
        \qquad R'_{\phi t \phi t}=-\frac{M F}{r}
        \label{coordinate frame Riemann components}
    \end{eqnarray}
Using this information we will now determine the form of the general
quantum modification; and from the polarization rule, this quantum
correction should satisfy the condition: $\delta k_1^2=-\delta
k_2^2$ for the two polarizations.

For vertical $(\theta)$ polarization we have, by using
(\ref{normalized general wavevector}) and (\ref{final theta
polarization}) in (\ref{quantum correction}):
    \begin{eqnarray}
       R_{abcd}k^ak^ca_1^ba_1^d & = & R_{a\theta c\theta} \frac{1}{r^2}
        k^ak^c
           =  R_{t\theta t \theta}\frac{1}{r^2}k^tk^t+R_{r\theta
         r\theta}\frac{1}{r^2}k^rk^r+R_{\phi\theta\phi\theta}\frac{1}{r^2}k^{\phi}k^{\phi}\nonumber\\
         & =&  -\frac{M F}{r} (\frac{1}{r^2}) (\frac{F^{-1} E_k}{\sqrt{3}})(\frac{F^{-1}
         E_k}{\sqrt{3}})\nonumber\\
         &+& \frac{M F^{-1}}{r} (\frac{1}{r^2}) (\frac{\sqrt{1-\frac{D^2F}{r^2}}
         E_k}{\sqrt{3}})(\frac{\sqrt{1-\frac{D^2F}{r^2}}E_k}{\sqrt{3}})\nonumber\\
         &-& 2Mr(\frac{1}{r^2})(\frac{DE_k}{r^2\sqrt{3}})(\frac{DE_k}{r^2\sqrt{3}})\nonumber\\
         &=&-\frac{ME_k^2F^{-1}}{3r^3}+\frac{ME_k^2F^{-1}(1-\frac{D^2F}{r^2})}{3r^3}-\frac{2ME_k^2D^2}{3r^5}\nonumber\\
         &=&-\frac{ME_k^2D^2}{r^5}
         \label{correction vertical}
    \end{eqnarray}
Similarly, for planar ($r-\phi$ plane) polarization we have, by
using (\ref{normalized general wavevector}) and (\ref{final r-phi
polarization}) in (\ref{quantum correction}):
    \begin{eqnarray}
         R_{abcd}k^ak^ca_2^ba_2^d & = &
         R_{arcr}k^{a}k^{c}a_2^ra_2^r+R_{arc\phi}k^{a}k^{c}a_2^ra_2^{\phi}
         +R_{a\phi c r}k^{a}k^{c}a_2^{\phi}a_2^{r}\nonumber\\
         &+&R_{a\phi c\phi}k^{a}k^{c}a_2^{\phi}a_2^{\phi}\nonumber\\
         &=&  R_{trtr}k^{t}k^{t}a_2^ra_2^r+R_{\phi r \phi
         r}k^{\phi}k^{\phi}a_2^ra_2^{r}+R_{\phi r r
         \phi}k^{\phi}k^{r}a_2^{r}a_2^{\phi}\nonumber\\
         &+&R_{r\phi\phi r}k^{r}k^{\phi}a_2^{\phi}a_2^{r}
         +R_{t \phi t \phi}k^{t}k^{t}a_2^{\phi}a_2^{\phi}
         +R_{r\phi r \phi}k^{r}k^{r}a_2^{\phi}a_2^{\phi}\nonumber\\
         &=&\frac{2M}{r^3}(\frac{F^{-1}E_k}{\sqrt{3}})^2(\frac{D
         F}{r})^2+\frac{M F^{-1}}{r}(\frac{D E_k}{\sqrt{3}
         r^2})^2(\frac{D F}{r})^2\nonumber\\
         &+&\frac{M F^{-1}}{r}(\frac{D E_k}{\sqrt{3}
         r^2})(\frac{E_k}{\sqrt{3}}\sqrt{1-\frac{D^2F}{r^2}})(\frac{D
         F}{r})(\frac{\sqrt{1-\frac{D^2F}{r^2}}}{r})\nonumber\\
         &+&\frac{M F^{-1}}{r}(\frac{E_k\sqrt{1-\frac{D^2F}{r^2}}}{\sqrt{3}})(\frac{D E_k}{\sqrt{3}
         r^2})(\frac{\sqrt{1-\frac{D^2F}{r^2}}}{r})(\frac{D
         F}{r})\nonumber\\
         &-&\frac{M
         F}{r}(\frac{F^{-1}E_k}{\sqrt{3}})^2(\frac{\sqrt{1-\frac{D^2F}{r^2}}}{r})^2\nonumber\\
         &+&\frac{M F^{-1}}{r}(\frac{E_k\sqrt{1-\frac{D^2F}{r^2}}}{\sqrt{3}})^2
         (\frac{\sqrt{1-\frac{D^2F}{r^2}}}{r})^2\nonumber\\
         &=&\frac{M E_k^2 D^2}{r^5}
         \label{correction planar}
    \end{eqnarray}
Therefore, the quantum modifications, $k^2=\delta k (a)$, for the
two polarizations $a_1$ and $a_2$ (vertical and planar respectively)
are:
    \begin{eqnarray}
        \delta k(a_1)&=&-(\frac{8c_{\alpha}}{m_e^2})\frac{ME_k^2 D^2}{r^5} \qquad \textrm{Vertical}\\
        \delta k(a_2)&=&(\frac{8c_{\alpha}}{m_e^2})\frac{ME_k^2
        D^2}{r^5} \qquad \textrm{Planar}
    \end{eqnarray}
where $-\delta k(a_1)=\delta k(a_2)$, as required.  Now, using
$k^{\mu}k_{\mu}-\delta k(a)$=0 we can write the general equations of
motion for the two polarizations:
    \begin{displaymath}
        k^{\mu}k_{\mu}\pm(\frac{8c_{\alpha}}{m_e^2})\frac{ME_k^2 D^2}{r^5}=0
    \end{displaymath}
    \begin{displaymath}
        (\frac{dr}{d\tau})^2=F^2(\frac{dt}{d\tau})^2-r^2F(\frac{d\phi}{d\tau})^2\pm F(\frac{8c_{\alpha}}{m_e^2})\frac{ME_k^2 D^2}{r^5}
    \end{displaymath}
As before, substituting for $\dot{t}$, $\dot{\phi}$ and transforming
$r\rightarrow\frac{1}{u}$, we have:
    \begin{equation}
        (\frac{du}{d\phi})^2=\frac{1}{D^2}-Fu^2\pm (\frac{D^2 M A}{3})Fu^5
        \label{genreal equation of motion dependent on phi}
    \end{equation}
We can also write $u$ as a function of time:
    \begin{equation}
        (\frac{du}{dt})^2=(Du^2F)^2(\frac{1}{D^2}-Fu^2\pm (\frac{D^2 M A}{3})Fu^5)
         \label{genreal equation of motion dependent on t}
    \end{equation}
where we have used the substitution for $A$ given in equation
(\ref{constant A}).  Now, the Eqns.~\ref{genreal equation of motion
dependent on phi} and \ref{genreal equation of motion dependent on
t} are the general equations of motion, $+$ for vertical ($\theta$)
polarization and $-$ for planar ($r$-$\phi$) polarization.  In these
equations we not only use the impact parameter of the form $1/D^2$
but also $D^2$, therefore the parameters for the two polarizations
are given as:
\begin{equation}
\frac{1}{D^2}=\frac{u_0^2}{3}(1\pm A Mu_0) \rightarrow
    D=\frac{\sqrt{3}}{u_0} \mp\frac{\sqrt{3}A M}{2}+\mathcal{O}(A)
\label{new impact}
\end{equation}
\subsection{General Trajectory to the Critical Orbit.}
Now that we have the general orbit equation (\ref{genreal equation
of motion dependent on phi}) we can first test whether, for the
critical impact perimeter $D$, the equation
$\frac{du}{d\phi}\rightarrow0$ for first order in $A$. Substituting
the critical orbits and the impact parameters given in
(\ref{modified planar solution}),(\ref{modified vertical solution})
and (\ref{new impact}) into (\ref{genreal equation of motion
dependent on phi}) and expanding up to first order in $A$ we have:
For planar polarization.
    \begin{eqnarray}
        (\frac{du}{d\phi})^2&=&\frac{u_0^2}{3}(1+A Mu_0)-[1-2M(u_0(1+\frac{A M}{6}u_0))[(u_0(1+\frac{A M}{6}u_0)]^2\nonumber\\
        &-&(\frac{\frac{3}{u_0^2(1+A M
        u_0)}A M}{3})[1-2M[u_0(1+\frac{A M}{6}u_0)]][u_0(1+\frac{A M}{6}u_0)]^5\nonumber\\
        &\rightarrow&0
    \end{eqnarray}
and similarly for vertical polarization.
       \begin{eqnarray}
        (\frac{du}{d\phi})^2&=&\frac{u_0^2}{3}(1-A Mu_0)-[1-2M[u_0(1-\frac{A M}{6}u_0)]][u_0(1-\frac{A M}{6}u_0)]^2\nonumber\\
        &-&(\frac{\frac{3}{u_0^2(1-A M
        u_0)}A M}{3})[1-2[(u_0(1-\frac{A M}{6}u_0)]][u_0(1-\frac{A M}{6}u_0)]^5\nonumber\\
        &\rightarrow&0
    \end{eqnarray}
Expanding these and eliminating all terms of order $A^2$ and higher,
we find that the right hand sides go to zero, as required.  Thus for
the appropriate impact parameters these equations behave as they
should.

The next step is to solve equation (\ref{genreal equation of motion
dependent on phi}) for the two polarizations.  This is most simply
done using numerical methods in Mathematica.  As a guide, we know
our solution will be of the form $u=u_0+k u_1$, where $u_0$ will be
the classical solution (\ref{u=critical sol}), $u_1$ will be a small
modification that pushes the critical orbit up or down depending on
photon polarization, and $k$ will be some constant that is first
order in $A$, i.e. of the form $k=A s$, where $s$ will be some
number given by the boundary condition: for $\phi\rightarrow\infty$
then $u(\phi)\rightarrow u_0(1\pm\frac{A}{6}u_0)$. If we substitute
for the critical impact parameter $D$ from (\ref{modified planar
solution}) and (\ref{modified vertical solution}) depending on the
polarization, and then transform to $u\rightarrow u_0+A s u_1$, and
expand to first order in $A$, we have non-linear first order
differential equations in $u_0$, $u_1$ and $\phi$.
\begin{itemize}
    \item For planar polarization:
    \begin{eqnarray}
        \frac{du_1(\phi)}{d\phi}&=&\frac{}{6s\sqrt{\frac{1}{27M^2}-u_0(\phi)^2+2Mu_0(\phi)^3}}
        [\frac{1}{27M^2}-27M^3u_0(\phi)^5\nonumber\\
        &+&54M^4u_0(\phi)^6- 6s u_0(\phi)u_1(\phi)+18 s Mu_0(\phi)^2u_1(\phi)]
        \label{u1(phi) equation for plane polarization}
    \end{eqnarray}
    \item For vertical polarization:
    \begin{eqnarray}
   \frac{du_1(\phi)}{d\phi}&=&\frac{1}{6s\sqrt{\frac{1}{27M^2}-u_0(\phi)^2+2Mu_0(\phi)^3}}
        [-\frac{1}{27M^2}+27M^3u_0(\phi)^5\nonumber\\
        &-&54M^4u_0(\phi)^6-6 s u_0(\phi)u_1(\phi)+18 s Mu_0(\phi)^2u_1(\phi)]
        \label{u1(phi) equation for theta polarization}
    \end{eqnarray}
\end{itemize}
These can be solved in one of two ways, (i) is to substitute the
classic solution (\ref{u=critical sol}) for $u_0$ and solve
analytically, (ii) is to solve $\frac{du_1}{d\phi}$ and the classic
equation for $\frac{du_0}{d\phi}$ simultaneously using numerical
techniques. We attempted to use method (i) to derive an analytic
solution for $u_1(\phi)$, however, due to the complex nature of the
equation we proceeded to use method (ii), that is solving by
numerical methods. To do this we set the constant $s=1$, and then
when the numerical values of $u_1(\phi)$ were determined we could
determine the constant $s$ so that $u(\phi)$ coincided with the
modified circular orbits given in (\ref{modified planar solution})
and (\ref{modified vertical solution}).  This technique was used for
reasons of convenience, because solving (\ref{u1(phi) equation for
plane polarization}) and (\ref{u1(phi) equation for theta
polarization}) for various values of $s$ would be time consuming as
each numerical calculation takes a significant amount of time;
therefore solving them once and then scaling the solution is a more
convenient method. The results of the equations were plotted as:
    \begin{equation}
        u(\phi)=u_0(\phi)+A s u_1(\phi)
        \label{general test equation}
    \end{equation}
where the constant $s$ was picked to satisfy the condition:
$\phi\rightarrow\infty$ $\Rightarrow $   $u(\phi)\rightarrow
u_0(1\pm\frac{A}{6}u_0)$ i.e. the trajectories tend to the critical
orbit, depending on polarization. In this way the constant $s$ was
determined to be $s=1/3$, which was tested for various values of
$M$. We have plotted the results of the numerical calculation in
Figs.~\ref{general critical trajectories small} and \ref{general
critical trajectories large}.  In Fig.~\ref{general critical
trajectories small}, you can see that the general critical orbits
follow a classic style path, however the orbit splitting is not
clearly visible.  But in Fig.~\ref{general critical trajectories
large} we have plotted a closer view of the critical orbits, and
here the splitting is highly visible.  It can be seen that the
general trajectories for the planar and vertically polarized photons
tend to the relevant critical orbits.
\begin{figure}
    \begin{center}
        \includegraphics[width=0.9\textwidth]{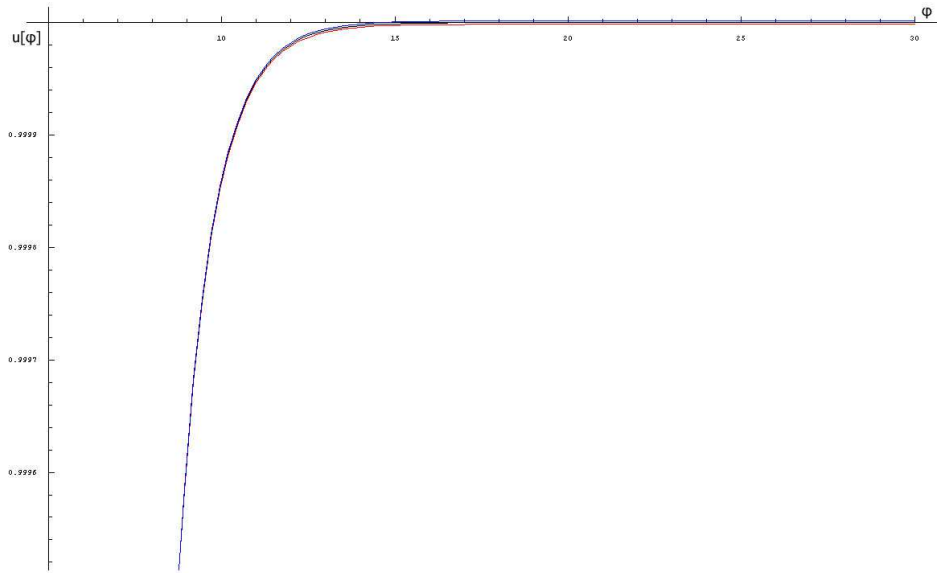}\\
        \caption{The quantum modified critical orbits follow classic
        type paths.  We have used a factor $A=0.00009$ in order to make the quantum corrections more visible,
        where a real value should be of order $\sim10^{-32}$, Eqn.~\ref{order of A}.
        The splitting of the orbits can be clearly seen in
        the Fig.~\ref{general critical trajectories large}, given below.}
        \label{general critical trajectories small}
    \end{center}
\end{figure}
\begin{figure}
    \begin{center}
        \includegraphics[width=0.9\textwidth]{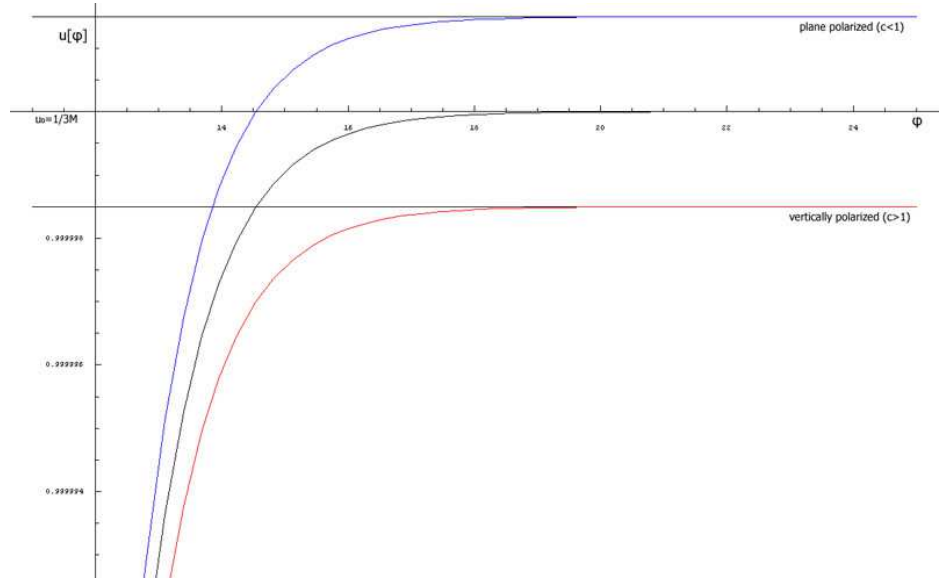}\\
        \caption{The quantum modified trajectories show clear splitting as they tend to the critical points.
        We used the constants $M=\frac{1}{3}$ and $A=0.00009$.}\label{general critical trajectories large}
    \end{center}
\end{figure}
\newpage
\section{Quantum Modification and the Event Horizon}
In Chapter.~\ref{null dynamics} it was shown that when we decreased
the impact parameter from the critical value ($D=3\sqrt{3}M$ to
$D-1/10$) the photon trajectory, given as $u(\phi)$, spiraled into
the singularity, Fig.~\ref{u vs phi with D=critical-0.1}; and when
we represent the trajectory as a function of coordinate time,
$u(t)$, it tends to the event horizon, $u=1/2M$. From the horizon
theorem it was seen that quantum modifications have no effect on
photon velocities directed normal to the event horizon.  So this
implies when a photon tends to the event horizon at an angle, e.g.
with an impact parameter $D=D_{critical}-1/10$, then the component
of velocity normal to the horizon should be unchanged, while the
component parallel to it is modified according to the quantum
correction; this modification should then result in a shift of the
photon trajectory, but the horizon should remain fixed. In order to
test this we used Eqn.~(\ref{genreal equation of motion dependent on
phi}) with an impact parameter $D=D_{critical}-1/10$ to show that
the photon trajectories still fall into the singularity. We then
used Eqn.~(\ref{genreal equation of motion dependent on t}), with
the new impact parameter, to study the behavior of the trajectories
around the event horizon.
\subsection{Trajectories to the singularity}
In order to construct quantum modified trajectories, which go past
the critical orbit and fall into the singularity, we require the
impact parameters:
    \begin{equation}
\frac{1}{D^2}=\frac{1}{(\frac{\sqrt{3}}{u_0\sqrt{(1\pm A
Mu_0)}}-\frac{1}{10})^2} \rightarrow
    D=\frac{\sqrt{3}}{u_0}\mp \frac{\sqrt{3}A M}{2}-\frac{1}{10}+\mathcal{O}(A)
    \label{d-less-crit}
    \end{equation}
where, as before, $+$ is for vertical polarization and $-$ is planar
polarization, in $1/D^2$.  For these impact parameters we
numerically solved Eqn.~(\ref{genreal equation of motion dependent
on phi}), and in Fig.~\ref{d-01} we can see that all the
trajectories follow a classic type path into the
singularity\footnote{The quantum modifications to the classic
trajectories are very small, and even if we use the hugely
exaggerated value of $A=0.00009$, as was used in Figs.~\ref{general
critical trajectories large} and \ref{general critical trajectories
small}, the modification is hardly visible. So, in order to magnify
the quantum correction even more we used $A=0.0009$.}.  However,
near the singularity you can see the splitting of the orbits as they
tend to $u\rightarrow\infty$. In Fig.~\ref{d-01-zoom} we have shown
a magnified view of the point where the trajectories cross the event
horizon.  In this figure it can be seen that the planar polarized
photon ($c<1$) crosses the event horizon at a point before the
classic trajectory and the vertically polarized photon ($c>1$)
crosses it at a point after the classic trajectory.  This makes
sense, as the planar polarized photons are pushed towards the black
hole and vertically polarized trajectories are pushed out, the
vertically polarized ones must spiral further around the black hole
to reach the event horizon compared to the classic or the planar
polarized trajectories.
\begin{figure}
     \begin{center}
        \includegraphics[width=0.9\textwidth]{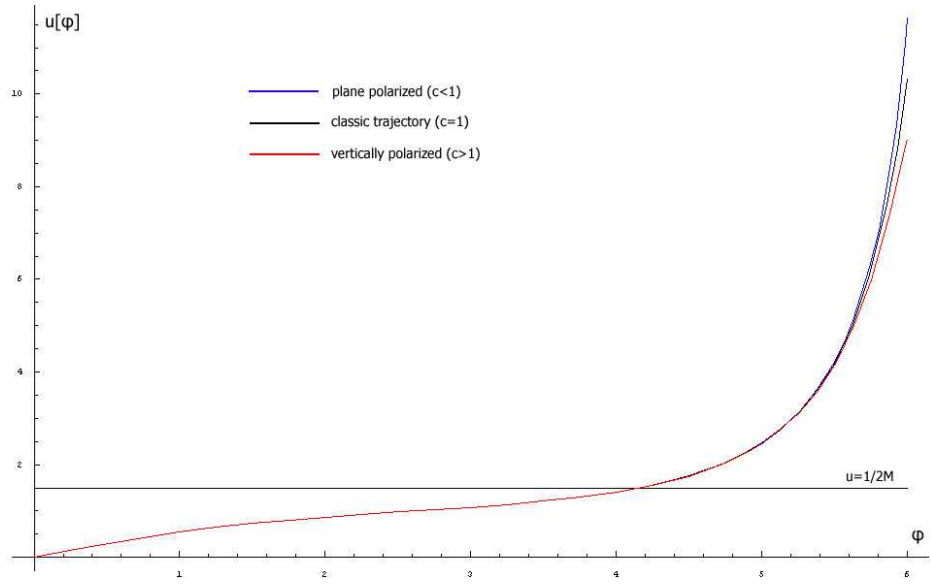}\\
        \caption{The quantum modified trajectories follow similar paths, as the classic trajectory, to the
        singularity, with $M=\frac{1}{3}$ and $A=0.0009$.}
        \label{d-01}
    \end{center}
\end{figure}
\begin{figure}
    \begin{center}
         \includegraphics[width=0.9\textwidth]{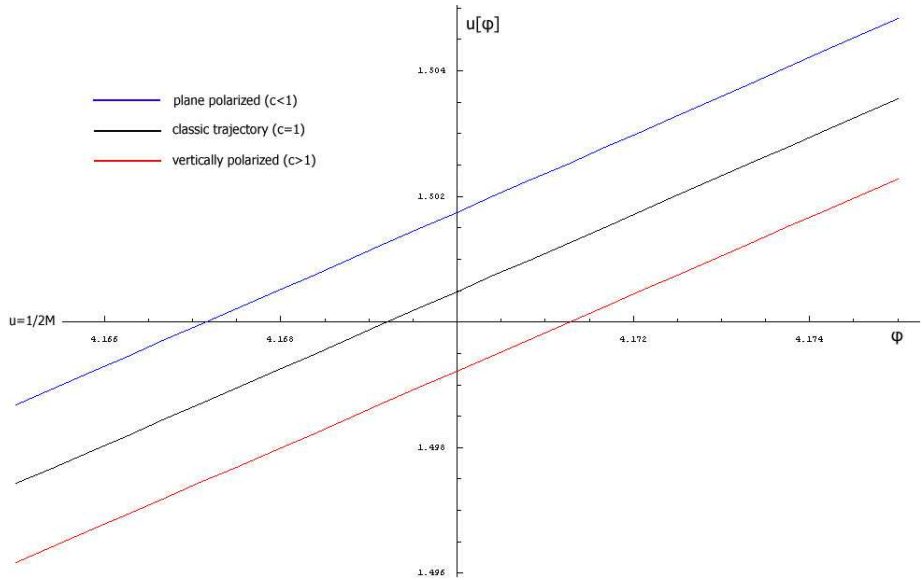}\\
        \caption{The classic and quantum modified trajectories crossing the event horizon $u=2M$, with $M=\frac{1}{3}$ and $A=0.0009$.}
        \label{d-01-zoom}
    \end{center}
\end{figure}
\subsection{Fixed Event Horizon}
To demonstrate the fact that the event horizon remains fixed at
$u=1/2M$ we applied the impact parameters given by
(\ref{d-less-crit}) to Eqn.(\ref{genreal equation of motion
dependent on t}).  Again, solving numerically we found that the
trajectories tend to the event horizon in the classic way,
Fig.~\ref{u(t)d-01}, however, they are again slightly shifted.  In
Fig.~\ref{u(t)d-01-zoom}, a close up of the point where the
trajectories tend to the event horizon at $u=1/2M$, you can clearly
see that the quantum modified orbits tend to the horizon before the
classic orbit.  This can be understood by the fact that the planar
polarized trajectory is pushed towards the black hole, hence it has
less of a distance to propagate before it reaches the horizon, and
even though the vertically polarized trajectory is pushed outwards,
the fact that it has a faster velocity than $c=1$ it reaches the
event horizon before the classic trajectory.
\begin{figure}
     \begin{center}
        \includegraphics[width=0.9\textwidth]{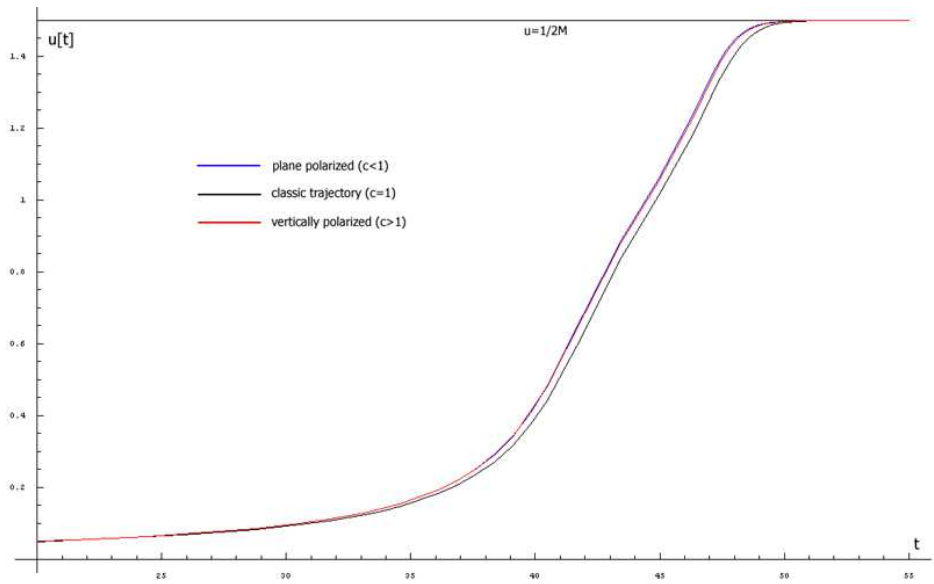}\\
        \caption{The quantum modified trajectories follow similar paths, as the classic trajectory, to the
        event horizon, with $M=\frac{1}{3}$ and $A=0.0009$.}
        \label{u(t)d-01}
    \end{center}
\end{figure}
\begin{figure}
    \begin{center}
         \includegraphics[width=0.9\textwidth]{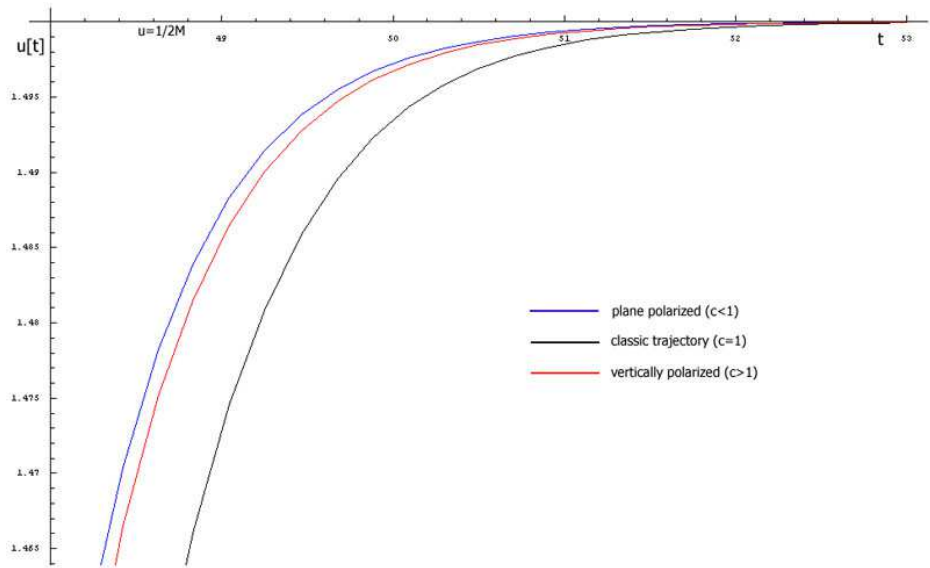}\\
        \caption{The classic and quantum modified trajectories reach the event horizon at differing times, with $M=\frac{1}{3}$ and $A=0.0009$.}
        \label{u(t)d-01-zoom}
    \end{center}
\end{figure}
The other more important thing we can note from this calculation is
that the event horizon remains fixed at the classic point $u=1/2M$,
due to the fact that, as was stated earlier, the quantum correction
has no effect on photon momentum normal to the horizon.  This means
even though the quantum correction implies velocities greater than
the speed of light, no photons can escape from the event horizon, so
the black hole remains black.
}

{\typeout{Quantum Modified Schwarzschild Metric}
\chapter{Quantum Modified Schwarzschild Metric}
\label{Quantum Modified Schwarzschild Metric} Now, using our
previous results, we can construct a new metric that encompasses the
quantum corrections due to vacuum polarization. This metric should
then be a sum of the classic Schwarzschild metric and the
polarization dependent quantum corrections we derived in
Eqns.~(\ref{correction vertical}) and (\ref{correction planar}):
\begin{eqnarray}
\mathcal{G}_{\mu\nu}k^{\mu}k^{\nu}&=&(g_{\mu\nu}+\mathfrak{g}_{\mu\nu})k^{\mu}k^{\nu}=0\nonumber\\
&=&(g_{\mu\nu}-\frac{8c_{\alpha}}{m_e^2}
R_{\mu\eta\nu\lambda}a^{\eta}a^{\lambda})k^{\mu}k^{\nu}=0
\end{eqnarray}
\section{Construction of the Metric}
Using the results in Sec.~\ref{quantum mod section} we can write the
first components of the quantum modified metric $\mathcal{G}_{tt}^1$
and $\mathcal{G}_{tt}^2$, for vertical and planar polarization
respectively, as:
\begin{eqnarray}
\mathcal{G}_{tt}^1&=&g_{tt}-\frac{8c_{\alpha}}{m_e^2}R_{t b t
d}a_1^b
a_1^d \nonumber\\
&=&F-\frac{8c_{\alpha}}{m_e^2}(-\frac{M
F}{r})(\frac{1}{r^2})\nonumber\\
&=&F+\frac{A M D^2}{9}F u^3\\
\mathcal{G}_{tt}^2&=&g_{tt}-\frac{8c_{\alpha}}{m_e^2}R_{t b t
d}a_2^b
a_2^d \nonumber\\
&=&F-\frac{8c_{\alpha}}{m_e^2}[\frac{2M}{r^3}(\frac{DF}{r})^2 - (\frac{MF}{r})(\frac{\sqrt{1-\frac{D^2F}{r^2}}}{r})^2]\nonumber\\
&=&F-\frac{A M D^2}{9}(3 D^2 F^2 u^5-Fu^3)
\end{eqnarray}
where, in the last lines of each component, we have used
Eqn~(\ref{constant A}) to replace the constants with $A$, and we've
made the transformation $r\rightarrow1/u$. Now, doing this for all
the other components we can construct the metrics for vertical and
planar polarizations:
\begin{itemize}
    \item Vertical Polarization: $a_1^{\mu}=u(0,0,1,0)$
        \begin{equation}
\mathcal{G}^1_{\mu\nu}=\left(
                         \begin{array}{cccc}
                           F-B F u^3 & 0 & 0 & 0 \\
                           0 & -\frac{1}{F}+\frac{B u^3}{ F} & 0 & 0 \\
                           0 & 0 & -\frac{1}{u^2} & 0 \\
                           0 & 0 & 0 & -\frac{1}{u^2}-B u\\
                         \end{array}
                       \right)
                       \label{quantum metric vertical}
\end{equation}
    \item Planar Polarization: $a_2^{\mu}=u(0,D F,0,D \frac{du}{d\phi})$
    \begin{equation}
\mathcal{G}^2_{\mu\nu}=\left(
                         \begin{array}{cccc}
                           F+B F u^3(1 -3D^2 F u^2) & 0 & 0 & 0 \\
                           0 & -\frac{1}{F}-\frac{B u^3 (1-D^2 F u^2)}{F} & 0 & Bu^3 D \frac{d u}{d\phi} \\
                           0 & 0 & -\frac{1}{u^{2}} & 0 \\
                           0 & Bu^3 D \frac{d u}{d\phi} & 0 & -\frac{1}{u^2}-B D^2 F u^3\\
                         \end{array}
                       \right)
                         \label{quantum metric planar}
\end{equation}
\end{itemize}
where $B=\frac{AM D^2}{9}$.  These are now the relevant quantum
modified Schwarzschild metrics for vertical and planar
polarizations\footnote{ In both the metric, ${G}^2_{\mu\nu}$, and
vector, $a_2^{\mu}$, we have made the substitution $r\rightarrow
1/u$ and $ \frac{du}{d\phi}=-\frac{1}{D}\sqrt{1- D^2 F u^2}$}.
\section{Dynamics with the Quantum Modified Metric} We can now
derive the same equations of motion as (\ref{genreal equation of
motion dependent on phi}), but now we can do it simply with the
quantum modified metrics.  Using the general wave vector in a plane:
\begin{eqnarray}
k^{\mu}&=&(\dot{t},\dot{r},\dot{\theta},\dot{\phi})\nonumber\\
&=&(\frac{J}{D}F^{-1},-J\frac{du}{d\phi},0,J u^2)
\end{eqnarray}
where we have transformed to $r\rightarrow 1/u$, and we substitute
for $\dot{t}$ and $\dot{\phi}$ as before.  Applying this wave vector
to (\ref{quantum metric vertical}) and(\ref{quantum metric planar})
we find:
\begin{eqnarray}
0&=&\mathcal{G}^1_{\mu\nu}k^{\mu}k^{\nu}\nonumber\\
 &\Rightarrow& (\frac{du}{d\phi})^2=\frac{1}{D^2}-Fu^2+
 (\frac{D^2A}{3})Fu^5
 \end{eqnarray}
 for vertical polarization, and
 \begin{eqnarray}
 0&=&\mathcal{G}^2_{\mu\nu}k^{\mu}k^{\nu}\nonumber\\
 &\Rightarrow& (\frac{du}{d\phi})^2=\frac{1}{D^2}-Fu^2- (\frac{D^2A}{3})Fu^5
\end{eqnarray}
for planar polarization.  These are identical to the equations we
derived in Sec.~\ref{quantum mod section}.
\subsubsection{Radial Geodesics}
We can now show that the metric for vertical polarization is
consistent with the fact that radially projected photon trajectories
are not modified. By using the vertical polarization metric,
(\ref{quantum metric vertical}), and a general radial wave vector,
($k^{\mu}=(\dot{t},\dot{r},0,0)$, we find:
\begin{eqnarray}
0&=&\mathcal{G}^1_{\mu\nu}k^{\mu}k^{\nu}\nonumber\\
\Rightarrow \qquad \frac{dt}{dr}&=&\pm\frac{1}{F}
\end{eqnarray}
which is identical to the classic radial geodesic equation,
(\ref{rearranged dr/dt J=0}).  However, the same is not true for the
planar polarization metric.  Due to our derivation of the
polarization vectors in Sec.~\ref{polarization vectors}, we
constructed a very general vertical polarization vector and then
normalized it; however, the one for planar polarization was
constructed for the case where $\frac{d\phi}{d\tau}\neq0$, as can be
seen in Eqns.~(\ref{dr/dtau}) and (\ref{wave and polarization});
therefore the planar polarization vector has $\phi$ dependence
"mixed" into it through the substitution of $\dot{\phi}$:
\begin{eqnarray}
(\frac{dr}{d\tau})^2&=&F^2(\frac{dt}{d\tau})^2-r^2F(\frac{d\phi}{d\tau})^2\nonumber\\
&=&(E^2-\frac{J^2 F}{r^2})^{\frac{1}{2}}=E(1-\frac{D^2
F}{r^2})^{\frac{1}{2}}\nonumber\\
\end{eqnarray}
We can note that it's a result of the substitution
$\frac{d\phi}{d\tau}=\frac{J}{r^2}$, in the polarization vectors,
that we acquire an extra parameter $D$ in our quantum correction
(apart from the one introduced through normalization, which was
incorporated into $B=A D^2/9$). We can then say that if we need to
use the metric for a radial trajectory with $\frac{d\phi}{d\tau}=0$
we can just set $D=0$, as this would lead to the removal of the
$\phi$ dependence. In this way, the planar polarized quantum
modified metric for radial trajectories is:
\begin{equation}
\mathcal{G}(\textsl{$\phi$=0})^2_{\mu\nu}=\left(
                         \begin{array}{cccc}
                           F+B F u^3 & 0 & 0 & 0 \\
                           0 & -\frac{1}{F}-\frac{B u^3 }{F} & 0 & 0 \\
                           0 & 0 & -\frac{1}{u^{2}} & 0 \\
                           0 & 0& 0 & 0\\
                         \end{array}
                       \right)
\end{equation}
Then applying this to a general radial wavevector, we again find the
classic radial geodesic equation:
\begin{eqnarray}
0&=&\mathcal{G}(\textsl{$\phi$=0})^2_{\mu\nu}k^{\mu}k^{\nu}\nonumber\\
\Rightarrow \qquad \frac{dt}{dr}&=&\pm\frac{1}{F}
\end{eqnarray}
Therefore, we have metrics for the Schwarzschild space time that
incorporate quantum corrections due to vacuum polarization; and
these metrics are consistent with classic results.
}

{\typeout{Summary}
\chapter{Summary}
\label{part1-summary} For Schwarzschild Spacetime we showed, in the
orthonormal frame, that, due to quantum corrections, the stable
circular orbit at $u_0=1/3M$ splits depending on the polarization of
the photon, and the new modified circular orbits are given by the
classical orbit plus a correction term to first order in a
dimensionless constant $A$: $u=u_0+ Au_1$, where
$A=\frac{24c_{\alpha}E_k^2}{J^2m_e^2}$ and has an order of
$10^{-32}$. As stated by the polarization sum rule this splitting of
the critical orbit is equal in magnitude but opposite in sign for
the two polarizations. We found that the vertically polarized photon
(c>1) is pushed out by a correction of $u_1=-\frac{M}{6}u_0^2$,
while the planar polarized photon is pulled in by a correction of
$u_1=\frac{M}{6}u_0^2$. It was also found that this orbit shift also
requires an appropriate modification of the classic impact parameter
$\frac{1}{D_0^2}=\frac{u_0^2}{3}$. This modification, for the
quantum corrected orbits, took the form
$\frac{1}{D^2}=\frac{1}{D_0^2}+A D_1$, and again as for the orbit
shift, the change was equal in magnitude but opposite in sign for
the two polarizations: $D_1=\frac{Mu_0^3}{3}$ for planar
polarization and $D_1=-\frac{Mu_0^3}{3}$ for vertical polarization.
Using this information, for the splitting of circular orbits, we
then constructed the quantum corrected general equations of motion
for the Schwarzschild spacetime.  Using these equation of motion it
was shown that a photon starting at $u=0$, with the appropriate
critical impact parameter, tends to the critical orbit associated
with that impact parameter - and the trajectory follows a similar
path to the classic case.

We then went on to show that, using the general quantum corrected
equations of motion, a photon projected towards the black hole with
an impact parameter less than the critical value falls into the the
singularity, in terms of the angular distance ($\phi$). Although the
photons follow a classic type path into the singularity, the
trajectories are slightly shifted according to polarization. The
planar polarized photon crosses the horizon before the classic
trajectory, and the vertically polarized one crosses it after, this
corresponds to the fact that planar polarized photons are pulled
towards the black hole and vertically polarized ones are pushed
away, hence the vertically polarized ones need to go a further
angular distance to reach the event horizon.

In terms of coordinate time ($t$) we found that the photon
trajectories tend to the event horizon. Therefore, the quantum
corrections do not shift the classic event horizon from
$u=\frac{1}{2M}$, which corresponds to the horizon theorem.  Also,
we found that, although the quantum corrected orbits follow a
classic type path to the event horizon, the point at which they hit
the horizon is, again, slightly shifted depending on polarization.
However, this time both polarizations hit the horizon before the
classic trajectory. The planar polarized photon tends to arrive at
the horizon first, then the vertically polarized one, and finally
the classic photon.  This could correspond to the fact that the
planar polarized photon, although it has a velocity lower than the
speed of light, has less of a distance to go, as its trajectory is
pulled towards the black hole.  For the vertically polarized case,
even though it has a faster than light velocity, it has a further
distance to go to reach the horizon as its trajectory is pushed away
from the black hole.

Having determined the equations of motion, with the quantum
correction, we then went on to construct a Schwarzschild metric that
incorporates the quantum correction:
$\mathcal{G}_{\mu\nu}=(g_{\mu\nu}+\mathfrak{g}_{\mu\nu})$, where the
correction: $\mathfrak{g}_{\mu\nu}$ was again first order in $A$. We
showed that with this metric and a general photon wave vector,
$k^{\mu}$, we obtain the quantum modified equations of motion, as
before.  Also, this new metric confirms the horizon theorem, that
is, when we use a wave vector indicating a radially projected photon
we obtain the classic equation of motions.  This was fine for
vertical polarization, however, in the planar polarization case we
had a problem; we had previously used a substitution that mixed a
"hidden" radial angle, $\phi$, into our polarization vector (as in
general orbits the planar polarization depends on $\phi$). By
tracing back to the origin of this substitution we found that, in
order to study radial trajectories, we need to set $D=0$, this then
removes the $\phi$ dependency; this then gives us the classic
equation of motion for planar polarized radially projected photons.

So in conclusion, after studying the dynamics of null trajectories
in  Schwarzschild spacetime we derived the polarization dependent
photon trajectories to first order in the constant $A$ (which is
dependent on the fine structure constant, the mass of the star, mass
of the electron, and the energy of the photon). We then incorporated
these modification into a general quantum modified metric, which
could also be used to derive the general quantum modified equations
of motion.  Also, the results of this work coincide with the
conditions of the horizon theorem and the polarization sum rule.
}

\part{Superfluid Behaviour of the 2+1d NJL Model at High Density}

{\typeout{Introduction}
\chapter{Introduction }
\label{chap:introduction}
\section{Quantum Chromodynamics}
\label{chap:introduction-sec:qcd} Since the 70's it has been
accepted that nucleons and other hadrons (baryons and mesons)
observed in particle accelerators are not fundamental particles
themselves, but are composed of fractionally charged fermions known
as quarks, which exchange bosons of the strong force known as
gluons.  In this description, known as the quark model, baryons and
mesons are depicted as bound states of three quarks and quark
anti-quark pairs respectively. In this way the quark model provides
a very natural explanation for the multiplicity and pattern of all
the strongly-interacting particles \cite{hands2,walters}.
Experimental tests of this theory (in a similar way to the classic
high-angle Rutherford scattering of $\alpha$-particles off atoms
demonstrating the existence of the nucleus) consists of high energy
inelastic scattering experiments of electrons off nucleons. The
evidence from such experiments is consistent with the presence of
pointlike spin-$\frac{1}{2}$ constituents called partons with a mass
one third that of the proton. These partons, which are able to move
freely within the
nucleon volume, are then identified with quarks.\\
\subsection{QCD: A Model of Strongly Interacting Matter}
Quantum Chromodynamics (QCD) was introduced in the early 70s as
the theoretical framework that translated the experimental and
conceptual description of the quark model into a quantitative
calculational scheme.  QCD describes quarks and anti-quarks as
quanta of the elementary fermion fields $\psi$ and $\bar{\psi}$,
each with an $SU(3)$ colour charge, and gluons as quanta of a
self-interacting non-abelian gauge field \textit{$A_{\mu}$}.  The
Lagrangian density of QCD is given by
\begin{equation}
    \mathcal{L}_{QCD}=\bar{\psi}^{\alpha}_{i}(i\Dslash-m_0)^{\alpha\beta}_{ij}\psi^{\beta}_{j}-\frac{1}{4}
    \mathcal{F}^{a}_{\mu\nu}\mathcal{F}_{a}^{\mu\nu}
\end{equation}
In the fermionic part $i$ and $j$ run over $N_f$ flavours of
quarks, $\alpha$ and $\beta$ run over the 3 colours, then $m_0$ is
an $N_f \times N_f$ mass matrix in flavour space.  The covariant
derivative
\begin{equation}
    \Dslash=\gamma^{\mu}
    D_{\mu}^{\alpha\beta}=\gamma^{\mu}(\delta^{\alpha\beta}\partial_{\mu}-\frac{i}{2}g(\lambda^a)^{\alpha
    \beta}A^a_{\mu})
\end{equation}
is introduced so the Lagrangian density is invariant under local
$SU(3)$ gauge transformations, where $g$ is the bare coupling
constant, $A^a_{\mu}$ is a vector (gauge) field with eight gluonic
degrees of freedom and $\lambda^a$ denote the generators of the
$SU(3)$ group. Due to the introduction of the eight gluon fields,
through gauge symmetry, we adjoin the free gluon lagrangian (the
final (gauge) part of the action) to give the full QCD lagrangian.
In the free gluon term of the action the field strength tensor is
\begin{equation}
    \mathcal{F}^a_{\mu\nu}=\partial_{\mu}A^a_{\nu}-\partial_{\nu}A^a_{\mu}+gf_{abc}A^b_{\mu}A^c_{\nu}
\end{equation}
where $f_{abc}$ are the structure constants of the $SU(3)$
group\footnote{$[T^l,T^m]=if^{lmn}T^n$, where $T^k$ are the
generators of the $SU(N)$ Lie group.}. As the QCD lagrangian is
symmetric under a non-Abelian gauge group, underlined by the
presence of the structure constants $f^{abc}$, the theory has some
non-trivial features that are not present in Abelian gauge theories
like quantum electrodynamics\cite{buballa}:
\begin{itemize}
    \item$\mathcal{L}_{QCD}$ contains gluonic self-coupling
    (three and four gluon vertices), which means the gluons themselves carry colour.
    \item At large momentum, $Q$, the QCD coupling behaves as:
    \begin{equation}
        \alpha_s\sim\frac{1}{\ln(Q^2/\Lambda^2_{QCD})}
        \label{running coupling}
    \end{equation}
    where $\Lambda_{QCD}\sim200MeV$ is the QCD scale parameter.
    \item Eqn.~(\ref{running coupling}) implies that  $\alpha_s(Q^2)\rightarrow0$
    as $Q^2\rightarrow\infty$, and $\alpha_s(Q^2)\rightarrow\infty$ when
    $Q^2=\Lambda^2_{QCD}$.
\end{itemize}
This behaviour of the strong force is known as asymptotic freedom,
and can be simply represented through the quark anti-quark potential
    \begin{equation}
        V(r)=-\frac{\alpha_s(r)}{r}+Kr
    \end{equation}
where $K$ is an experimentally determined constant, called the
string tension, with an estimated value of
$\simeq(420MeV)$\footnote{in units of energy over length we have
$K\simeq880MeV/fm$.}\cite{hands2}; and $\alpha_s$ is the coupling
given in Eqn.~(\ref{running coupling}), which varies with distance
as $1/\ln(r^{-1})$. Now it can be seen that for small distances the
first term dominates, in which case the strong force behaves like an
attractive Coulomb potential, and in the limit $r\rightarrow0$ the
quarks can be considered as free non-interacting
particles\cite{hands2}. However, with greater separations the
potential scales approximately linearly due to the self interaction
of the gluons, as seen through the second term. This is related to
the phenomenon of "confinement", i.e. to the empirical fact that
coloured objects, like quarks and gluons, do not exist as physical
degrees of freedom in the vacuum. So, as the coupling constant
(related to the potential) becomes larger for greater separations
perturbative treatments of QCD become less and less effective.
\subsection{Lattice QCD}
\label{chapter 1: lattice qcd}
 A perturbative treatment of QCD leads to
a successful description of the force between quarks at small
distances .  However, as already stated, at large distances a
perturbative treatment of QCD is less fruitful.  One method to
address the non-perturbative nature of QCD at large distances is
that of Lattice Gauge Theory, proposed by Wilson in 1974. In this
method all the fields are defined on a discrete Euclidean space-time
lattice with a nonzero lattice spacing $a$. Thus, in this way the
lattice QCD calculations can be numerically carried out without the
use of perturbative expansion.  Even with its advantages, lattice
field theory still has problem, such as:
\begin{itemize}
    \item It's difficult to discretise the fermion field in a
    chirally symmetric way.
    \item Simulations with reasonable light current quark masses are
    computationally very expensive.
    \item To simulate a smooth space time by implementing a large
    lattice volume and small lattice spacing requires a very large
    number of lattice sites.
    \item Simulations for non-zero chemical potential ($\mu\neq0$) are next to impossible due to the
    sampling weight (used in Monte-Carlo methods) becoming complex (Appendix~\ref{lattice qcd mu}).
\end{itemize}
However, with the continual improvement of lattice algorithms and
the advances in computing power, lattice QCD is the driving force in
our understanding of strongly interacting matter.
\section{Chiral Symmetry in QCD} An important feature of QCD is its
chiral symmetry for fermions with a vanishing mass (or approximate
symmetry as is the case for physical quarks).  Chiral symmetry is
related to the symmetries associated with a particle's handedness,
which in turn is defined by its helicity.  Before we go into chiral
symmetry we'll briefly discuss the concept of helicity and
chirality.
\subsection{Helicity and Chirality}
A particle propagating with spin $\vec{s}$ has helicity
$h=\vec{s}\cdot\vec{k}/|k|$, which is the projection of the spin
axis along the direction of its motion $\vec{k}$, where positive
helicity is right-handed and negative is left-handed\footnote{So for
spin $\frac{1}{2}$ particles there are two possible helicity
eigenstates given as: $h=\pm\frac{1}{2}$, known as left- and
right-handed states.}. Also, if the particle has a vanishing mass
its helicity would then be invariant under Lorentz transformations,
which (for the massless case) leads to two good quantum numbers
$B_L$ and $B_R$, referring to the separate conservation of left- and
right-handed particle numbers in the absence of external effects.
However, in the case of massive particles these quantum numbers are
not separately conserved, but their sum is a good quantum number:
$B=B_L+B_R$. So, through this concept of helicity we can define the
chirality operators that project out left- and right-handed field
states\footnote{The field operators which create and destroy a quark
are $\bar{\psi}$ and $\psi$ respectively.}
\begin{eqnarray}
\psi_L&=&\frac{1}{2}(1-\gamma_5)\psi=P_L\psi \qquad
\psi_R=\frac{1}{2}(1+\gamma_5)\psi=P_R\psi\nonumber\\
\bar{\psi}_L&=&\bar{\psi}\frac{1}{2}(1+\gamma_5)=\bar{\psi}P_R
\qquad \bar{\psi}_R=\bar{\psi}\frac{1}{2}(1-\gamma_5)=\bar{\psi}P_L
\label{chirality operator}
\end{eqnarray}
and these chiral field states, $\psi_L$ and $\psi_R$, satisfy the
equations
\begin{equation}
\gamma_5\psi_L=-\psi_L \qquad \gamma_5\psi_R=+\psi_R
\end{equation}
where $\gamma_5=i\gamma^0\gamma^1\gamma^2\gamma^3$
  and its eigenvalues are called
"chirality"\footnote{The gamma matrices also satisfy
$\gamma_5\gamma_5=1$ and $\{\gamma^{\mu},\gamma^5\}=0$.}. In
general, unlike helicity, chirality is not directly measurable. We
find that in the limit where $m\rightarrow 0$ (or for $E\gg m$) the
helicity and chirality of a particle are in one-to-one
correspondence, and the chiral fields associated with the massless
particles thus represent physical states. However, as helicity is
not Lorentz invariant massive particle fields must be expressed as a
sum of left- and right-handed chiral fields,
\begin{equation}
\psi=P_L\psi+P_R\psi=\psi_L+\psi_R
\end{equation}
which is the covariant formulation of massive particles. So in
this case, of massive particles, chirality and helicity are
distinct things, and thus the chiral states are not physical
states, but represent internal degrees of freedom.
\subsection{Chiral Symmetry}
\label{chiral symmetry}
 With this discussion of handedness and chirality we can go on to
discuss what is meant by chiral symmetry.  In general, chiral
symmetry is the symmetry associated with the independent
transformations of the left- and right-handed chiral states of a
particle. So, when we say QCD, for $N_f$ quark flavours, possesses
chiral symmetry under $SU(N_f)_L$ $\times$ $SU(N_f)_R$, we
actually refer to the lagrangian of the theory being invariant for
the separate transformations of the left and right-handed chiral
fields\cite{smit}
\begin{eqnarray}
&\psi&\rightarrow V\psi,\qquad V=V_LP_L+V_RP_R,\nonumber\\
&\bar{\psi}&\rightarrow \bar{\psi} \bar{V},\qquad
\bar{V}=V^{\dag}_LP_R+V^{\dag}_RP_L \label{chiral transformation}
\end{eqnarray}
where $V_L$, $V_R$ $\in$ $SU(N_f)$. This is equivalent to the
symmetry under the $SU(N_f)_V\times SU(N_f)_A$, with the
transformations
\begin{eqnarray}
    &SU(N_f)_V:& \psi\rightarrow e^{-\frac{i}{2}\tau_a\theta_a}\psi
    \qquad \bar{\psi}\rightarrow
\bar{\psi}e^{\frac{i}{2}\tau_a\theta_a}\\
    &SU(N_f)_A:& \psi\rightarrow
    e^{-\frac{i}{2}\gamma_5\tau_a\vartheta_a}\psi\nonumber \qquad
    \bar{\psi}\rightarrow
    \bar{\psi}e^{-\frac{i}{2}\gamma_5\tau_a\vartheta_a}
\end{eqnarray}
where $\tau_a$ are the generators of flavor $SU(N_f)$\footnote{For
$N_f=2$ the generators are the Pauli matrices.}. Chiral symmetry
would be exact in the limit of $N_f$ massless flavours, but for
non-vanishing mass ($m_0\neq0$) it is explicitly broken from
$SU(N_f)_V\times SU(N_f)_A$ to $SU(N_f)_V$\footnote{It can be shown
that $\mathcal{L}_{QCD}$ is invariant under $SU(N_f)_V$, but not
under $SU(N_f)_A$.}, as the mass term in $\mathcal{L}_{QCD}$ mixes
the chiral states $\psi_L$ and $\psi_R$\cite{koch}. However, though
quarks have non-vanishing masses, chiral symmetry is still a useful
concept for the up/down quarks $(N_f=2)$ as the masses are very
small and could be considered negligible compared to the QCD scale
parameter $\Lambda_{QCD}$ (and to a lesser extent for the inclusion
of the strange quark i.e. $N_f=3$). Thus, as long as the masses are
small compared to the relevant scale of the theory one may treat
$SU(N_f)_A$ as an approximate symmetry, so that predictions based
upon the assumptions of the symmetry should be reasonably close to
the actual results\cite{koch}.  As this is an approximate symmetry
we could ask: what are the observable phenomena related to chiral
symmetry being approximate rather than an exact symmetry? One of the
most obvious effects is seen through the non-zero (but small) mass
of the pions, i.e. the Goldstone bosons associated with the
spontaneous breaking of chiral symmetry for $N_f=2$ (which will be
discussed in the next section).
\subsubsection{Total Chiral Symmetry Group of QCD}
It can be shown that the QCD Lagrangian is also invariant under
phase transformations of the left- and right-handed quarks ($u_L$
and $d_L$), which is the $U(1)_L \times U(1)_R$ symmetry.  So the
total chiral symmetry group of the QCD Lagrangian for $N_f=2$
flavors can be written as
\begin{eqnarray}
U(2)_L\times U(2)_R&=&SU(2)_L\times SU(2)_R\times U(1)_L\times
U(1)_R\nonumber\\
&=&SU(2)_V\times SU(2)_A\times U(1)_V\times U(1)_A\nonumber\\
 \label{full
chiral symmetry group}
\end{eqnarray}
where the axial and vector $U(1)$ rotations can be written as
\begin{eqnarray}
  &U(1)_V:& \psi\rightarrow e^{-i\theta}\psi
    \qquad \bar{\psi}\rightarrow
\bar{\psi}e^{i\theta}\\
    &U(1)_A:& \psi\rightarrow
    e^{i\gamma_5\vartheta}\psi\nonumber \qquad
    \bar{\psi}\rightarrow
    \bar{\psi}e^{i\gamma_5\vartheta}
\end{eqnarray}
Having the full chiral symmetry group of QCD we can discuss the
physical manifestations of theses symmetries in nature.  The
$U(1)_A$ symmetry is (said to be) explicitly broken due to quantum
fluctuations\cite{glozman}, but the $U(1)_V$ symmetry is responsible
for baryon number conservation, and hence is  labeled as $U(1)_B$.
The pure unitary transformation $SU(2)_V$ corresponds to isospin
conservation, where the axial transformation, $SU(2)_A$, alters the
parity that is associated with a state.  This means the
manifestation of $SU(2)_A$ in nature would require that each isospin
multiplet be accompanied by a mirror multiplet that has opposite
parity.  However, as no such multiplets are observed in nature it is
an accepted view that $SU(2)_A$ is a broken symmetry of QCD, where
the resulting massless Goldstone bosons are associated with the
pions.  We can also write the conserved currents associated with
each of the symmetries (as given by Klevansky \cite{klevansky}):
\begin{eqnarray}
SU(2)_V&:& \qquad I^k_{\mu}=\bar{\psi}\gamma_{\mu}\tau^{k}\psi\Rightarrow Isospin\nonumber\\
U(1)_V&:&  \qquad I_{\mu}=\bar{\psi}\gamma_{\mu}\psi\Rightarrow Baryonic\nonumber\\
SU(2)_A&:& \qquad I^k_{5\mu}=\bar{\psi}\gamma_{\mu}\gamma_5\tau^{k}\psi\Rightarrow Chiral\nonumber\\
U(1)_A&:&  \qquad I_{5\mu}=\bar{\psi}\gamma_{\mu}\gamma_5\psi\Rightarrow Axial\nonumber\\
\label{chiral conserved currents}
\end{eqnarray}
We have used the notation $I$ for the currents, this is due to the
fact that in this thesis $J$ is reserved for source terms, as is the
convention in condensed matter physics.
\subsection{The QCD Vacuum and Spontaneous Symmetry Breaking}
\label{chap:intorduction-sec:qcd-subsec:spontaneous}As discussed,
due to the smallness of the current quark masses, QCD is said to
possess approximate chiral symmetry.  However, in the world around
us this symmetry is spontaneously broken due to dynamical mass
generation, occurring when quarks interact with vacuum quark
condensates. In this section we will discuss this phenomenon of
spontaneous breaking of chiral symmetry as it will help in
building a conceptual picture of the QCD phase diagram, which will
then lead to a greater understanding of the
superconducting/superfluid phases of strongly-interacting
particles at high baryon density.
\subsubsection{Dynamic Mass Generation}
In metallic superconductivity it is seen that a small
electron-electron attraction (due to effects of the surrounding
lattice) leads to bosonic particles of bound electron pairs, known
as Cooper pairs. These bosons then form a condensate in the ground
state of the metal leading to superconductivity.  In a similar way
we can assert that the ground state of QCD, or vacuum, is unstable
with respect to the formation of a quark condensate.

Quark-anti-quark pairs are created in the vacuum\footnote{These
fermion pairs have zero total momentum and angular momentum, thus
they contain net chiral charge i.e pairing left-handed quarks with
the antiparticles of right-handed quarks\cite{peskin}.}, near the
surface of the Dirac sea, as the binding energy of the
$\bar{\psi}\psi$ pair exceeds the energy needed to excite the quark
anti-quark pair.  Once excited the strong attractive interaction
between them causes the bound fermion pairs to condense, which leads
to the creation of an energy gap. The resulting vacuum quark
condensate is quantitatively characterized by a nonzero vacuum
expectation value
\begin{equation}
<\bar{\psi}\psi>=<0|\bar{\psi}_L\psi_R+\bar{\psi}_R\psi_L|0>\neq0
\label{chiral quark condensate}
\end{equation}
This nonzero expectation value signals that the vacuum mixes the
quark chiral states leading to them acquiring an effective mass.
This can be seen conceptually if you consider that a left-handed
quark\footnote{for $E\gg m$ we can assume that $m\rightarrow 0$ so
its helicity and chirality will be equivalent.} propagating through
a vacuum can be annihilated by $\psi_L$, leaving $\bar{\psi}_R$ to
create a right handed quark chiral state with the same momentum. As
this continually happens, with the quark traveling through the QCD
vacuum, its chiral state will flip at a rate proportional to
$<\bar{\psi}\psi>$, which implies it would propagate just as if it
had a mass\footnote{A naive, but some what helpful, picture is that
of a spoon being dragged through honey, in such a case it would seem
to have a greater apparent mass (or inertia) due to the viscous drag
of the honey\cite{griffiths}.} \cite{hands2}.
\subsubsection{Spontaneously Broken Chiral Symmetry}
Through this dynamically-generated mass, called the constituent mass
$\Sigma$ as opposed to the current mass $m$, chiral symmetry is
spontaneously broken. This spontaneous breaking of symmetry,
occurring due to QCD's own dynamics, leads to massless Goldstone
bosons. For the case of $N_f=2$\footnote{From Goldstones Theorem it
can be shown that the breaking of the symmetry $SU(2)_V\times
SU(2)_A$ (3+3 generators) to $SU(2)_V$ (3 generators) gives rise to
3 massless Goldstone bosons.} we end up with three massless
Goldstone Bosons, which are identified with the isospin triplet of
relatively light mesons, the pions, $\pi^{\pm}$ and $\pi^0$. These
pions are light, but not massless, which (as previously discussed)
is a consequence of the fact that chiral symmetry was initially not
an exact symmetry as quarks have a nonzero, but small, mass to begin
with.

This generation of (constituent) rest mass, and the associated
chirally broken phase, is portrayed in Fig.~\ref{E-K chirality}.
\begin{figure}
    \begin{center}
        \includegraphics[width=1\textwidth]{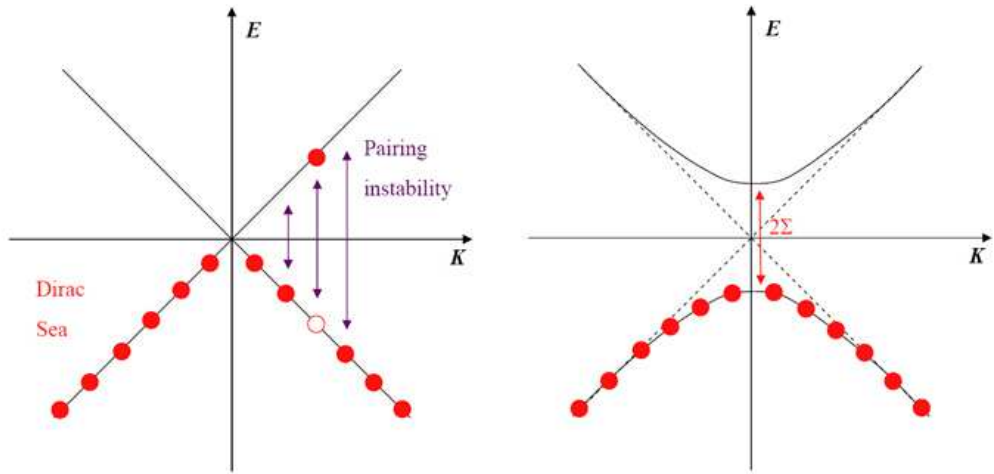}\\
        \caption{$\bar{\psi}\psi$ pairing instability leading to a chirally broken phase of the vacuum.}
        \label{E-K chirality}
    \end{center}
\end{figure}
In this energy-momentum diagram it can be seen that, due to the
pairing of the quark anti-quark pairs, an energy gap of $2\Sigma$ is
created between the highest (quark) and lowest (anti-quark) states.
This gap then represents a rest mass that is far greater than the
current mass i.e. $\Sigma\gg m$, which can be interpreted as a
physical representation of a vacuum with broken chiral
symmetry\footnote{The vacuum with restored chiral symmetry also has
a gap, equal to $2m$, however as it is very small we talk about
approximate chiral symmetry.}.

In Sec.~\ref{chap:introduction-sec:phase} we will go onto discuss
how a variation in thermodynamic conditions could lead to the
restoration of chiral symmetry, and how extreme variation then leads
to exotic strongly interacting matter i.e. colour
superconductivity/superfluidity.
\section{The QCD Phase Diagram}\label{chap:introduction-sec:phase}
\footnote{This section is a condensed summary of 'The Phase Diagram
of QCD' by Simon Hands \cite{hands2}.} When it had become clear that
hadrons were indeed a state of confined quarks and gluons, it was
suggested that they should become deconfined at high temperatures or
densities when the hadrons strongly overlap and lose their
individuality\cite{buballa}. In this picture, we then have two
phases, the "hadronic phase" where quarks and gluons are confined,
and the quark-gluon plasma where they are deconfined. Such phases of
strongly interacting matter and the associated transitions between
them, with external thermodynamic control parameters like
temperature and density, make up a map of The QCD Phase Diagram.
\subsection{A Classic Phase Diagram of $H_2O$}
A classical example of a phase diagram is that of $H_2O$,
Fig.~\ref{fig-h20}.
\begin{figure}
    \begin{center}
        \includegraphics[width=0.7\textwidth]{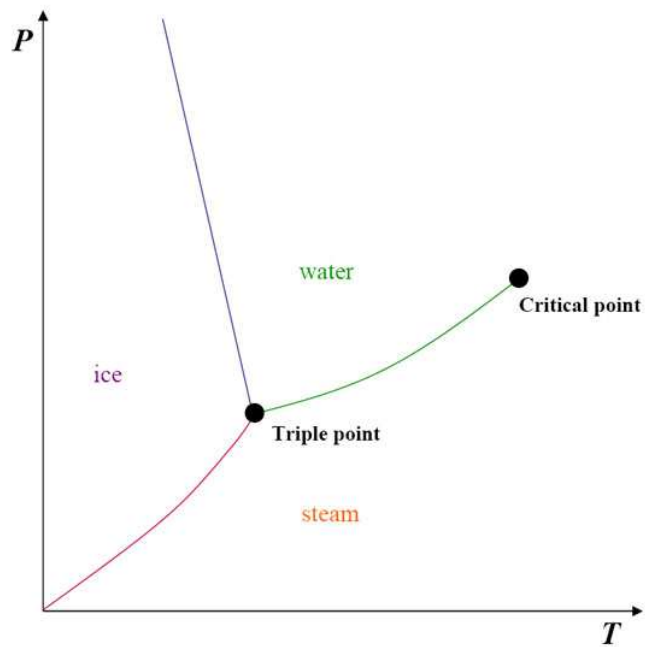}\\
        \caption{The phase diagram of $H_2O$ (not to scale).}
        \label{fig-h20}
    \end{center}
\end{figure}
In this diagram the thermodynamic control parameters are temperature
$T$ and pressure $P$, and the different manifestations (or phases)
of $H_2O$ are separated into three regions of ice, water and steam,
with the separating boundaries between them marked by the
equilibrium coexistence curves $P(T)$.  Thus, a first order phase
transition such as melting or boiling is observed when moving along
a path in the $(T,P)$ plane that crosses the equilibrium boundaries.
In this diagram there are two special points: the critical point
($T_c=650K$, $P_c=2.21\times10^7Nm^{-2}$) where the phases of water
and vapour become indistinguishable, and the triple point
($T_{tr}=273.16K$,$P_{tr}=600Nm^{-2}$) where all three phases
coexist.
\subsection{A Map of the QCD Phase Diagram}
 A possible schematic of the QCD phase diagram is presented
in Fig.~\ref{fig-qcd}, with the thermodynamic control parameters
temperature, $T$, and baryon chemical potential, $\mu$.
\begin{figure}
    \begin{center}
        \includegraphics[width=0.9\textwidth]{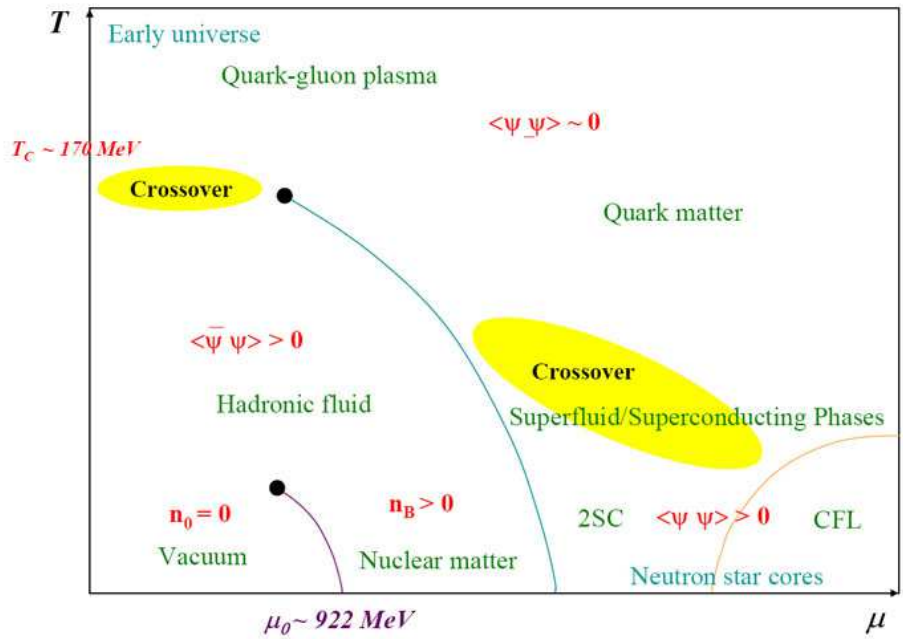}\\
        \caption{The proposed phase diagram of QCD (not to scale).}
        \label{fig-qcd}
    \end{center}
\end{figure}
In this diagram we can note several locations which have analogues
in condensed matter physics.  The first location is the bottom left
hand corner of the phase diagram where $T$ and $\mu$ are both small.
Here the thermodynamic behavior of QCD can be described as that of a
vapour of hadrons, i.e. composite states of quarks and/or
anti-quarks.  In this location extensive work has been carried out
to classify and quantify the bound states of strongly interacting
matter, hence, this area has a certain amount of resemblance to
relativistic atomic physics.  It is then argued, and has been shown
through numerical calculations\footnote{For a discussion of this see
\cite{hands2} and references therein.} that as we increase $T$ the
hadronic vapour phase cannot persist. Eventually there comes a
point, found to be $T_c\simeq170MeV$\cite{karsch}, where the
dominant degrees of freedom are no longer hadrons, but quarks and
gluons themselves.  This "gas" of strongly interacting matter is
known as the quark-gluon plasma (QGP), thus the upper left region of
the phase diagram has the classic analogue of relativistic plasma
physics.  Two possible candidates for the formation of the QGP phase
are the early universe, immediately after the Big Bang, and high
energy collisions between nuclear particles.

The final area of the phase diagram, and also the most elusive, lies
along the $\mu$-axis.  Unlike simulations along the $T$-axis,
lattice gauge theory simulation become ineffective when applied to
QCD with $\mu\neq0$.  However, through extrapolation methods it is
seen that as you increase $\mu$, for zero temperature, the ground
state (i.e. the vacuum) persists until $\mu$ reaches the value of
nuclear matter (nucleon rest mass minus the binding energy), at
which point it becomes energetically favorable to fill the ground
state with a bound nucleonic fluid. This onset value is estimated at
$\mu_0=922MeV$, at which point baryon density $n_B$ jumps from zero
to nuclear density $n_{B0}\simeq 0.16fm^{-3}$. This is referred to
in \cite{hands2} as 'room chemical potential' as the vacuum and
nuclear matter coexist at this point.   Extrapolation methods, can
be employed for densities up to $2-3n_{B0}$, however beyond this we
are forced to rely on approximate treatments such as the MIT bag
model or the NJL model.  Through such treatments it is estimated
that as $n_B$ increases we again expect a transition from a phase
where matter exists in the form of nucleons to one where the
dominant degrees of freedom are quarks and gluons.  This phase of
dense, strongly interacting, matter is believed to exist at the
cores of neutron stars.  It is also speculated that under such
conditions a phenomenon similar to the Bardeen-Cooper-Schrieffer
(BCS) instability, that leads to superconductivity in metals and
superfluidity in liquid $^3$He at low temperature, results in the
formation of diquark pairs and the onset of a colour
superconducting/superfluid phase (see Sec. \ref{coloured matter}).
Therefore, the lower right region of the diagram has similarities to
a branch of condensed matter physics; and it is this region of the
phase diagram with which this thesis will be dealing.
\subsection{Restoration of Chiral Symmetry}
Before we go on to coloured BCS phenomenon (the area with which this
thesis is concerned) we will discuss the restoration of chiral
symmetry in the various areas of the phase diagram, in particular
the areas of high $T$ and $\mu$, i.e. the upper part around to the
bottom right of the phase diagram.
\subsubsection{$<\bar{\psi}\psi>\sim0$ For Extreme $T$}
It has been seen through numerical simulations that the onset of QGP
coincides with the restoration of chiral symmetry\cite{karsch}. As
you move from the phase of bound strongly interacting matter, at low
temperature and density, up to the high temperature region dominated
by QGP there is seen to be a drop in the vacuum quark condensate
i.e. the order parameter $<\bar{\psi}\psi>$ tends to zero in the QGP
phase. However, to be precise it has been shown \cite{hands2} that
the order parameter $<\bar{\psi}\psi>$ doesn't strictly vanish, but
rather drops very steeply in the transition region, and this
transition is seen to occur at $T_c\simeq170MeV$. Thus, in the QCD
phase diagram the formation of QGP along the $\mu=0$ axis is
naturally referred to as a crossover rather than a true first or
second order phase transition. In a hand waving way we can interpret
the "destruction" of the vacuum quark condensate as a result of the
high temperature influence on the strong coupling constant, $g(T)$.
We can say that at very high temperatures, where QGP occurs, large
energies are exchanged in inter-particle collisions, which results
in the weakening of the strong interaction due to asymptotic
freedom. Therefore, as the energy needed to excite quark anti-quark
pairs is no longer below the binding energy between them the quark
condensate does not form, which in turn leads to the restoration of
chiral symmetry.
\subsubsection{$<\bar{\psi}\psi>\sim0$ For Extreme $\mu$}
As previously discussed, to study the high density region of the QCD
phase diagram, i.e. above $2-3n_{B0}$ (and $T\approx0$), one must
employ alternative models to lattice gauge theory such as the MIT
bag model and the NJL model. Each model has its pros and cons: the
most important characteristic of the bag model is its property of
quark confinement, and that of the NJL model is chiral symmetry and
its spontaneous breakdown in the vacuum; on the other hand the bag
model violates chiral symmetry and the NJL model does not possess
confinement.  So one must decide which property is of greater
importance and employ the appropriate model. Through simulations,
using the NJL model, it is seen that as you increase baryon density
beyond the nuclear density $n_{B0}$, i.e. move from the hadronic
phase to that of strongly interacting matter, a phase transition
takes place from the chirally broken phase to one of restored chiral
symmetry \cite{hands1}. This restoration of chiral symmetry in the
high density regime can be considered due to the quarks becoming the
dominant degrees of freedom at this density.  At extreme densities
($\mu\sim1200MeV$ \cite{hands2}) hadrons strongly overlap, lose
their individuality and start to behave like one big mass of quark
matter; which is thought to occur in neutron stars. Being fermions,
the quarks occupy the Fermi sea up to the level $E_F$. This leads to
$\bar{\psi}\psi$ vacuum excitations, for $|k|\ll k_F$, becoming
Pauli-blocked. Therefore, at some point, these pairs require  so
much energy to excite that it becomes preferable to revert to a
chirally symmetric ground state. In this way, at high density,
chiral symmetry is restored.
\section{Coloured BCS Phenomenon}
\label{coloured matter}
\subsection{BCS Processes}
In BCS theory the pairing of Fermions (known as Cooper pairs) leads
to the formation of Bosonic particles, which condense and leave an
energy gap $2\Delta$ that is equal to the energy of the binding
between the pair, Fig. \ref{E-K super-conductor}.  In metals this
BCS condensation of electrons leads to superconductivity, i.e. the
flow of electric current without resistance.  It is also seen that
BCS condensation is accompanied by the breaking of the $U(1)$ local
electromagnetic gauge symmetry.  Such BCS processes are also
relevant in the study of superfluidity (the flow of fluid with
negligible viscosity).  One type of superfluid, $^4He$, is a Bosonic
fluid, and simply undergoes Bose-Einstein condensation at low
temperatures. However, $^3He$, a Fermionic fluid, undergoes a BCS
type process which leads to the condensation of bound Cooper pairs.
As both superconductor and superfluid have an energy gap, the way to
distinguish between the two is that a superfluid is characterised by
a ground state which does not respect a global symmetry of the
underlying action.
\begin{figure}
    \begin{center}
        \includegraphics[width=1\textwidth]{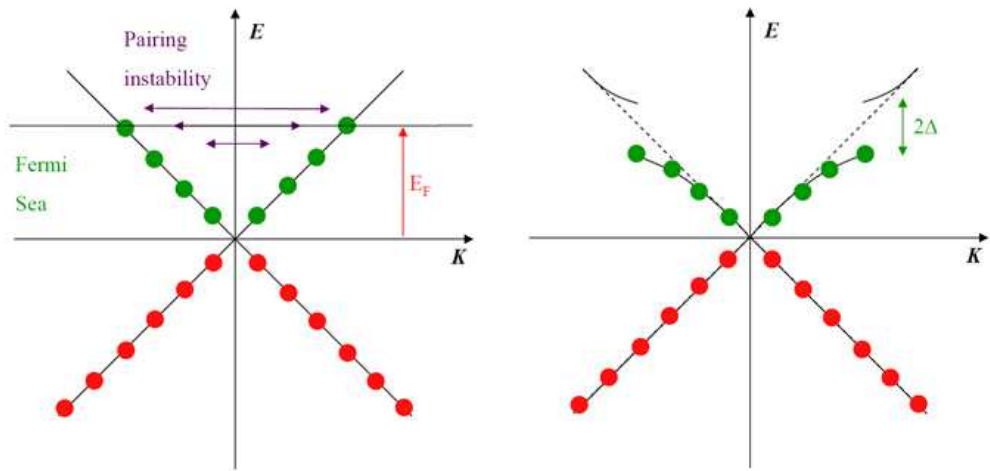}\\
        \caption{$\psi\psi$ pairing instability leading to super-conductivity.}
        \label{E-K super-conductor}
    \end{center}
\end{figure}
\subsection{BCS in QCD}
Due to the strong attraction between quarks the BCS process leads to
some interesting results in QCD.  One of these, as we have already
seen in sec. \ref{chap:intorduction-sec:qcd-subsec:spontaneous}, is
the condensation of quark-antiquark pairs.  This leads to
spontaneous Chiral symmetry breaking, and a resulting effective
quark mass which is much higher than the current mass.  Also, it is
believed that at high densities, when chiral symmetry is restored,
and quarks are deconfined, condensation of diquark pairs occurs; and
the resulting $qq$ wave function is a gauge non-singlet. This
condensation leads to the formation of an energy gap separating the
ground and excited states by $2\Delta$.  As in "orthodox" BCS
processes, this condensation of diquark pairs is accompanied by the
breaking of local colour gauge symmetry, $SU(3)_c$; and in analogy
with electromagnetic superconductivity this high density phenomenon
is known as colour superconductivity (CSC).

If we were able to conduct Lattice QCD simulations at $\mu\neq0$ it
could be possible that we'd see the colour superconducting phase.
However, due to the sampling weight, used in Monte-Carlo methods,
becoming complex, to study quark matter in the high density phase we
require effective field theories, such as the NJL model. The NJL
model is a purely Fermionic model, containing no gauge degrees of
freedom, and interactions are simply represented by a four-point
interaction. One of the most important features that makes the NJL
model an ideal effective theory of QCD at high density is that it
not only observes the same (Chiral) symmetries of QCD, but the
breaking of these symmetries in the vacuum is analogous to BCS
superconductivity.  For this reason it is viewed as an appropriate
model to study coloured BCS phenomena. However, in the NJL model the
$qq$ wave function is gauge singlet, which implies the ground state
is not superconducting, but rather superfluid\footnote{Another
method to simulate the high density region of QCD is using 2 colour
QCD, which also has a diquark condensate gauge singlet state.}.
}

{\typeout{The Nambu-Jona-Lasino Model}
\chapter{The Nambu-Jona-Lasino Model }
\label{nabu-model} Quantum chromodynamics is the accepted theory of
strong interactions and at short distances is highly successful.
However, as discussed in the previous chapter, for larger distances
where perturbative techniques break down lattice gauge theory is
required to calculate the QCD expectation values.  This method
itself has problems such as the requirement of huge computing power,
problems associated with including fermions on a discretized
lattice, problems for $\mu\neq0$, etc (See Sec. \ref{chapter 1:
lattice qcd}). For this reason it is reasonable to isolate the
relevant physics associated with a process of interest and construct
an approximate model of the exact theory that accentuates the main
features of the theory; to study strongly interacting matter at high
density (i.e. $\mu\neq0$) we do just this.  In this chapter we start
off by discussing the Nambu-Jona-Lasino (NJL) model as a model of
QCD, its chiral symmetry and other associated technical issues, and
also the benefits of its use for lattice calculations at high
density. We then go onto discuss the lattice transcription of the
NJL model in $2+1$ dimensions. Finally we discuss QCD calculations
involving the lattice NJL model in 2+1 dimensions, referring to the
results of \cite{hands1} and how they relate to the investigations
of this thesis.

\section{The NJL Model of QCD}
Non-perturbative QCD calculations in the high density regime are
next to impossible due to the measure of the Euclidean path integral
becoming complex for baryon chemical potential $\mu\neq0$.  This
phenomenon, known as the sign problem, involves the QCD path
integral effectively generating meaningless complex probabilities.
One way to overcome this problem is by the use of the pre-QCD model
of Nambu and Jona-Lasino \cite{nambu1,nambu2}. In this model
nucleons were considered to interact through a two body interaction,
and the resulting lagrangian was symmetric under the full chiral
symmetry group given in Eqn.~(\ref{full chiral symmetry group}).
However, in accordance with experimental observation the $U(1)_A$
symmetry was later excluded.

As this theory was conceived before the advent of quarks, Nambu and
Jona-Lasino used nucleons as the elementary building blocks of the
model.  In analogy to the effective electron-electron interaction in
BCS theory the nucleon-antinucleon attractive interaction was
considered to be responsible for the formation of Cooper pairs of
nucleons and antinucleons, which results in a nucleon condensate.
This condensate would then lead to a vacuum mass gap and the
breaking of chiral symmetry, where the pion could then be identified
as the Goldstone boson occurring in the theory due to the breaking
of the axial symmetry.

With the possibility of quarks the NJL model can be reinterpreted as
a simple model of QCD, with the quarks taking the place of the
nucleons. However in this simplified model the strong interaction,
mediated by the gluons, is excluded; instead the interaction between
quarks is assumed to be a point-like four-point interaction (for
$N_f=2$). This means it does not describe quark confinement - the
phenomenon whereby single isolated quarks are never observed.  On
the other hand the NJL model has other interesting properties which
make it a suitable substitute for QCD, i.e. the model's global
symmetries and the patterns of their breaking are similar to those
of QCD \cite{klevansky}.

\subsection{NJL Lagrangian}
In Euclidean space, the original lagrangian density describing the
NJL model with $N_f=2$ quark flavours in the isospin representation
of $SU(2)$ can be written as
\begin{equation}
\mathcal{L}=\overline{\psi}^p(\dslash+m_0)\tau^{pq}_0\psi^q
-\frac{g^2}{2}[(\overline{\psi}^p\tau^{pq}_0\psi^q)^2
-(\overline{\psi}^p\gamma_5\vec{\tau}^{pq}\psi^q)^2] \label{simple
NJL in Euclidean space}
\end{equation}
where $\psi$ and $\overline{\psi}$ are independent Grassmann Dirac
4-spinors representing the quark fields, $m_0$ is the current, or
bare-mass of the quark, $\vec{\tau}\equiv(\tau_1,\tau_2,\tau_3)$ is
a vector of the $2\times2$ Pauli matrices, which run over internal
isospin or flavour degrees of freedom, and $\tau_0$ is the unit
matrix.

When the NJL model was reinterpreted as a quark model, with $\psi$
as the quark field with two flavour and 3 colour degrees of freedom,
it was kept in the original NJL form (\ref{simple NJL in Euclidean
space}), and in this thesis we will work with this form of the
model. However, Eqn.~(\ref{simple NJL in Euclidean space}) is not
unique and other chirally symmetric interaction terms can be
included.  A more general form of the NJL model is given as
\cite{buballa}:
\begin{equation}
\mathcal{L}=\overline{\psi}^p(\dslash+m_0)\tau^{pq}_0\psi^q
-\frac{g^2}{2}[(\overline{\psi}^p\tau^{pq}_0\psi^q)^2-(\overline{\psi}^p\vec{\tau}^{pq}\psi^q)^2+(\overline{\psi}^p\gamma_5\tau^{pq}_0\psi^q)^2
-(\overline{\psi}^p\gamma_5\vec{\tau}^{pq}\psi^q)^2] \label{NJL in
Euclidean space}
\end{equation}

\subsection{Bosonization}
\label{bosonization} Through the introduction of auxiliary scalar
and pseudo-scalar fields, $\sigma$ and $\vec{\pi}$, the NJL model
can be rewritten in a bosonized form, which makes it easier to treat
both numerically and analytically. Starting with the NJL Lagrangian
(\ref{simple NJL in Euclidean space}) we can introduce the
generating functional
\begin{equation}
Z\sim\int\mathcal{D}\psi\mathcal{D}\bar{\psi}\exp[\int
d^4x\mathcal{L}(\psi,\bar{\psi})] \label{path integral njl}
\end{equation}
Then using the fact that path integrals of Gaussian
functions can be performed exactly \cite{klevansky},
\begin{equation}
\int \mathcal{D}\Phi \exp [ \int d^4x(\pm A\Phi-
B\Phi^2)]\sim\exp[\int d^4x \frac{A^2}{4B}]
\end{equation}
we can write:
\begin{eqnarray}
\exp[\int d^4x G (\bar{\psi}\psi)^2]&\sim&\int\mathcal{D}\sigma
\exp\{\int
d^4x[(\bar{\psi}\psi)\sigma-\frac{1}{4G}\sigma^2]\}\nonumber\\
\exp[\int d^4x G
(\bar{\psi}i\gamma_5\vec{\tau}\psi)^2]&\sim&\int\mathcal{D}\pi
\exp\{\int d^4x[(\bar{\psi}i\gamma_5\vec{\tau}
\cdot\vec{\pi}\psi)-\frac{1}{4G}\vec{\pi}^2]\}\nonumber\\
\end{eqnarray}
where we have $G=-\frac{g^2}{2}$.  Therefore, we can rewrite
(\ref{path integral njl}) in a bosonized form as
\begin{eqnarray}
Z&\sim&\int\mathcal{D}\psi\mathcal{D}\bar{\psi}\mathcal{D}\sigma\mathcal{D}\vec{\pi}
\exp\{\int d^4x[
\bar{\psi}[\dslash+m_0+(\sigma+i\gamma_5\vec{\tau}\cdot\vec{\pi})]\psi\nonumber\\
&-&\frac{1}{4G}(\sigma^2+\vec{\pi}\cdot\vec{\pi})]\}
\label{bosonized path integral njl}
\end{eqnarray}
Now using $\Phi=\sigma+i\vec{\pi}\cdot\vec{\tau}$ and
$\sigma^2+\vec{\pi}\cdot\vec{\pi}=\frac{1}{2}Tr \Phi^{\dagger}\Phi$
we have the bosonized form of the NJL lagrangian:
\begin{eqnarray}
\mathcal{L}=\bar{\psi}[\dslash+m_0+(\sigma+i\gamma_5\vec{\tau}\cdot\vec{\pi})]\psi
-\frac{1}{8G}Tr\Phi^{\dagger}\Phi \label{bosonized path integral njl
simple}
\end{eqnarray}

\subsection{Chiral Symmetry of the NJL Model}
\label{chiral njl model} The simplest way to show that the NJL
lagrangian is symmetric under chiral transformations (\ref{chiral
transformation}) is to use the bosonized lagrangian in
(\ref{bosonized path integral njl simple}). For vanishing mass the
kinetic term can be easily shown to be
symmetric\footnote{$\bar{V}V=V^{\dagger}_RV_LP_L+V^{\dagger}_LV_RP_R$
implies mass term breaks chiral symmetry (Sec.~\ref{chiral
symmetry}).}:
\begin{eqnarray}
\bar{\psi}\gamma^{\mu}\partial_{\mu}\psi\rightarrow
\bar{\psi}\bar{V}\gamma^{\mu}\partial_{\mu}V\psi&=&\bar{\psi}\gamma^{\mu}
(V^{\dagger}_LV_L P_L+V^{\dagger}_RV_RP_R)\partial_{\mu}\psi\nonumber\\
&=&\bar{\psi}\gamma^{\mu}
(P_L+P_R)\partial_{\mu}\psi\nonumber\\
 &=&\bar{\psi}\gamma^{\mu}\partial_{\mu}\psi
\end{eqnarray}
However, the symmetry of the bosonic terms is a little more subtle.
Bosonic fields, as in the last term of (\ref{bosonized path integral
njl simple}), transform as\cite{smit}:
\begin{equation}
\Phi\rightarrow V_L\Phi V^{\dagger}_{R} \label{boson chiral
transform}
\end{equation}
Using this we can show that the first bosonic term is symmetric, by
first writing it in the form:
\begin{eqnarray}
\bar{\psi}(\sigma+i\gamma_5\vec{\tau}\cdot\vec{\pi})\psi&=&
\bar{\psi}_L(\sigma+i\vec{\tau}\cdot\vec{\pi})\psi_R+\bar{\psi}_R(\sigma-i\vec{\tau}\cdot\vec{\pi})\psi_L\nonumber\\
&=&\bar{\psi}_L\Phi\psi_R+\bar{\psi}_R\Phi^{\dagger}\psi_L
\end{eqnarray}
where
\begin{equation}
\Phi=\left(
    \begin{array}{cc}
      \sigma+i\pi_3 & \pi_2+i\pi_1 \\
      -\pi_2+i\pi_1 & \sigma-i\pi_3 \\
    \end{array}
  \right)
\end{equation}
Now under the chiral transformation we have:
\begin{eqnarray}
\bar{\psi}_LV^{\dagger}_L\Phi
V_R\psi_R+\bar{\psi}_RV^{\dagger}_R\Phi^{\dagger}V_L\psi_L
\end{eqnarray}
which is symmetric, as the bosonic field $\Phi$ transforms as
(\ref{boson chiral transform}).
\section{The Lattice Transcription of the NJL Model} In this thesis
we use the lattice transcription of the NJL model in $2+1$
dimensions, as used in \cite{hands1}. We replace the space-time
continuum with a 3-dimensional lattice, where each site is separated
from its nearest neighbours by an arbitrary lattice spacing $a$. In
dimensions $d > 2 + 1$ this lattice spacing must be chosen for the
effect of introducing an ultra-violet (UV) cutoff $\Lambda\sim
a^{-1}$, which then regularises the divergences and makes the theory
mathematically well-defined, however, in $d = 2 + 1$ dimensions the
theory is renormalizable and well defined on the lattice for an
arbitrary cutoff\cite{hand-fermi}\footnote{As the interaction
between quarks is assumed to be a point-like in character, the
theory is not renormalizable for dimensions greater than $3+1$.}.

The structure of our model consists of the fermion fields being
defined on the lattice sites $x$ and the bosonic fields being
defined on the dual lattice sites $\tilde{x}$, translated from the
original lattice by $(\frac{1}{2} , \frac{1}{2} ,\frac{1}{2} )$. The
space-time integrals and differentials are represented by sums and
finite differences, so that in the staggered fermion notation the
noninteracting quark action
\begin{equation}
S = \int d^3x \bar{\psi}(\dslash + m_0)\psi
\end{equation}
is given by
\begin{equation}
S =\sum_x a^3\{{\frac{1}{
2a}\sum^2_{\nu=0}\eta_{\nu}(x)[(\bar{\chi}_x \chi_{x+\hat{\nu}})
-(\bar{\chi}_x \chi_{x-\hat{\nu}})] + m_0(\bar{\chi}_x \chi_x)}\}
\end{equation}
where $\bar{\chi}$ and $\chi$ are the Grassmann-valued staggered
fermion fields defined on the lattice sites, and $\eta_{\nu}(x)$ is
the Kawamoto-Smit phase $(-1)^{x_0+...x_{\nu-1}}$. Now, using this
definition of the free action, we can go onto describe the
construction of the full NJL model as used in this thesis.

\subsubsection{Lattice NJL Model for $\mu=0$}
The NJL model on the lattice for $\mu=0$, in its bosonised form,
is defined by the Euclidean action
\begin{equation}
S=\sum_x \bar{\chi}M[\Phi]\chi+\frac{1}{g^2}\sum_{\tilde{x}}
tr\Phi^{\dag}\Phi\nonumber\\
\end{equation}
where $\chi$, $\overline{\chi}$ are, as before, the fermionic
fields defined on the lattice sites, and
$\Phi\equiv\sigma+i\overrightarrow{\pi}\cdot\overrightarrow{\tau}$
is a $2\times2$ matrix of bosonic auxiliary fields, again as
before, defined on the dual lattice sites. Also, $M[\Phi]$ is the
kinetic operator, and is in the form for staggered lattice
fermions interacting with scalar fields,
\begin{eqnarray}
M^{pq}_{xy}&=&\frac{1}{2}\delta^{pq}\sum_{\nu=0}^2[\eta_\nu(x)(\delta_{y
x+\hat{\nu}} - \delta_{y x-\hat{\nu}})
+2m\delta_{xy}]\nonumber\\
&+&\frac{1}{8}\delta_{xy}\sum_{<\tilde{x},x>}
[\sigma(\tilde{x})\delta^{pq}+i\varepsilon(x)\vec{\pi}
(\tilde{x})\cdot\vec{\tau}^{pq}] \label{kinetic operator mu=0}
\end{eqnarray}
Here the parameters are bare fermion mass $m$ and coupling constant
$g^2$. The Pauli matrices, $\vec{\tau}$, are normalized to
$\textrm{tr}(\tau_i \tau_j)=2\delta_{ij}$ and act on the internal
$SU(2)$ isospin indices $p,q=1,2$, and $\varepsilon(x)$ denotes the
phase $ (-1)^{x_0+x_1+x_2}$.  Finally we have $<\widetilde{x},x>$,
which represents the sum over the set of $8$ dual lattice sites
neighbouring $x$.  In this bosonized form we can integrate over the
auxiliary $\Phi$ fields which leads to an equivalent action in terms
of fermions that self-interact via a four-point contact term
proportional to $g^2$, as shown in Sec. \ref{bosonization}.

\subsubsection{Lattice NJL Model for $\mu\neq0$}
To introduce a chemical potential $\mu$ into a fermionic field
theory we must incorporate a term of the form
$\mathcal{L}_{\mu}=\mu\bar{\psi}\gamma_0\psi$ into the fermionic
Lagrangian.  The inclusion of this term can be understood in a
simple way by considering that to introduce a source into a
Lagrangian we must incorporate a source term of the type
$\mathcal{L}_{source}=J\phi$, where $J$ is the source for the field
$\psi$.  Therefore, as
$\bar{\psi}\gamma_0\psi\sim\psi^{\dagger}\psi$ is the quark number
operator $N_q$, we can think of $\mu$ as a source for this operator,
which leads to the creation of fermions over anti-fermions as a more
energetically favourable process.  Therefore, the chemical potential
term would then be incorporated into the partition function in the
usual way, as $e^{-\mu N_q}$. However, introducing a chemical
potential in a discretized field theory is not as simple.  We cannot
simply implement a term of the form given above and naively
discretise it, as it can be shown that this leads to divergences in
the energy density of the theory \cite{karsch,gavai}. This can be
resolved if we consider that incorporating a chemical potential in
the form given above is like the zeroth component of a vector
potential interaction term in a fermionic Lagrangian,
$\bar{\psi}\gamma_{\mu}\psi A^{\mu}\Rightarrow
\bar{\psi}\gamma_{0}\psi A^{0}$.  Thus, due to this similarity to a
gauge field, we must incorporate the chemical potential term in a
"gauge-invariant" way. This is done by introducing $\mu$ into the
zeroth derivative by multiplying the forward time difference by
$e^{\mu}$ and the backward time difference by $e^{-\mu}$.  In this
way the kinetic operator (\ref{kinetic operator mu=0}) becomes
\begin{eqnarray}
M^{pq}_{xy}&=&\frac{1}{2}\delta^{pq}[(e^{\mu}\delta_{y x+\hat{0}}
-e^{-\mu} \delta_{y x-\hat{0}})\nonumber\\
 &+&\sum_{\nu=1,2}\eta_\nu(x)(\delta_{y x+\hat{\nu}} - \delta_{y x-\hat{\nu}})
+2m\delta_{xy}]\nonumber\\
&+&\frac{1}{8}\delta_{xy}\sum_{<\tilde{x},x>}
[\sigma(\tilde{x})\delta^{pq}+i\varepsilon(x)\vec{\pi}
(\tilde{x})\cdot\vec{\tau}^{pq}] \label{M}
\end{eqnarray}

\subsubsection{Introduction of the Diquark Sources}
Due to limitations of computer memory we are forced to work on a
finite volume system, and working under such conditions leads to
certain complications.  One such issue is that due to finite volume
effects on spontaneous symmetry breaking. A simple example of this
phenomenon is seen in the O(N) model. One of the most important
properties of this model is the effect of spontaneous spin symmetry
breaking at the critical temperature, Appendix~\ref{condensed
matter}.  In the 2-dimensional O(N) model, above the critical
temperature, the spins at each lattice site are randomly orientated
due to thermal fluctuations, hence we have the average value
$<S_i>=0$. When we hit the critical temperature, thermal
fluctuations become less dominant compared to spin interactions and
the correlation length increases, this in turn leads to spontaneous
magnetisation (i.e. magnetisation with a zero external field). Then,
at $T=0$ the magnetisation is at its maximum, and the correlation
length is comparable to the size of the lattice. However, it is seen
that such effects only occur in the thermodynamic limit
\begin{equation}
M=\lim_{N\rightarrow \infty} \frac{1}{N}\sum_i <S_i>\neq0
\end{equation}
\begin{figure}
    \begin{center}
        \includegraphics[width=0.7\textwidth]{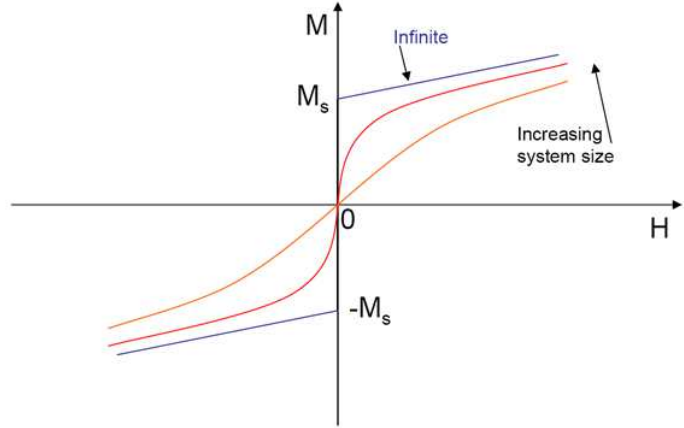}\\
        \caption{Magnetisation $M$ against external field $H$.}
        \label{Mag}
    \end{center}
\end{figure}
These finite volume effects are shown in Fig.~\ref{Mag}, which shows
magnetisation verses external magnetic field.  It can be seen that
spontaneous symmetry breaking only occurs in the limit of
$N\rightarrow \infty$.  So how can we simulate an infinite lattice?
This problem can be overcome in computational calculations by
including a symmetry breaking term that represents the external
field. Therefore, for a finite volume system we can conduct
simulations with various values of the external field, such as the
magnetic field $H$ in the O(N) model, and then extrapolate to the
limit of the field going to zero, $H\rightarrow0$. This then gives
the spontaneous magnetisation value of the system at temperature
$T$. A similar problem is seen in the study of chiral symmetry
breaking, where we include a finite bare mass $m$ to allow the
measurement of the chiral condensate $<\bar{\psi}\psi>$ on a finite
volume. It is for a similar reason we include the diquark and
anti-diquark symmetry breaking terms, which allows us to measure the
values for $<\psi\psi>$ and $<\bar{\psi}\bar{\psi}>$ condensates.
The implementation of such diquark symmetry breaking terms requires
the introduction of diquark and anti-diquark fields $j$ and
$\bar{j}$ that can be taken to zero for the infinite volume system.
Therefore, in analogy with the mass term (for the measurement of the
chiral condensate), and the external field term (in the O(N) model
for the measurement of spontaneous magnetisation) we include the
following terms into the NJL Lagrangian:
\begin{equation}
j\chi^{tr}\tau_{2}\chi+\bar{j}\bar{\chi}\tau_{2}\bar{\chi}^{tr}
\label{diquark source of the njl model}
\end{equation}

\subsubsection{Final Form of the Model and its Symmetries}
We can now write down the full NJL model, as studied in this thesis:
\begin{equation}
S_{NJL}=S_{fer}+S_{bos}
 \label{NJL-final}
\end{equation}
where the fermionic and bosonic parts of the action are:
\begin{equation}
S_{fer}=\sum_x
\bar{\chi}M[\Phi]\chi+j\chi^{tr}\tau_2\chi+\bar{j}\bar{\chi}\tau_2\bar{\chi^{tr}}
\label{njl discrete fermion}
\end{equation}
\begin{equation}
S_{bos}=\frac{1}{g^2}\sum_{\tilde{x}} tr\Phi^{\dag}\Phi \label{njl
discrete boson}
\end{equation}
where $M[\Phi]$ is given in Eqn.~(\ref{M}).  To construct the
partition function of this system we rewrite the fermion part of the
action by defining a bispinor $\Psi^{tr}=(\bar{\chi}^{tr},\chi)$. In
this, Gor'kov, basis we can write the fermion action as
\begin{equation}
S_{fer}= \Psi^{tr}\mathcal{A}\Psi
\end{equation}
where the Nambu-Gor'kov matrix $\mathcal{A}$ is
\begin{equation}
\mathcal{A}^{pq}_{xy}=
\left(%
\begin{array}{cc}
  \bar{j}\tau_2^{pq}\delta_{xy} & \frac{1}{2}M^{pq}_{xy} \\
  -\frac{1}{2}M^{qp}_{yx} & j\tau_2^{pq}\delta_{xy} \\
\end{array}%
\right) \label{A}
\end{equation}
In this basis we can integrate out the fermion fields, which leaves
us with the following Euclidean path integral:
\begin{equation}
Z=\int  D\Phi^{\dagger}D\Phi \sqrt{\det
2\mathcal{A}}\exp(-S_{bos}(\Phi)) \label{partition function for the
njl model in bosonic form}
\end{equation}
The discretized NJL model, given by
Eqns.~(\ref{NJL-final})-(\ref{njl discrete boson}) and (\ref{M}),
still has the full QCD symmetry $SU(2)_L\times SU(2)_R\times
U(1)_B$, as discussed in sec.~\ref{chiral njl model}.  But now the
projection operators (\ref{chirality operator}) become
$P_{R/L}\rightarrow P_{e/o}=\frac{1}{2}[1\pm\varepsilon(x)]$, which
project onto even and odd sublattices, respectively; and the
transformations take the form:
\begin{eqnarray}
\label{left and right njl symmetry} &\chi&\rightarrow(V_eP_e
+V_oP_o)\chi \qquad
\bar{\chi}\rightarrow\bar{\chi}(P_eV_e^{\dagger} +P_oV_o^{\dagger})\nonumber\\
&\Phi&\rightarrow V_o\Phi V_e^{\dagger} \qquad [V_e, V_o \in
SU(2)]\\
\label{baryon njl symmetry} &\chi&\rightarrow e^{i\alpha} \qquad
\bar{\chi}\rightarrow\bar{\chi}e^{-i\alpha} \qquad[e^{i\alpha}\in
U(1)_B]
\end{eqnarray}
The symmetry of the diquark source terms, Eqn.~(\ref{diquark source
of the njl model}), requires an extra explanation.  These terms were
added to break baryon symmetry, $U(1)_B$, hence allowing the
measurement of the diquark condensate as $j\rightarrow0$. Therefore,
under the transformation (\ref{baryon njl symmetry}) it is not
symmetric,
\begin{equation}
j\chi^{tr}\tau_{2}\chi+\bar{j}\bar{\chi}\tau_{2}\bar{\chi}^{tr}\rightarrow
j\chi^{tr}\tau_{2}\chi
e^{i2\alpha}+\bar{j}\bar{\chi}\tau_{2}\bar{\chi}^{tr}e^{-i2\alpha},
\end{equation}
where the symmetry is restored for vanishing $j$ and $\bar{j}$.
Under the $SU(2)$ symmetry, (\ref{left and right njl symmetry}), the
terms are still symmetric.  We can show this by considering only the
$j$ term:
\begin{eqnarray}
j\chi^{tr}\tau_{2}\chi&\rightarrow& j\chi^{tr}(V_e^{tr}P_e
+V_o^{tr}P_o)\tau_{2}(V_eP_e +V_oP_o)\chi\nonumber\\
&=&j\chi^{tr}(V_e^{tr}\tau_{2}V_eP_e+V_o^{tr}\tau_{2}V_oP_o)\chi\nonumber\\
\end{eqnarray}
Now using the identity: $\tau_2 U \tau_2=U^*$ we can
write\footnote{Transpose of this identity gives: $(-\tau_2) U^{tr}
(-\tau_2)=U^{\dagger}$  $\Rightarrow$   $U^{tr}
\tau_2=\tau_2U^{\dagger}$.}:
\begin{eqnarray}
j\chi^{tr}(\tau_{2}V_e^{\dagger}V_e
P_e+\tau_{2}V_o^{\dagger}V_oP_o)\chi&=&
j\chi^{tr}\tau_{2}(P_e+P_o)\chi\nonumber\\
&=&j\chi^{tr}\tau_{2}\chi
\end{eqnarray}
which shows the symmetry of the diquark source term (similarly for
the anti-diquark source term).
\section{QCD Dynamics using the NJL Model}
\label{qcd thermodynamics} Previous studies of the $2+1$d NJL model
have shown that a strong first order phase transition, at
$\mu_c\simeq\Sigma\simeq0.65$, takes place to a chirally symmetric
state \cite{hands1}.  This transition is seen to be accompanied by
the onset of a non-vanishing density of baryon charge in the ground
state.  However, in this high density region it is also seen, for
$2+1$ dimensions, that there is no diquark condensation and the
$U(1)_B$ symmetry is not broken, and there's no evidence of a
non-zero energy gap.  More interestingly, it is seen that the
diquark condensate scales non-analytically at high density, $<\psi
\psi>\propto j^{\frac{1}{\delta}}$.

In this section we will look at how these results were determined,
and more importantly (for our work) how the non-analytic behaviour
of the diquark condensate could imply superfluidity.
\subsection{Phase Transitions at High Density}
\subsubsection{Chiral Condensate}
Chiral symmetry restoration is signalled by the first order
transition of the chiral condensate, $<\bar{\psi}\psi>\rightarrow
0$, which is determined as:
\begin{eqnarray}
<\bar{\psi}\psi>&=&\frac{1}{V}\frac{\partial \ln Z}{\partial
m}=\frac{1}{Z V}\frac{\partial Z}{\partial m}\nonumber\\
&=&\frac{1}{Z V}\frac{\partial}{\partial m}\int
\mathcal{D}\Phi^{\dagger}\mathcal{D}\Phi
\sqrt{\det 2\mathcal{A}}\exp(-S_{bos}(\Phi))\nonumber\\
&=&\frac{1}{Z V}\int
\mathcal{D}\Phi^{\dagger}\mathcal{D}\Phi\frac{1}{2}
\frac{1}{\sqrt{\det 2\mathcal{A}}}\frac{\partial (\det 2\mathcal{A})}{\partial m}\exp(-S_{bos}(\Phi))\nonumber\\
\end{eqnarray}
using $\det 2\mathcal{A}=\exp(\textrm{tr} \ln(2\mathcal{A}))$, we
have:
\begin{eqnarray}
<\bar{\psi}\psi>&=&\frac{1}{Z V}\int
\mathcal{D}\Phi^{\dagger}\mathcal{D}\Phi\frac{1}{2}\textrm{tr}(\frac{2\frac{\partial
\mathcal{A}}{\partial m}}{2\mathcal{A}})
\frac{\det 2\mathcal{A}}{\sqrt{\det 2\mathcal{A}}}\exp(-S_{bos}(\Phi))\nonumber\\
&=&\frac{1}{Z V}\int
\mathcal{D}\Phi^{\dagger}\mathcal{D}\Phi\frac{1}{2}\textrm{tr}(
\mathcal{A}^{-1}\frac{\partial \mathcal{A}}{\partial m})
\sqrt{\det 2\mathcal{A}}\exp(-S_{bos}(\Phi))\nonumber\\
\label{qbarq}
\end{eqnarray}
From (\ref{M}) and (\ref{A}) we have
\begin{equation}
\frac{\partial \mathcal{A}}{\partial m}=\left(
                                          \begin{array}{cc}
                                            0 & \frac{1}{2}\delta \\
                                            -\frac{1}{2}\delta & 0   \\
                                          \end{array}
                                        \right)
\end{equation}
therefore (\ref{qbarq}) becomes
\begin{equation}
<\bar{\psi}\psi>=\frac{1}{Z }\int
\mathcal{D}\Phi^{\dagger}\mathcal{D}\Phi\frac{1}{4
V}\textrm{tr}(\mathcal{A}^{-1} \left(
                                          \begin{array}{cc}
                                            0 & \frac{1}{2}\delta \\
                                            -\frac{1}{2}\delta & 0   \\
                                          \end{array}
                                        \right))
                                        \sqrt{\det 2\mathcal{A}}\exp(-S_{bos}(\Phi))
\end{equation}
Then, the expectation we calculate is:
\begin{equation}
<\bar{\psi}\psi>=<\frac{1}{4 V}\textrm{tr}(\mathcal{A}^{-1} \left(
                                          \begin{array}{cc}
                                            0 & \frac{1}{2}\delta \\
                                            -\frac{1}{2}\delta & 0   \\
                                          \end{array}
                                        \right))>\label{chiral expectation}
\end{equation}

\subsubsection{Baryon Charge Density}
As chiral symmetry is restored at high density, and nucleons
dissociate into quarks, there is seen to be an onset of a
non-vanishing density of baryon charge in the ground state,
$n_B=<\bar{\psi}\gamma_0\psi>$, which is determined using Eqn.
(\ref{partition function for the njl model in bosonic form}),
\begin{eqnarray}
n_B&=&<\bar{\psi}\gamma_0\psi>=\frac{1}{V}\frac{\partial \ln
Z}{\partial \mu}=\frac{1}{Z V}\frac{\partial Z}{\partial
\mu}\nonumber\\
&=&\frac{1}{Z V}\int
\mathcal{D}\Phi^{\dagger}\mathcal{D}\Phi\frac{1}{2}\textrm{tr}(
\mathcal{A}^{-1}\frac{\partial \mathcal{A}}{\partial \mu})
\sqrt{\det 2\mathcal{A}}\exp(-S_{bos}(\Phi))\nonumber\\
\end{eqnarray}
where, as before, we find using (\ref{M}) and (\ref{A})
\begin{equation}
\frac{\partial \mathcal{A}}{\partial \mu}=\left(
                                          \begin{array}{cc}
                                            0 & \frac{1}{2}\frac{\partial M}{\partial \mu} \\
                                            -\frac{1}{2}\frac{\partial M^{tr}}{\partial \mu} & 0   \\
                                          \end{array}
                                        \right)
\end{equation}
where
\begin{eqnarray}
\frac{\partial M}{\partial
\mu}&=&\frac{1}{2}\delta^{pq}[\exp(\mu)\delta_{y,x+\hat{0}}+\exp(-\mu)\delta_{y,x-\hat{0}}]\nonumber\\
\frac{\partial M^{tr}}{\partial
\mu}&=&\frac{1}{2}\delta^{pq}[\exp(\mu)\delta_{x+\hat{0},y}+\exp(-\mu)\delta_{x-\hat{0},y}]\nonumber\\
&=&\frac{1}{2}\delta^{pq}[\exp(\mu)\delta_{y,x-\hat{0}}+\exp(-\mu)\delta_{y,x+\hat{0}}]\nonumber\\
\end{eqnarray}
Now, defining:
\begin{eqnarray}
\mathcal{X}&=&\frac{\partial \mathcal{A}}{\partial \mu}\nonumber\\
&=&\frac{1}{4}
    \left(
        \begin{array}{cc}
            0 &\delta^{pq}[e^{\mu}\delta_{y,x+\hat{0}}+e^{-\mu}\delta_{y,x-\hat{0}}] \\
            -\delta^{pq}[e^{\mu}\delta_{y,x-\hat{0}}+e^{-\mu}\delta_{y,x+\hat{0}}] & 0 \\
        \end{array}
    \right)\nonumber\\
    \label{baryon x matrix}
\end{eqnarray}
we then have:
\begin{equation}
<\bar{\psi}\gamma_0\psi>=\frac{1}{Z}\int \mathcal{D}\Phi
\frac{1}{2V} \textrm{tr}(\mathcal{A}^{-1} \mathcal{X})   \sqrt{\det
2\mathcal{A}}\exp(-S_{bos}(\Phi))
\end{equation}
We then determine the baryon density by the expectation:
\begin{equation}
n_B=<\frac{1}{4V}\textrm{tr}(\mathcal{A}^{-1} \mathcal{X})>
\label{baryon density expectation}
\end{equation}
These transitions in the chiral condensate and the baryon density
can be seen in Fig. \ref{condensate}, from which we can note the
strong first order transition at $\mu=0.65$.
\begin{figure}
    \begin{center}
        \includegraphics[width=0.6\textwidth]{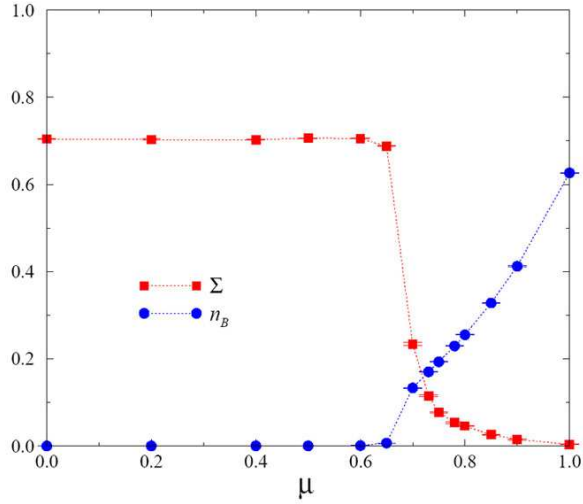}\\
        \caption{Chiral condensate $\Sigma$ and baryon charge density $n_B$ as a function of $\mu$ (from \cite{hands1}).}
        \label{condensate}
    \end{center}
\end{figure}

\subsection{Diquark Condensation}
\label{diquark and superfluid state} The diquark condensates
$<\psi\psi>$ are calculated in a similar way to the previous
expectation values, but this time the derivatives of the partition
function are taken with respect to the diquark source (or the
anti-diquark source for $<\bar{\psi}\bar{\psi}>$):
\begin{equation}
<\psi\psi>=\frac{1}{V}\frac{\partial \ln Z}{\partial j}
\end{equation}
In order to effectively study diquark condensation the operators
$\psi\psi_{\pm}$ are defined as:
\begin{equation}
<\psi\psi_{\pm}>=\frac{1}{2V}\left(
                         \begin{array}{c}
                           \frac{\partial \ln Z}{\partial j}\pm \frac{\partial \ln Z}{\partial \bar{j}} \\
                         \end{array}
                       \right)
\end{equation}
with corresponding source strengths $j_{\pm}=j\pm \bar{j}$, so when
$j=\bar{j}\Rightarrow j_-=0$. Then diquark condensation is given by:
\begin{equation}
<qq_+>=\frac{1}{V}\frac{\partial \ln Z}{\partial
j_+}=\frac{1}{4V}<\textmd{tr}\tau_2\mathcal{A}^{-1}> \label{diquark
qq+}
\end{equation}
Then, the non-vanishing of (\ref{diquark qq+}) in the limit
$j_+\rightarrow 0$ is a criterion for the formation of a diquark
condensate and the resulting spontaneous breakdown of the $U(1)_B$
symmetry.
\subsubsection{Criticality and Superfluidity}
However, in \cite{hands1} it was seen that the diquark condensate
does not extrapolate to a non-zero value for high density, but
scales non-analytically to zero as
\begin{equation}
<\psi\psi>\propto j^{\frac{1}{\delta}},
\end{equation}
implying no breakdown of $U(1)_B$ symmetry.  It was then suggested
that this behaviour could be due to working in 2+1 dimensions. The
Mermin--Wagner theorem predicts: there can be no phase transition to
a state with long-range order at finite temperature for a two
dimensional system with a continuous global symmetry\cite{mermin}.
This was supported in \cite{walters}, where diquark condensation and
an energy gap were observed in simulations of the 3+1d NJL model.

It is also seen that the exponent $\delta$ varies continuously with
chemical potential, $\mu$, taking the value $\delta\approx3$ at
$\mu=0.8$ and $\approx5$ at $\mu=0.9$. This suggests a line of
critical points for $\mu>\mu_c$. This type of behaviour is similar
to that observed in other 2-dimensional critical thermodynamic
systems. For a spin system (XY model) at its critical temperature
the spontaneous magnetisation, $M$, scales with applied magnetic
field, $H$, as $M\propto H^{\frac{1}{\delta}}$, where again the
exponent forms a line of critical points.\footnote{A similar
non-analytic behaviour is seen for a fermionic model exhibiting
chiral symmetry breaking: $<\bar{\psi}\psi>\propto
m^{\frac{1}{\delta}}$.}

The simplest description of a critical system is given by the XY
model, for which long range order would also spontaneously break a
$U(1)$ global symmetry.  Using this system Kosterlitz and Thouless
(KT) described the critical behaviour, of a 2 dimensional system, as
a consequence of binding and unbinding of vortex pairs at a critical
temperature $T_{KT}$ (Appendix~\ref{condensed matter}). Applying
this to a 2-dimensional superfluid, the critical temperature is seen
to be the point where superfluid vortices bind and unbind. Thus,
above the critical temperature when the vortices are free they
experience long-rang mutual interactions leading to the behaviour of
a viscous fluid; but below the critical temperature
vortex-antivortex pairs bind, reducing long range interaction,
resulting in superfluid behaviour. The jump from the spin XY model
to superfluids comes from the fact that we replace the total spin
angle (of vortex-antivortex pairs) in the XY model by the phase
angle of the possible superfluid condensate wave function
(Appendix~\ref{kt transition and supefluid}). Therefore, (as was
conjectured in \cite{hands1}) in analogy to the non-analytic
behaviour displayed by the XY model the high density (and low
temperature) critical phase of the 2+1d NJL model can be considered
to be a gapless thin film BCS superfluid.

In the rest of this thesis we will investigate this possible
superfluid behaviour of 2+1d NJL model.  This will be done by
extending the baryon density to a baryon three current by the use of
Ward Identities. Then, by the use of a spatially varying diquark
source we will introduce a gradient in the diquark pair wave
function, which should force a flow of the baryon current, and hence
result in a measurable signal. Using this current signal we will
explore its behaviour (using the helicity modulus, $\Upsilon$,
Sec.~\ref{helicity}) with variations in spatial volume, temperature,
and diquark source.  We will also attempt to isolate the superfluid
(and non-superfluid) states of the 2+1d NJL model with respect to
chemical potential and temperature.  The variation of superfluidity
with respect to temperature will be of most importance.  The
temperature where superfluidity changes from zero to non-zero should
be the critical point of the system, i.e. the point where vortex and
antivortex pairs come together to form the superfluid phase.  If
this critical point of the NJL model coincides with the critical
point predicted by the KT theory, this would then be strong evidence
that the critical phase of the 2+1d NJL model could be a gapless
thin film BCS superfluid.
}

{\typeout{Baryon 3-Current and the NJL Model with a Twisted
Diquark Source}
\chapter{Forced Baryon Current Flow}
\label{forced baryon current flow} Chiral symmetry restoration of
the $2+1$d NJL model is accompanied by the onset of baryon charge
density $n_B$, as discussed in Sec. \ref{qcd thermodynamics}; in the
first part of this chapter we will extend this baryon density to a
baryon 3-current $I_{\mu}$, where $n_B$ is the timelike component,
$I_0$, of the current. We will then implement a twisted diquark
source i.e. a periodically varying source spanning the whole system.
In the final section we will then conduct simulations with this
twisted source; these will involve comparing the induced baryon
3-current with the constant source to one with a twisted source.  We
will show that by introducing a twisted source we introduce a
diquark phase gradient between the system sites, this then results
in a non-zero superfluid current. We will then present the results
of the effects on this baryon 3-current, as a result of variations
in spatial volume, temperature (i.e changes in temporal volume) and
number of cycles of the diquark source.
\section{Simulating the Baryon 3-Current}
In Sec. \ref{qcd thermodynamics} we saw that the baryon charge
density is given by:
\begin{equation}
I_0\equiv
n_B\equiv<\bar{\psi}\gamma_0\psi>=\frac{1}{V}\frac{\partial \ln
Z}{\partial \mu}=<\frac{1}{4V}\textrm{tr}(\mathcal{A}^{-1}
\mathcal{X})>
\end{equation}
This was easily derived, as the NJL lagrangian contains a chemical
potential term, which allows simple measurements of the baryon
charge density. This charge density can be denoted as the timelike
component of a baryon 3-current, which is the conserved Noether
current associated with baryon number conservation $U(1)_B$, Eqn.
(\ref{chiral conserved currents}). The association of a conserved
Noether current to a global symmetry in continuum field theory has
an equivalent in Euclidean quantum field theory: known as the
Ward-Takahashi identities.  These identities can be derived for a
general action (Using the naive fermions):
\begin{equation}
S\sim\frac{1}{2}\sum_{x\nu}(\bar{\psi}_x\gamma_{\nu}\psi_{x+\nu}-\bar{\psi}_x\gamma_{\nu}\psi_{x-\nu})
\label{general action}
\end{equation}
with a partition function:
\begin{equation}
\mathcal{Z}=\int\mathcal{D}\psi\mathcal{D}\bar{\psi} e^{-S}
\end{equation}
If this partition function is invariant under a simple $U(1)$
transformation:
\begin{equation}
\psi_x\rightarrow\psi_x e^{i\alpha} \qquad
\bar{\psi}_x\rightarrow\bar{\psi}_x e^{-i\alpha}
\end{equation}
we can derive the Ward-Takahashi identities, and hence the conserved
currents of the symmetry.  Under the $U(1)$ transformation, the
above action becomes:
\begin{eqnarray}
S\rightarrow
\frac{1}{2}&\sum_{x\nu}&[e^{-i\alpha}(\bar{\psi}_x\gamma_{\nu}\psi_{x+\nu}-\bar{\psi}_x\gamma_{\nu}\psi_{x-\nu})\nonumber\\
&+&e^{i\alpha}(\bar{\psi}_{x-\nu}\gamma_{\nu}\psi_{x}-\bar{\psi}_{x+\nu}\gamma_{\nu}\psi_{x})]
\end{eqnarray}
which can be written as $S\rightarrow S+\delta S$, where
\begin{eqnarray}
\delta S \sim
-&\frac{i\alpha}{2}&[(\bar{\psi}_{x+\nu}\gamma_{\nu}\psi_{x}+\bar{\psi}_x\gamma_{\nu}\psi_{x+\nu})\nonumber\\
&-&(\bar{\psi}_{x}\gamma_{\nu}\psi_{x-\nu}+\bar{\psi}_{x-\nu}\gamma_{\nu}\psi_{x})]+\mathcal{O}(\alpha^2)\nonumber\\
&=&-i\alpha\nabla_{\nu}^- I_{\nu}(x)
\end{eqnarray}
where the final line, known as the Ward identity, is written in
terms of the backwards operator,$\nabla_{\nu}^{-}$, and $I_{\nu}(x)$
is given as:
\begin{equation}
I_{\nu}(x)=\frac{1}{2}(\bar{\psi}_{x+\nu}\gamma_{\nu}\psi_{x}+\bar{\psi}_x\gamma_{\nu}\psi_{x+\nu})
\label{ward current}
\end{equation}
This can be shown to be the conserved current of the $U(1)$
symmetry, by writing the transformed partition function
as\footnote{$\mathcal{D}\psi \mathcal{D}\bar{\psi}\rightarrow
\mathcal{D}\psi' \mathcal{D}\bar{\psi}'$ as the Jacobian=1.}:
\begin{eqnarray}
\mathcal{Z}'&=&\int\mathcal{D}\psi '\mathcal{D}\bar{\psi}'
e^{-S(\psi ',\bar{\psi}')}e^{-\delta S(\psi
',\bar{\psi}')}\nonumber\\
&=&\int\mathcal{D}\psi
'\mathcal{D}\bar{\psi}'(1-i\alpha\nabla_{\nu}^{-}
I_{\nu}(x))e^{-S(\psi ',\bar{\psi}')}\nonumber\\
&=&\mathcal Z\nonumber\\
&\Rightarrow&<\nabla_{\nu}^{-}I_{\nu}(x)>=0
\end{eqnarray}
The last line shows a zero expectation value of the Ward identity.
This then implies that the current $I_{\nu}(x)$ is conserved.

Now the conserved current in Eqn.~(\ref{ward current}) can be used
to determine the current components $I_1$ and $I_2$ associated with
the $U(1)_B$ global symmetry.  In order to do this we can compare
our NJL action, Eqn.~(\ref{M}) and (\ref{NJL-final}), with the above
action, Eqn.~(\ref{general action}), so we can write the Ward
current components for our NJL model as:
\begin{eqnarray}
I_{0}(x)&=&\frac{1}{2}(e^{\mu}\bar{\chi}_x
\chi_{x+\hat{0}}+e^{-\mu}\bar{\chi}_x\chi_{x-\hat{0}})\nonumber\\
I_{i}(x)&=&\frac{1}{2}(\eta_{i}\bar{\chi}_x
\chi_{x+\hat{i}}+\eta_{i}\bar{\chi}_x\chi_{x-\hat{i}})
\end{eqnarray}
where $i=1,2.$  Now comparing the $I_0$ component with the baryon
charge density expectation value, Eqn.~(\ref{baryon density
expectation}), we can adjust the matrix (\ref{baryon x matrix}) to
correspond with the $I_i$ components.  Then the equivalent matrix
for the $I_i$ components is given by:
\begin{equation}
\mathcal{X^{\star}}_{i}=\frac{1}{4}
    \left(
        \begin{array}{cc}
            0 &\delta^{pq}\eta_{i}[\delta_{y,x+\hat{i}}+\delta_{y,x-\hat{i}}] \\
            \delta^{pq}\eta_{i}[\delta_{y,x-\hat{i}}+\delta_{y,x+\hat{i}}] & 0 \\
        \end{array}
    \right)
    \label{general baryon current x matrix}
\end{equation}
where the expectation value of the baryon current density, $I_i$,
are given as\footnote{These expectation values are calculated using
the Hybrid Molecular Dynamics (HMD) code, Appendix~\ref{hmd}.}:
\begin{equation}
I_{i}\equiv<\bar{\psi}\gamma_{i}\psi>=<\frac{1}{4V}\textrm{tr}(\mathcal{A}^{-1}
\mathcal{X^{\star}}_{i})> \label{baryon current}
\end{equation}
where $i=1,2$.
\section{Introducing the Twisted Diquark Source}
\label{helicity} Previous simulations, as in \cite{hands1}, were
conducted using a constant diquark source,
\begin{equation}
j\psi\psi+\bar{j}\bar{\psi}\bar{\psi}
\end{equation}
where $j=\bar{j}=j_0$.  This was sufficient to force diquark
condensation and derive physically meaningful results in the limit
$j\rightarrow0$.  Such simulations have shown, as the Mermin-Wagner
theorem states, there is no observable diquark condensation for the
$2+1$d NJL system.  However, the diquark condensate behaves in a
non-analytic way, which could imply the existence of a type of
superfluid phase at high density (discussed in Sec.~\ref{diquark and
superfluid state}). Such a two dimensional superfluid state, without
an obvious diquark condensation phase, can be associated with the
binding [and unbinding] of vortex pairs that leads to the critical
phase, as in the XY model (Appendix \ref{condensed matter}). For our
NJL system the vortices can be associated with the phase of the
current carrying superfluid particles, i.e. the phase of the diquark
pairs:
\begin{equation}
<\psi\psi>(\vec{x})=\varphi(\vec{x})_{diquark}=\varphi_0
\exp(i\theta(\vec{x})) \label{diquark wave}
\end{equation}
Where $\theta(\vec{x})$ is the phase corresponding to the external
diquark source; where for a constant real source, as used in
\cite{hands1}, the phase is $\theta=0$.

We can now substitute Eqn.~(\ref{diquark wave}) into the current
density:
\begin{equation}
\vec{I}\sim
(\varphi\vec{\nabla}\varphi^*-\varphi^*\vec{\nabla}\varphi),
\label{current}
\end{equation}
which will give us the baryon current density over the $x$-$y$
plane, i.e. the components $I_1$ and $I_2$:
\begin{equation}
\vec{I}=\Upsilon\vec{\nabla}\theta(\vec{x}), \label{baryon current}
\end{equation}
where $\Upsilon$ is defined as the constant of proportionality
between the current density and the gradient of the phase of the
superfluid components, $\Upsilon=(\phi_0)^2$, and is known as the
helicity modulus(Appendix \ref{condensed matter})\cite{hasen}. From
this we can see that using the constant source above, or a complex
constant diquark source:
\begin{equation}
j=j_0\exp(i\theta) \qquad \bar{j}=j_0\exp(-i\theta)\textrm{,}
\end{equation}
where $\theta$ is a real constant, we have no gradient over the
$x$-$y$ plane:
\begin{equation}
\vec{\nabla}\theta=0
\end{equation}
This zero gradient leads to a zero current density. For this reason,
we must implement a source that generates a non zero gradient over
the $x$-$y$ plane, which would then lead to a spatially varying
diquark wave function, which in turn implies a non-zero current
density.  If we introduce a "twisted" (or periodic) source:
\begin{equation}
j=j_0\exp (i\theta(\vec{x})) \qquad
\bar{j}=j_0\exp(-i\theta(\vec{x})) \label{twisted source}
\end{equation}
where the phase, dependent on spatial position, is given
as\footnote{$L_1$ and $L_2$ are the x-space and y-space dimensions
of the lattice.}:
\begin{equation}
\theta(\vec{x})=2\pi\frac{x_1}{L_1}+2\pi\frac{x_2}{L_2}
\label{phase}
\end{equation}
then, depending on the direction of the twist, this leads to a
non-zero current density, given by inserting (\ref{phase}) into
(\ref{baryon current}):
\begin{equation}
I_1=\Upsilon \frac{2\pi}{L_1} \qquad I_2=\Upsilon \frac{2 \pi}{L_2}
\label{baryon super current}
\end{equation}
So, in an attempt to force a superfluid current flow, we implemented
a generalised version of (\ref{twisted source}),
\begin{equation} j=j_0\exp (i 2\pi n_1 \frac{x_1}{L_1}) \qquad
\bar{j}=j_0\exp(-i 2 \pi n_2 \frac{x_2}{L_2}) \label{twisted source
with n cycles}
\end{equation}
where $n_1$ and $n_2$ are the number of cycles in the source
spanning the lattice.  Thus, similarly to (\ref{baryon super
current}), the baryon current should become:
\begin{equation}
I_1=\Upsilon \frac{2\pi}{L_1}n_1 \qquad I_2=\Upsilon \frac{2
\pi}{L_2}n_2 \label{baryon current with n}
\end{equation}
Also, as the baryon charge density would only depend on the
magnitude of the diquark source, the signal for $I_0$ should be
unaffected with this twisting of the source.
\section{Simulations of the 3-current}
Running simulations, as is often the case, is a time consuming task.
Each run of the Hybrid Molecular Dynamics (HMD) code\footnote{See
Appendix \ref{hmd}.}, with particular magnitudes of diquark source
$|j|=j_0$ and density (chemical potential) $\mu$, generates results
for one value of $\vec{I}$. So, in order to create a plot of baryon
current as a function of $j$ with 9 points, we had to run 9
simulations - for one particular value of chemical potential.  Then,
to plot the extrapolated value of baryon current (as
$j\rightarrow0$) as a function of $\mu$, with 10 points, would mean
we need to run 90 simulations. Taking into account the fact that
these simulations were conducted with variations in spatial
dimensions $L_s$ (to study finite volume effects) and variations in
the temporal dimension $L_t$ (to study effects due to temperature
variation), we were looking at over 1000 runs.  So in an attempt to
eliminate unnecessary runs we conducted our study in two parts. This
section represents the preliminary study of the superfluid phase.
Here we start off first by testing the baryon three current with the
twisted source. Then we go on to study the general thermodynamic
behaviour of the superfluid state; that is, we analyse how the
baryon current behaves at high and low temperatures and chemical
potentials, without running detailed simulations in the intermediate
transition period. In this way, if we see a jump in the superfluid
current from high temperature to low temperature, we could then go
and study the transition period, and relate it to the $KT$
transition as described by the XY model. Then in a similar way we
could study the effect of the current with respect to changes in
chemical potential.
\subsubsection{Physical Units}
The simulations in this work were conducted with a strong coupling
$g^2=2.0$ corresponding to $\Sigma a=0.71$, as in \cite{hands1};
where the result $\Sigma a$ is the dimensionless measure of the
constituent quark mass. In the same way our measurements of the
helicity modulus, from the HMD simulations, will be in the
dimensionless form $\Upsilon=\Upsilon_0 a$, determined by a straight
line extrapolation of the baryon current against diquark source.
Therefore, in order to obtain physically meaningful results we will
need to take the ratio:
\begin{equation}
\frac{\Upsilon_0}{\Sigma}=\frac{\Upsilon_0 a}{\Sigma
a}=\frac{\Upsilon}{0.71}
\end{equation}
This will become necessary in Chapter.~\ref{thin film}, when we
relate our measured value of $\Upsilon$ to temperature.
\subsection{Constant and Twisted Diquark Source} In Fig.~\ref{i-v-j}
we have plotted all the components of the baryon 3-current as a
function of the constant diquark source $j=\bar{j}=j_0$.  In this
plot it can be seen that the only non zero parts are the baryon
charge density $I_0$.  This is as we expected, as the constant
source produces no gradient, hence there is no superfluid baryon
current flow, i.e. $I_1=I_2=0$.
\begin{figure}
    \begin{center}
        \includegraphics[width=0.8\textwidth]{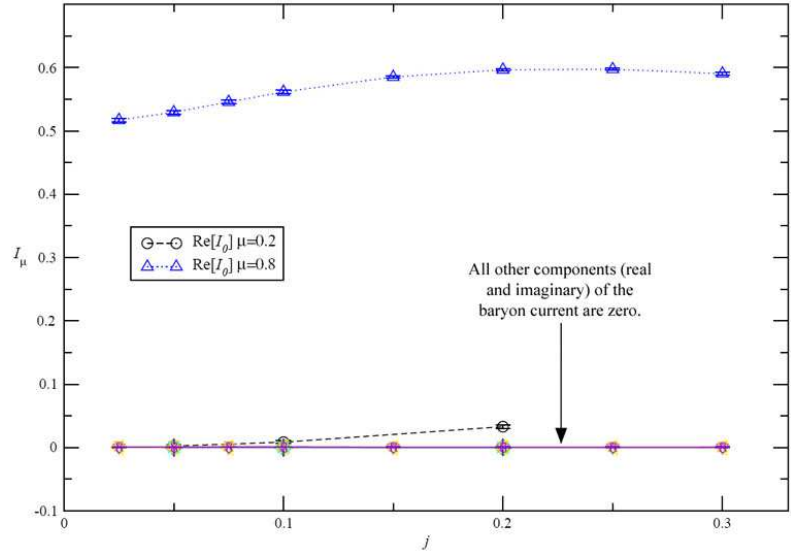}\\
        \caption{Real and Imaginary parts of the baryon 3-current, $I_{\mu}$, as a function of constant diquark source, $j$.  (Lattice Size=$16\times 16\times 64$)}
        \label{i-v-j}
    \end{center}
\end{figure}
In Fig.~\ref{non-zero-i-v-j} we have plotted the non-zero components
of the baryon 3-current with the twisted diquark source
(\ref{twisted source with n cycles}).  We conducted two sets of
simulation, the first was with the source only twisted in the $x$
direction, and the second with the source only twisted in the $y$
direction.  We found, as expected, that for a twist in the $x$
direction we had a super current signal $I_1\neq0$ and for a twist
in the $y$ direction we had a signal $I_2\neq0$.  These simulations
were conducted for the same magnitudes of the diquark source, so
that both simulations gave the same signal for the baryon charge
density $I_0=n_B$.
\begin{figure}
    \begin{center}
        \includegraphics[width=0.8\textwidth]{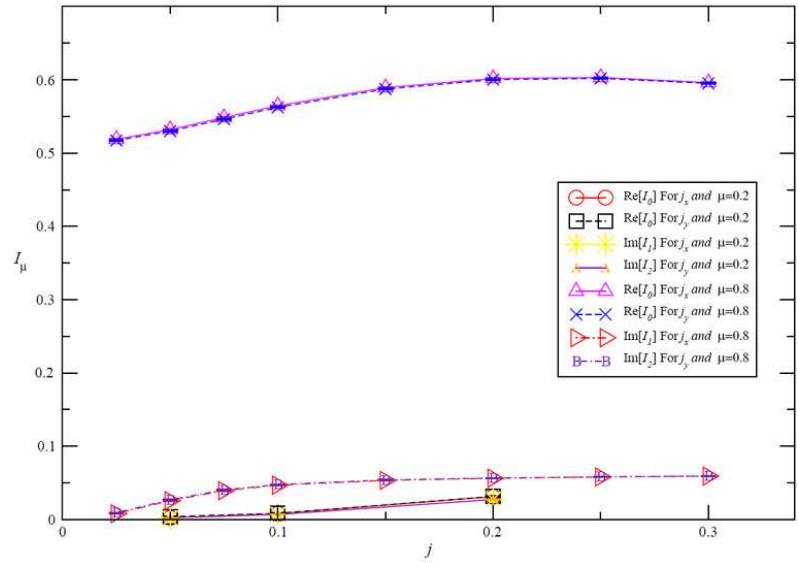}\\
        \caption{Non-zero components of baryon 3-current, $I_{\mu}$, as a function of twisted diquark source $j_{\mu}$, with $n_1=n_2=1$. (Lattice Size=$16\times 16\times 64$)}
        \label{non-zero-i-v-j}
    \end{center}
\end{figure}
It is also seen that the signals generated for the super currents
$I_1$ and $I_2$ are identical; this is obvious from eqn.~\ref{baryon
current with n} if we have $L_1=L_2$ and $n_1=n_2$. Therefore in the
rest of the simulations we only conducted simulations for a twist in
the $y$ direction, and assumed that a twist in the $x$ direction
would give identical results.
\subsection{Twisted Source with n-Cycles}
\label{twisted with n} We conducted simulations to test how the
baryon current, $I_2$, changed as we varied the number of cycles (or
twists) in the diquark source, $n_{2}$. This was done for high and
low chemical potential $\mu=0.8$ and $0.2$, and the results have
been plotted in Fig.~\ref{i2-v-j2-n-cycles}.
\begin{figure}
    \begin{center}
        \includegraphics[width=0.8\textwidth]{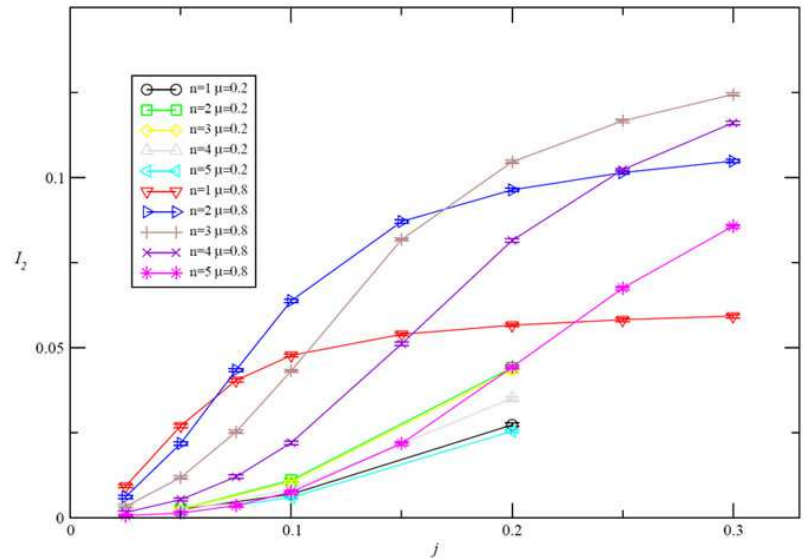}\\
        \caption{Baryon current, component $I_{2}$, as a function of twisted diquark source, $j_{2}$, for various values of $=n_2$. (Lattice Size=$16\times 16\times 24$)}
        \label{i2-v-j2-n-cycles}
    \end{center}
\end{figure}
You can note that as you increase the number of cycles for the low
chemical potential, $\mu=0.2$, the $I_{\mu}\propto j^2$ behaviour,
which is more prominent in Figs.~\ref{i2-v-j2-mu-t=64} and
\ref{i2-v-j2-mu-t=4}, becomes more pronounced.

However, at high $\mu$, the behaviour shifts from an approximate
linear fit ($I_{\mu}\propto j$) to a parabolic fit ($I_{\mu}\propto
j^2$). This shift then leads to a zero extrapolation of the baryon
current as $j\rightarrow0$. So as you increase the number of cycles
the baryon current ceases to flow. This shift and the resulting
destruction of the baryon current can be considered due to finite
volume effects. From Eqn.~(\ref{baryon current with n}) it can be
seen that increasing $n$ from $1$ to $2$ cycles has an effect of
decreasing the wavelength from $L$ to $L/2$. Therefore, increasing
the lattice size would counteract the effect of increasing the
number of cycles. This can be seen in Fig.~\ref{i2-v-j2-n=2-lx=32}
where we have plotted the baryon current, $I_2$, against diquark
source with $n=2$ cycles. We have also plotted the a baryon current
for $n=1/2$. However, due to periodic boundary conditions the
results for non-integer values of $n$ do not coincide with those of
the physically relevant current signals, corresponding to integer
values of $n$.  This is seen clearly from the plots for $L_s=16,
n=0.5$ and $L_s=32, n=1$, in Fig.~\ref{i2-v-j2-n=2-lx=32}.
\begin{figure}
    \begin{center}
        \includegraphics[width=0.8\textwidth]{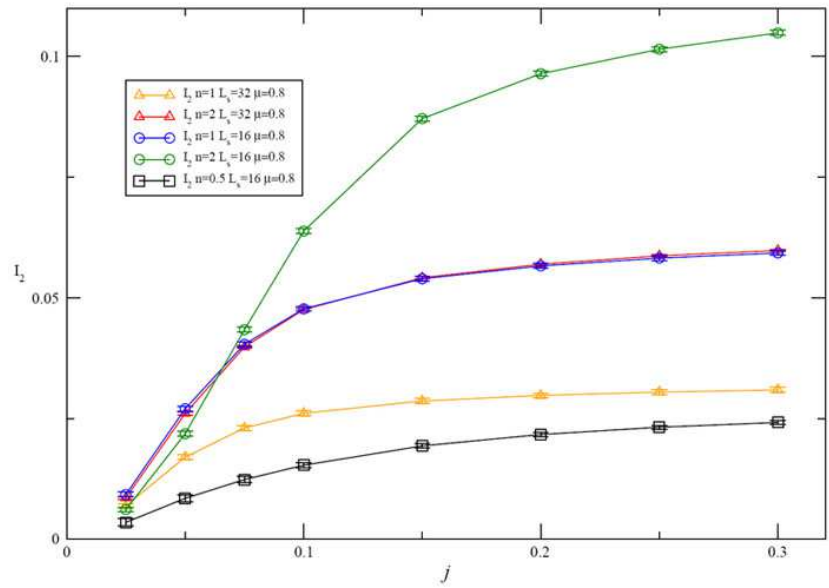}\\
        \caption{Baryon current, component $I_{2}$, as a function of twisted diquark source, $j_{2}$, for $=n_2=1$, $2$ and $0.5$ and lattice size $L_s=16$, $32$, and $L_t=24$}
        \label{i2-v-j2-n=2-lx=32}
    \end{center}
\end{figure}

It can also be noted that by using different size spatial dimensions
we obtain very different baryon currents, so the baryon current is
dependent on the spatial lattice size.  In order to obtain a result
that is independent of spatial dimensions, we need to work with the
helicity modulus, which will only vary depending on spatial size as
a result of finite volume effects.  The use of the Helicity modulus
will be discussed in further detail in Sec.~\ref{finite-vol-effect}.
\subsection{Effects of Temperature and Density Variation}
\begin{figure}
    \begin{center}
        \includegraphics[width=0.85\textwidth]{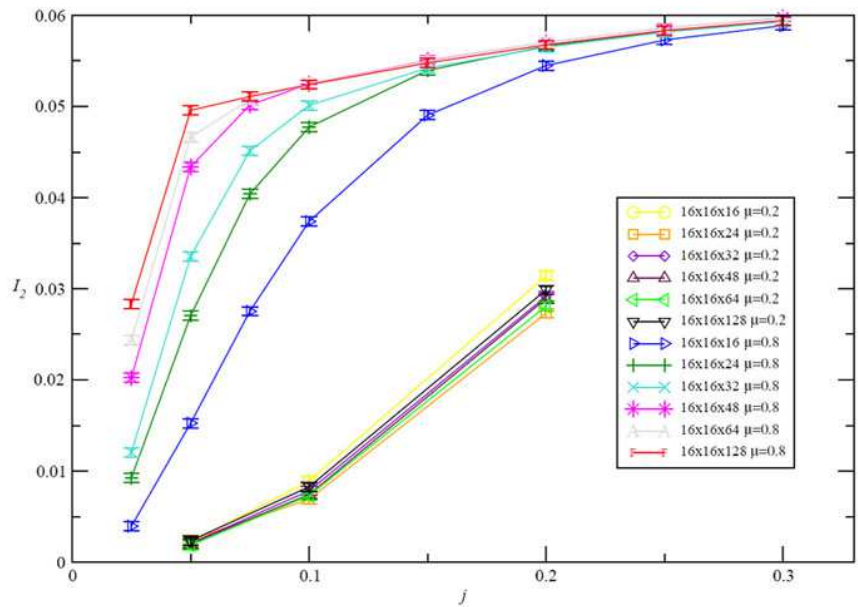}\\
        \caption{Baryon current, $I_2$, as a function of twisted diquark source, $j$, for various time dimensions and fixed space, $L_s=16$.}
        \label{i2-v-j2-time}
    \end{center}
\end{figure}
In Fig.~\ref{i2-v-j2-time} we have plotted the baryon current,
$I_2$, as a function of diquark source, for a fixed spatial volume
($L_s=16$), two values of chemical potential and various time
dimensions ($L_t$). At low chemical potential, $\mu=0.2$, it can be
seen that there is not much variation in $I_2$ with changes in
temperature $(T=1/L_t)$. However, at high chemical potential,
$\mu=0.8$, the current can be seen to collapse onto a straight line
as you decrease temperature.  By $L_t=128$ it becomes clear that
$I_2$ extrapolates to a non-zero value as $j\rightarrow0$. From
these results we decided to conduct further simulations using
$L_t=64$, as this gives us results comparable to $L_t=128$, however
in a third of the CPU time.

In Fig.~\ref{i2-v-j2-mu-t=64} we have plots of various $\mu$ at low
temperature ($L_t=64$).  It can be seen that the current behaves in
a fundamentally different way at high density ($\mu\geq0.68$)
compared to low density ($\mu\leq0.65$).  As we go from high to low
$\mu$ the negative curvature of the high density phase shifts
sharply to a strong quadratic type, which can then be extrapolated
to $I_2=0$. In this way we can determine the critical chemical
potential for the onset of superfluidity as $0.65<\mu_c<0.68$, which
is also the region of chiral symmetry restoration i.e.
$\mu_c\sim0.65$ (Fig. \ref{condensate}). On the other hand, it can
be seen from Fig.~\ref{i2-v-j2-mu-t=4} that at high temperature
($L_t=4$), all the chemical potential curves follow a quadratic
extrapolation to $I_2=0$ at $j=0$.  So, at high temperature no
superfluid state exists.
\begin{figure}
    \begin{center}
        \includegraphics[width=0.85\textwidth]{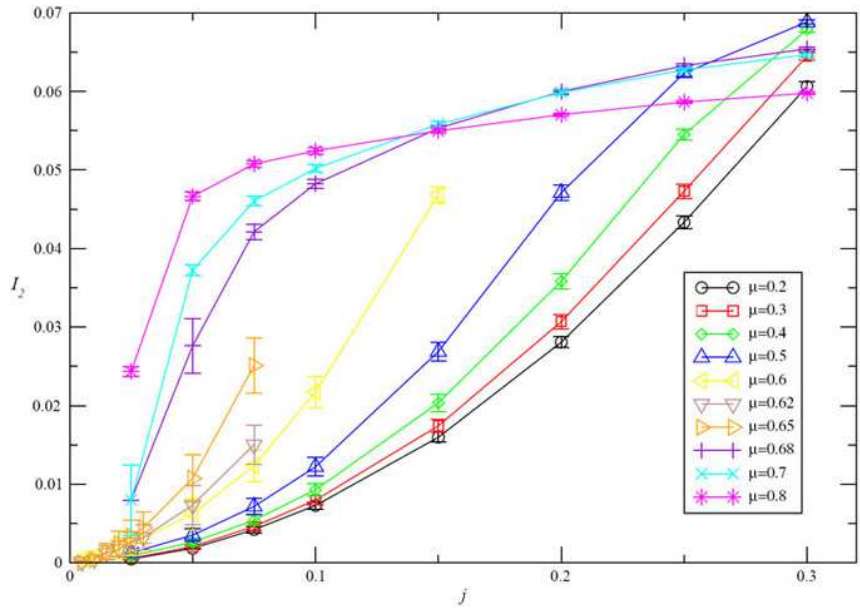}\\
        \caption{Baryon current, $I_2$, as a function of twisted diquark source, $j$, for various chemical potential ($\mu$) at low temperature ($L_t=64$). ($L_s=16$)}
        \label{i2-v-j2-mu-t=64}
    \end{center}
\end{figure}
\begin{figure}
    \begin{center}
        \includegraphics[width=0.85\textwidth]{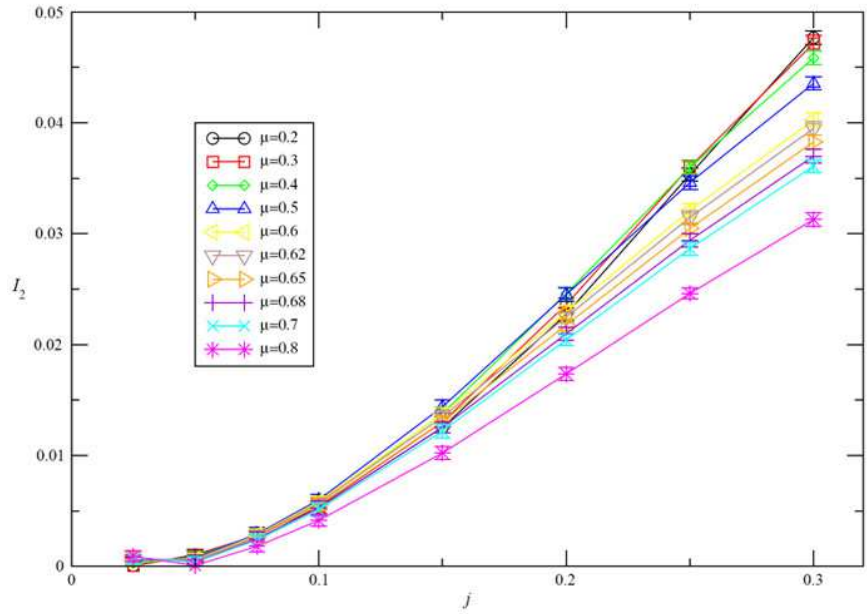}\\
        \caption{Baryon current, $I_2$, as a function of twisted diquark source, $j$, for various chemical potential ($\mu$) at high temperature ($L_t=4$). ($L_s=16$)}
        \label{i2-v-j2-mu-t=4}
    \end{center}
\end{figure}
Also, even though we have a possible linear extrapolation for the
low temperature and high density phase, as seen in
Fig.~\ref{i2-v-j2-time} (for $L_t=128)$, we do not have a systematic
method of extrapolating to $j\rightarrow0$ for $\mu\geq\mu_c$ in
general, therefore we are unable to make any decisive statement
about the nature of the transition.

\subsubsection{$I_2$ as a Function of $j$}
The behaviour of the baryon current, $I_2$, as a function of source,
$j$, can be summarised as:
\begin{equation}
I_2(j,\mu,T)=I_2(0,\mu,T)+A(\mu,T)j+B(\mu,T)j^2
\label{extrapolation}
\end{equation}
At low temperature, and for high $\mu$ (as for the case $\mu=0.8$ in
Figs.~\ref{i2-v-j2-time}~ and \ref{i2-v-j2-mu-t=64}) we have: $B=0$
and $A\neq0$, and a possible straight line fit extrapolates (as
$j\rightarrow0$) to $I_2(0,\mu)\neq0$.  However, for low $\mu$ (as
for $\mu\leq0.65$ in Fig.~\ref{i2-v-j2-mu-t=64}) we have: $A=0$ and
$B\neq0$ which leads to a $j^2$ fit, extrapolating (as
$j\rightarrow0$) to $I_2(0,\mu)=0$.  On the other hand, at high
temperature this behaviour changes (Fig.~\ref{i2-v-j2-mu-t=4}), here
the extrapolation is of the form $j^2$ for all values of $\mu$, i.e.
$I_2(0,\mu,T)=A(\mu,T)=0$. The extrapolation shift from $j$ to $j^2$
can be understood if we have different wave functions for the
current carrying superfluid particles at different chemical
potentials:
\begin{eqnarray}
\phi(\vec{x})&=& \phi_0 j_0 \exp(i\theta(\vec{x})) \qquad
\mu<\mu_c\nonumber\\
\phi(\vec{x})&=& \phi_0 \exp(i\theta(\vec{x})) \qquad \mu>\mu_c
\label{j and jsquared}
\end{eqnarray}
The linear $j_0$ dependence, at low chemical potential, is expected
due to the fact that there should be no diquark condensate for
$\mu<\mu_c$; so $\phi \rightarrow0$ as $j\rightarrow0$. In this way,
using Eqn.~(\ref{current}), the baryon current is:
\begin{eqnarray}
\vec{I}&\sim& \phi_0^2 j_0^2 \vec{\nabla}\theta(\vec{x}) \qquad
\mu<\mu_c\nonumber\\
\vec{I}&\sim& \phi_0^2 \vec{\nabla}\theta(\vec{x}) \qquad \mu>\mu_c
\label{j and jsquared currents}
\end{eqnarray}
and in the limit $j\rightarrow0$, we have:
\begin{eqnarray}
\vec{I}&\sim & 0 \qquad \mu<\mu_c \nonumber\\
\vec{I}&\sim &\phi_0^2 \vec{\nabla}\theta(\vec{x})=\textrm{constant}
\qquad  \mu>\mu_c \label{j and jsquared currents for j->0}
\end{eqnarray}
which is the behaviour of the baryon current as seen throughout our
simulations.
\section{Finite Volume Effects}
\label{finite-vol-effect}We have, so far, been working with the
baryon current, $I_2$, for example in Fig.~\ref{i2-v-j2-mu-t=64} we
have plotted the current against diquark source. However, this is
not the most appropriate variable to work with, due to its
dependence on spatial volume, as we saw briefly in Sec.~\ref{twisted
with n}. This dependence on spatial volume can be seen clearly when
we conduct simulations for a fixed temperature, but differing
spatial sizes: in Fig.~\ref{i2-v-jy-space-lt=64} it can be seen
that, at low temperature, the baryon current differs hugely for the
three different spatial volumes $L_s=16$, $24$ and $32$, and any
linear extrapolation of the baryon current with $j\rightarrow0$ will
give very different results. However, at low $\mu$ and/or high
temperature, when the extrapolation is of the $j^2$ form, finite
volume effects do not matter too much as all extrapolations tend to
$I_2\rightarrow0$, Figs.~\ref{i2-v-jy-space-lt=64} and
\ref{i2-v-j2-ls-lt=12}.
\begin{figure}
    \begin{center}
        \includegraphics[width=0.8\textwidth]{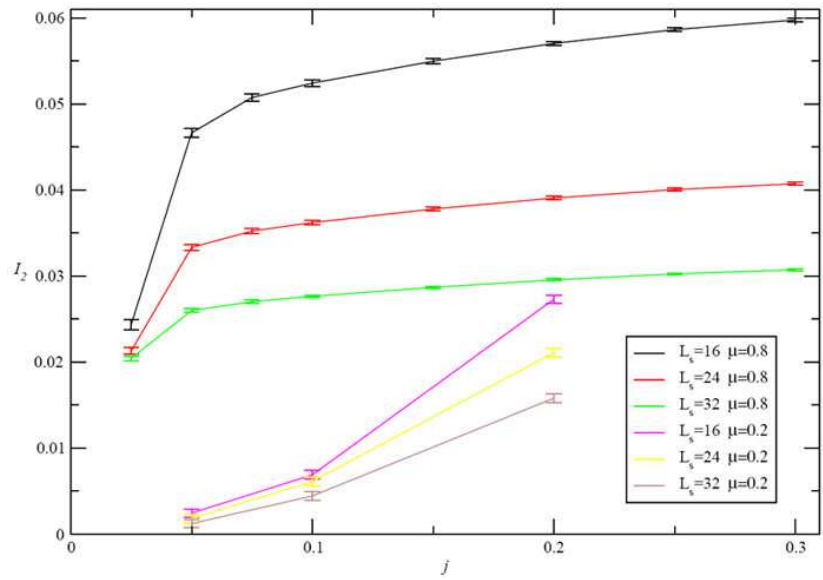}\\
        \caption{Baryon current, $I_2$, as a function of twisted diquark source, $j$, for various spatial dimensions, $\mu=0.2$ and $0.8$, and low temperature, $L_t=64$.}
        \label{i2-v-jy-space-lt=64}
    \end{center}
\end{figure}

\begin{figure}
    \begin{center}
        \includegraphics[width=0.8\textwidth]{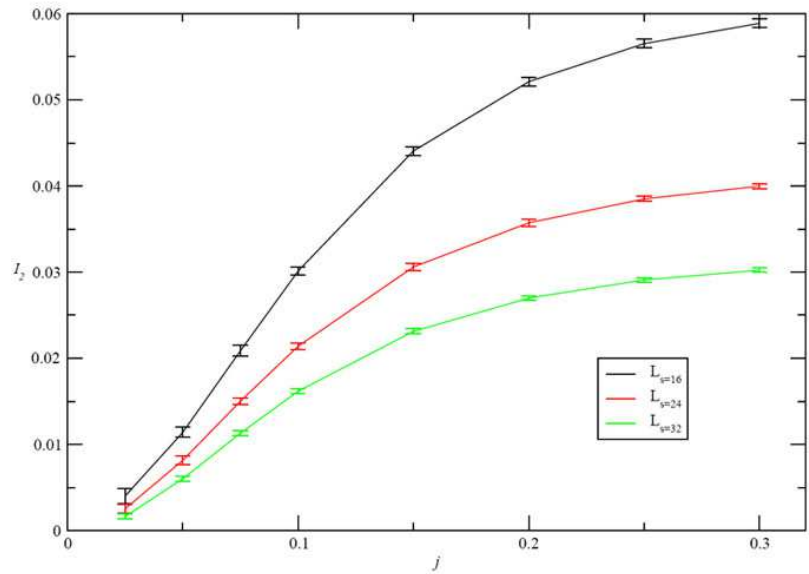}\\
         \caption{Baryon current, $I_2$, as a function of twisted diquark source, $j$, for various space dimensions, $\mu=0.8$, and high temperature, $L_t=12$.}
        \label{i2-v-j2-ls-lt=12}
    \end{center}
\end{figure}
\subsection{Working with the Helicity Modulus} We can overcome
effects of spatial dependence by working with the helicity modulus,
$\Upsilon$, which, as seen earlier, is the constant of
proportionality between the baryon current and the gradient of the
source's phase, Eqn.~(\ref{baryon current}).  In this way we are
able to represent, in a volume independent manner, the superfluid
behaviour of the system.  So, when we plot $\Upsilon=I_2 L_s / 2
\pi$ against $j$ for varying spatial volumes, $L_s$, but constant
temperature and chemical potential, any variations in the plots
could be put down to finite volume effects.
\subsubsection{Low Temperature Finite Volume Effects}
In Fig.~\ref{y-v-j-ls-lt=64} we have plotted $\Upsilon$ against $j$
for low temperature, $L_t=64$; this is only done for high $\mu$, as
for lower values of $\mu$ the extrapolation is always of the $j^2$
type, which implies the current, and thus, the helicity modulus
extrapolates to zero as $j\rightarrow0$.
\begin{figure}
    \begin{center}
        \includegraphics[width=0.9\textwidth]{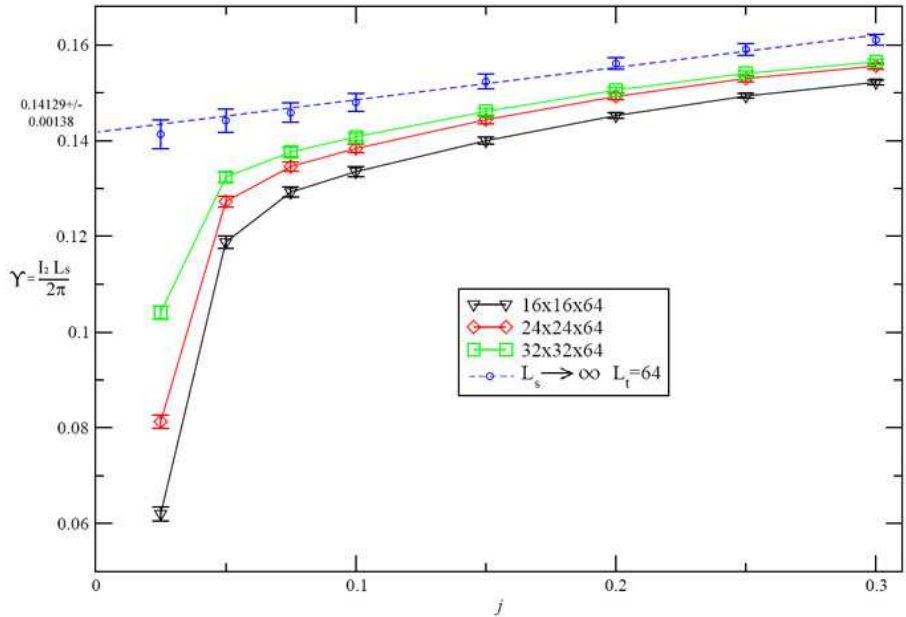}\\
        \caption{Helicity modulus, $\Upsilon$, as a function of twisted diquark source, $j$, for various spatial dimensions, $\mu=0.8$ and low temperature, $L_t=64$.}
        \label{y-v-j-ls-lt=64}
    \end{center}
\end{figure}
We can note that the dip in $\Upsilon$, at low $j$, decreases as the
spatial volume increases; thus, if this dip is a finite volume
effect it should tend to a straight line in the infinite volume. In
order to analyse this dip at larger volumes, we extrapolated
$\Upsilon$ to the limit of $L_s\rightarrow\infty$ by plotting
$\Upsilon$ for each value of $j$ against $1/L_s$,
Fig.~\ref{y-v-1-over-ls-t=64}, and then the infinite volume value of
$\Upsilon$ was deduced by extrapolating the plots
$1/L_s\rightarrow0$.  The infinite volume plot of $\Upsilon$ is also
plotted in Fig.~\ref{y-v-j-ls-lt=64}; and it can be seen that the
dip vanishes in this limit, hence this effect can be put down to
finite volume effects.
\begin{figure}
    \begin{center}
        \includegraphics[width=0.7\textwidth]{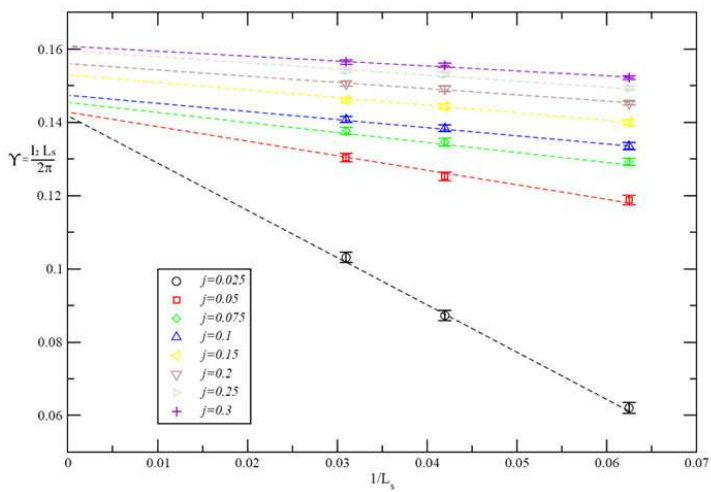}\\
        \caption{Helicity modulus, $\Upsilon$, as a function of $1/L_s$, for various $j$, $\mu=0.8$ and low temperature, $L_t=64$.}
        \label{y-v-1-over-ls-t=64}
    \end{center}
\end{figure}
From the low temperature, infinite volume plot, we have conducted an
extrapolation of $\Upsilon$ for $j\rightarrow0$, which should
represent the dimensionless value of $\Upsilon$ at low temperature:
$\Upsilon=0.1413\pm0.0014$.
\subsubsection{High Temperature Finite Volume Effects}
Using the same techniques as before we studied the superfluid
behaviour at high temperature, $L_t=4$. In
Fig.~\ref{y-v-j-ls-lt=4-extrapolated} we have plotted $\Upsilon$ for
$\mu=0.8$, $L_t=4$ and, again, for various spatial volumes. As
before we extrapolate to the infinite volume limit, where the
extrapolation lines are shown in Fig.~\ref{y-v-1-over-ls-t=4}.
\begin{figure}
    \begin{center}
        \includegraphics[width=0.9\textwidth]{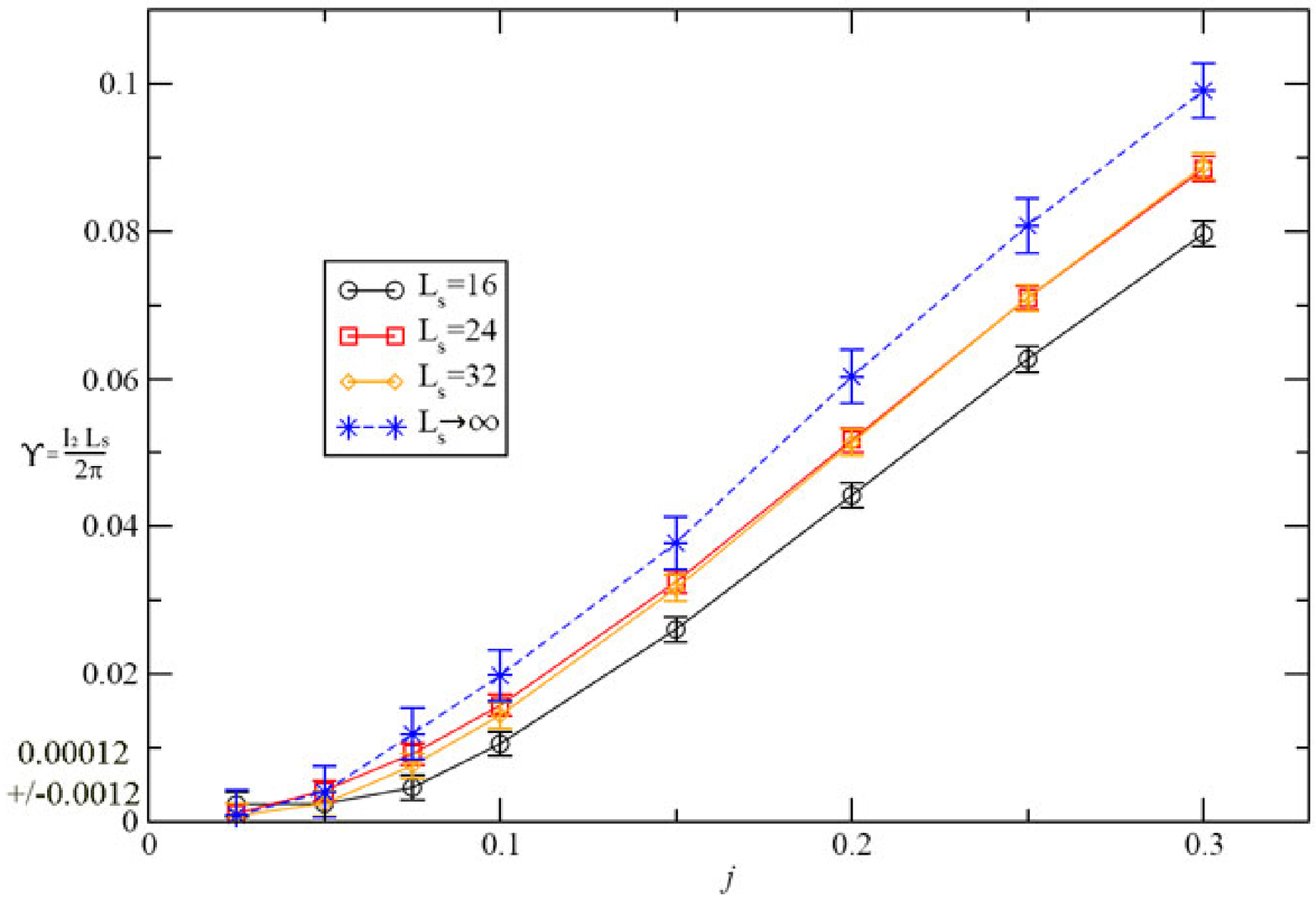}\\
        \caption{Helicity modulus, $\Upsilon$, as a function of twisted diquark source, $j$, for various spatial dimensions, $\mu=0.8$ and high temperature, $L_t=4$.}
        \label{y-v-j-ls-lt=4-extrapolated}
    \end{center}
\end{figure}
\begin{figure}
    \begin{center}
        \includegraphics[width=0.7\textwidth]{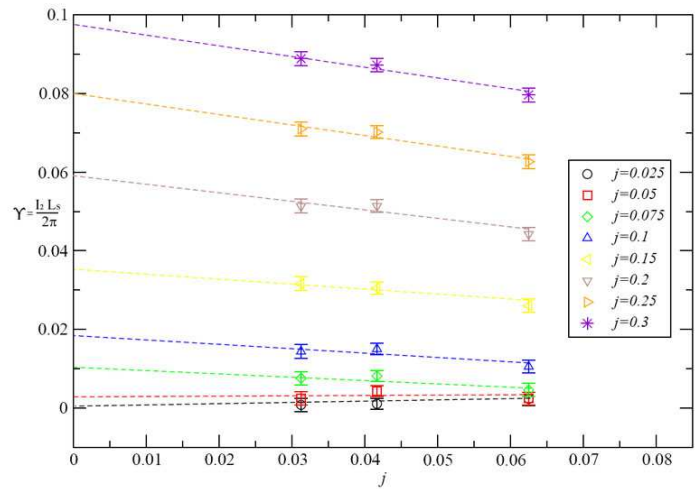}\\
        \caption{Helicity modulus, $\Upsilon$, as a function of $1/L_s$, for various $j$, $\mu=0.8$ and high temperature, $L_t=4$.}
        \label{y-v-1-over-ls-t=4}
    \end{center}
\end{figure}
In Fig.~\ref{y-v-j-ls-lt=4-extrapolated} we have also plotted the
curve for $L_s\rightarrow\infty$.  From this we see that at high
temperature and in the infinite volume limit the $\Upsilon\propto
j^2$ behaviour does not change significantly, if anything it becomes
more pronounced at lower values of $j$. From this extrapolation we
find $\Upsilon=0.00012\pm0.0012\sim0$.
\subsubsection{Behaviour at Intermediate Temperature}
We have shown that at low temperature there are significant, but
controllable, finite volume effects, which are removed by
extrapolating to the infinite volume limit.  In the infinite volume
limit the low temperature linear behaviour, $\Upsilon(j)\propto
\Upsilon(0)+ Aj$, becomes obvious and straight line extrapolations
for $j\rightarrow0$ lead to $\Upsilon\neq0$.  At high temperature,
where we expect a $\Upsilon(j)\propto j^2$ behaviour, working on a
finite volume does not have much of an effect, that is the finite
volume plots do not differ from the infinite volume extrapolation.
So, we can state that the superfluid current drops from a non-zero
value at low $T$ to a zero value at high $T$.

However, if we attempt to analyse the intermediate temperature
region we encounter some difficulties. In
Fig.~\ref{y-v-j-ls-lt=12-extrapolated} we have plotted $\Upsilon$
for $\mu=0.8$, $L_t=12$.
\begin{figure}
    \begin{center}
        \includegraphics[width=0.9\textwidth]{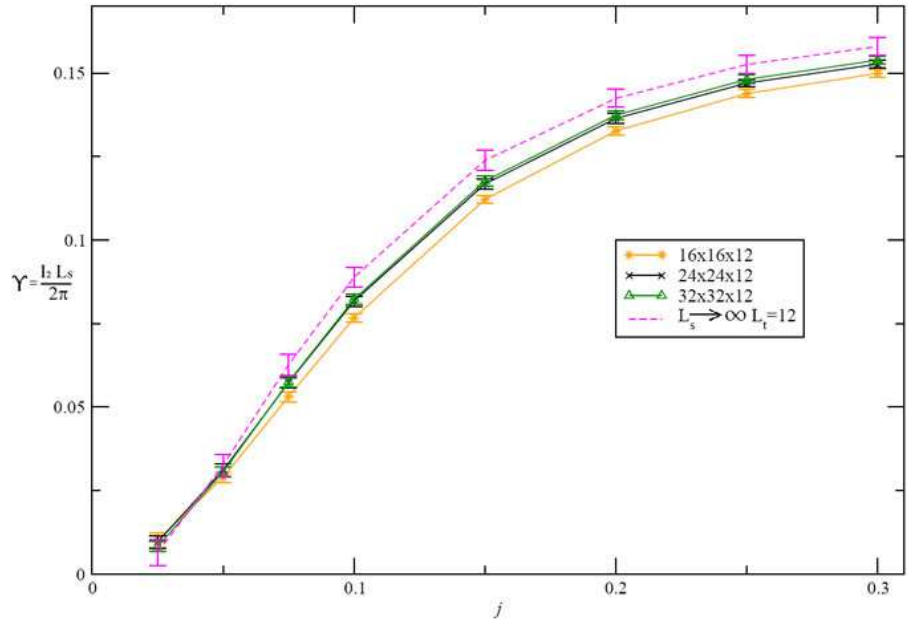}\\
        \caption{Helicity modulus, $\Upsilon$, as a function of twisted diquark source, $j$, for various spatial dimensions, $\mu=0.8$ and high temperature, $L_t=12$.}
        \label{y-v-j-ls-lt=12-extrapolated}
    \end{center}
\end{figure}
The extrapolation of this, to the thermodynamic limit, is given in
Fig.~\ref{y-v-1-over-ls-t=12}. We see that the various volume plots
have an "s" shaped form, which remains even in the infinite volume
limit.
\begin{figure}
    \begin{center}
        \includegraphics[width=0.7\textwidth]{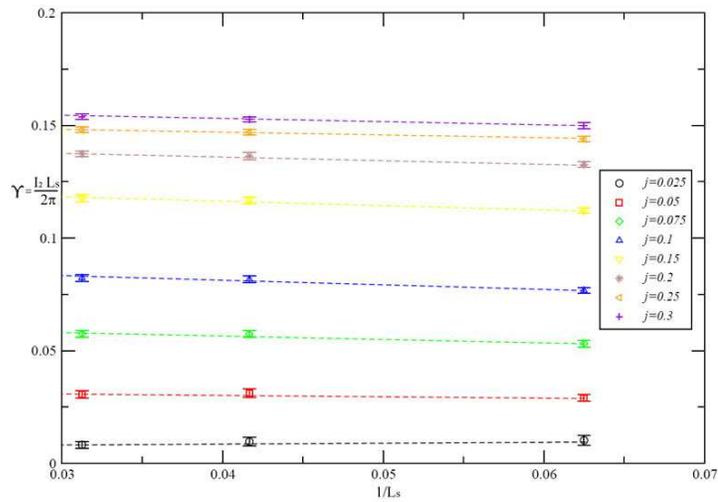}\\
        \caption{Helicity modulus, $\Upsilon$, as a function of $1/L_s$, for various $j$, $\mu=0.8$ and high temperature, $L_t=12$.}
        \label{y-v-1-over-ls-t=12}
    \end{center}
\end{figure}
In this case we could implement a $j^2$ extrapolation for the first
four points, however this does not seem very conclusive as the last
4 points lie well outside any $j^2$ behaviour.  On the other hand,
in Fig.~\ref{y-v-j-ls=40-lt=12}, where we have again plotted a curve
for $L_t=12$, but now we have used a larger volume, $(L_s=40)$, and
obtained further points at low $j$, a $j^2$ extrapolation does seem
appropriate.  From this extrapolation $\Upsilon$ can be deduced to
be $\Upsilon=0.00016\pm0.0014\sim0$.
\begin{figure}
    \begin{center}
        \includegraphics[width=0.9\textwidth]{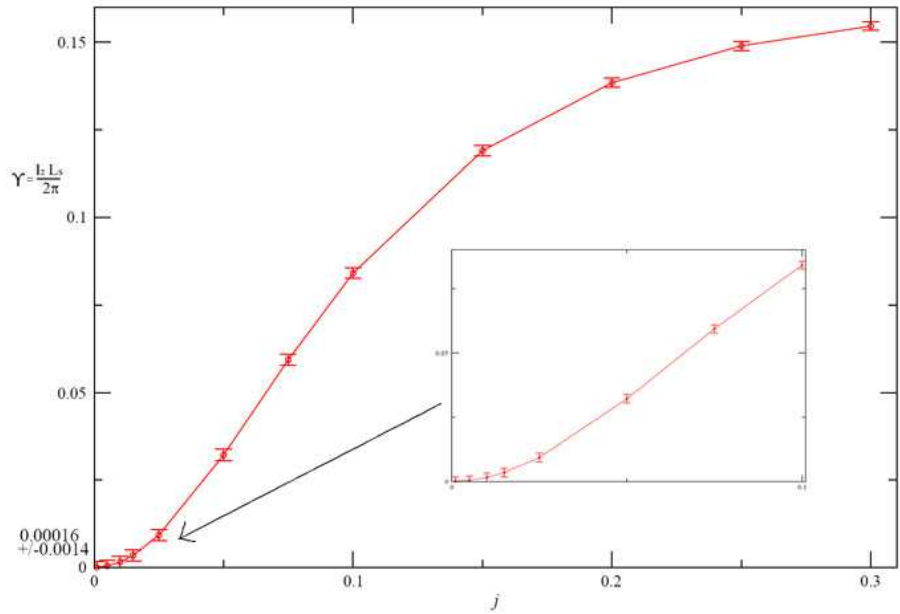}\\
        \caption{Helicity modulus, $\Upsilon$, as a function of twisted diquark source, $j$, for $L_s=40$, $\mu=0.8$ and high temperature, $L_t=12$.}
        \label{y-v-j-ls=40-lt=12}
    \end{center}
\end{figure}

A similar situation occurs as you decrease the temperature further.
In Fig.~\ref{y-v-j-ls-lt=24-extrapolated} we have a plot of
$L_t=24$, and its extrapolation lines in
Fig.~\ref{y-v-1-over-ls-t=24}. At this temperature we can see that
it is not obvious if this plot falls under a linear extrapolation
(giving $\Upsilon\neq0$) or a parabolic extrapolation (giving
$\Upsilon=0$).  At low $j$ the curve must tend to $\Upsilon=0$ hence
it must follow a $j^2$ curve to the origin.  But, more plausibly, we
could apply a straight line extrapolation to the last 5 points. Even
in the infinite volume limit it does not obviously fall under either
extrapolation criteria.
\begin{figure}
    \begin{center}
        \includegraphics[width=0.8\textwidth]{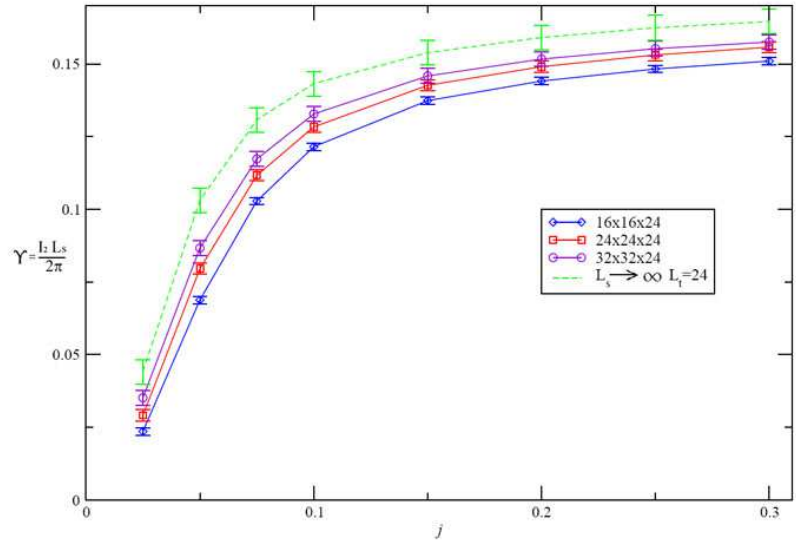}\\
        \caption{Helicity modulus, $\Upsilon$, as a function of twisted diquark source, $j$, for various spatial dimensions, $\mu=0.8$ and temperature, $L_t=24$.}
        \label{y-v-j-ls-lt=24-extrapolated}
    \end{center}
\end{figure}
\begin{figure}
    \begin{center}
        \includegraphics[width=0.8\textwidth]{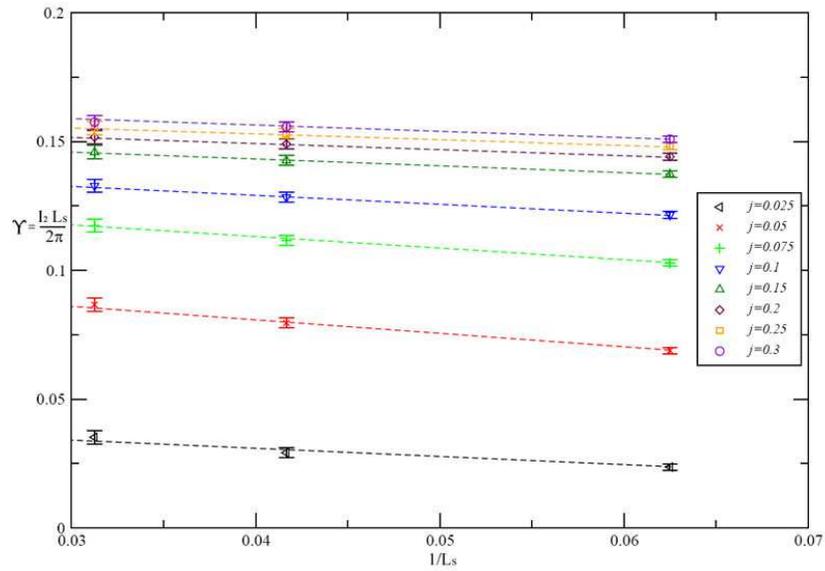}\\
        \caption{Helicity modulus, $\Upsilon$, as a function of $1/L_s$, for various $j$, $\mu=0.8$ and temperature, $L_t=24$.}
        \label{y-v-1-over-ls-t=24}
    \end{center}
\end{figure}

In the next chapter we will go on to investigate the intermediate
temperature range.  We will attempt to deduce some criterion for the
extrapolation of these temperatures, which will then help in
isolating the critical temperature i.e. the temperature at which
$\Upsilon$ falls from the non-zero superfluid value at low $T$ to a
zero value in the non-superfluid phase at high $T$.
}

{\typeout{Superfluidity}
\chapter{Thin Film Superfluid Dynamics}
\label{thin film} In the previous chapter it was shown that the
baryon current-density has a strong transition in the region
$0.65<\mu_c<0.68$ - the region corresponding to chiral symmetry
restoration.  Also, working with the spatially independent helicity
modulus, $\Upsilon$, it was shown that at low temperature, $L_t=64$,
and high chemical potential, $\mu=0.8$, the helicity modulus
extrapolates to $\Upsilon=0.1413\pm0.0014$; and at high temperature,
$L_t=4$, it extrapolates to zero.  However, at intermediate
temperatures, for example $L_t=12$ and $L_t=24$, extrapolation
techniques are not obvious - even in the infinite volume limit. In
this chapter we will investigate the behaviour of $\Upsilon$ in this
intermediate region, and attempt to provide some extrapolation
criteria for this puzzling domain. The aim of this chapter will be
to try and isolate the critical temperature and see if it
corresponds with the predicted $T_{KT}$ temperature, given by the
theory of Kosterlitz and Thouless (KT).

\section{$T_{KT}$ given by Kosterlitz and Thouless' Theory}
If the phase transition, for the 2+1d NJL model at high density, is
described by the KT theory (Appendix~\ref{condensed matter}) then
the critical temperature should be given by
Eqn.~(\ref{tkt-superfluid}), or in lattice units:
\begin{equation}
a L_t=\frac{1}{T_{KT}}=\frac{2}{\pi}\frac{1}{\Upsilon_0} \qquad
\rightarrow \qquad L_t=\frac{1}{a
T_{KT}}=\frac{2}{\pi}\frac{1}{\Upsilon},
\end{equation}
where $\Upsilon=\Upsilon_0 a$ is the dimensionless measured value.
Using our value of $\Upsilon$ we find:
\begin{equation}
L_t=\frac{2}{\pi}\frac{1}{0.14129\pm0.00138}=4.51\pm0.44
\label{kt-derive}
\end{equation}
Therefore, according to the KT theory the critical $L_t$ should be
approximately $6$, which is the nearest even lattice length where
the fluid phase jumps to that of a superfluid state. This implies
that everything above (and including) this critical length
(temperature) should be in the  superfluid phase (the bound vortex
phase)i.e. $\Upsilon\neq0$. This suggests that the $L_t=6$ data
should follow a linear extrapolation to the origin. However, from
our previous findings it was shown that $L_t=12$ follows more of a
$j^2$ extrapolation than a linear one, but even this is not very
clear. We will now investigate the intermediate domain, ranging from
$L_t=6 - 64$, in order to determine, whether or not, the KT critical
point exists in our $2+1d$ NJL model.
\section{Extrapolation Criteria and the Critical Point}
In order to investigate the change in $\Upsilon$ from the high
temperature ($\Upsilon(j)\propto j^2$) phase to the low temperature
linear ($\Upsilon(j)\propto\Upsilon(0)+Aj$) phase we have plotted,
in Fig.~\ref{y-v-j-all-lt}, $\Upsilon$ against $j$ for temperatures
ranging from $L_t=2$ to $L_t=64$.
\begin{figure}
    \begin{center}
        \includegraphics[width=1\textwidth]{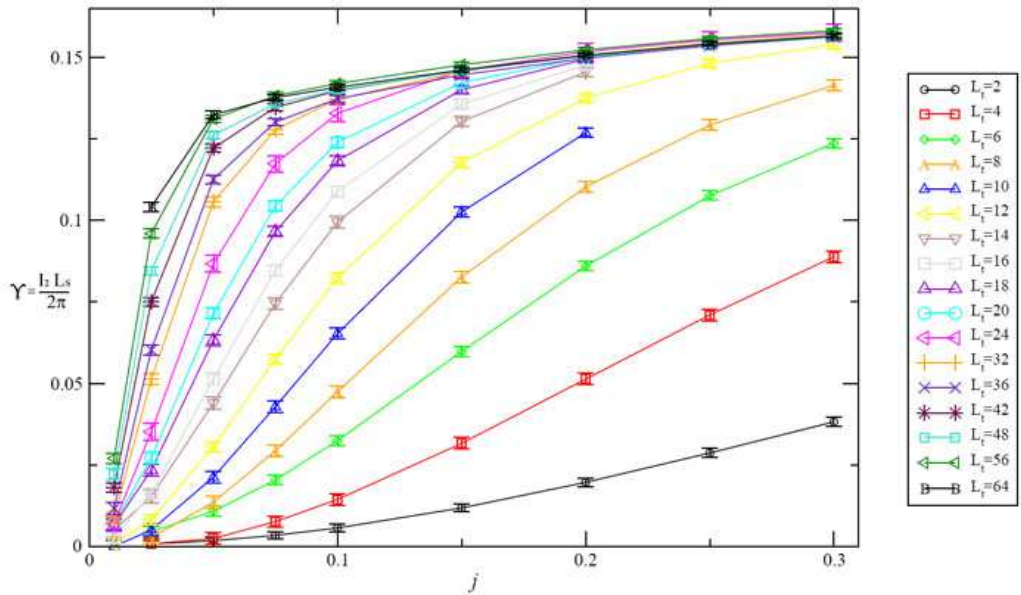}\\
        \caption{$\Upsilon$ as a function of $j$ for various $L_t$ ($\mu=0.8$ and $L_s=32$).}
        \label{y-v-j-all-lt}
    \end{center}
\end{figure}
In this plot we can see a steady transition from the obvious $j^2$
form at $L_t=2$ to an "s" shaped form between $L_t=6-24$ and a
straight line form above $L_t=24$ (which dips to the origin at low
$j$). From this we cannot really distinguish any real, abrupt, phase
change.  So, to obtain a different perspective we have plotted the
same data as $\Upsilon$ against $L_t$ for the various $j$ values in
Fig.~\ref{y-v-lt-ls=32--j=all}; and in
Fig.~\ref{y-v-lt-ls=32--j=0.025} we have a more detailed plot of the
$j=0.025$ data.
\begin{figure}
    \begin{center}
        \includegraphics[width=0.9\textwidth]{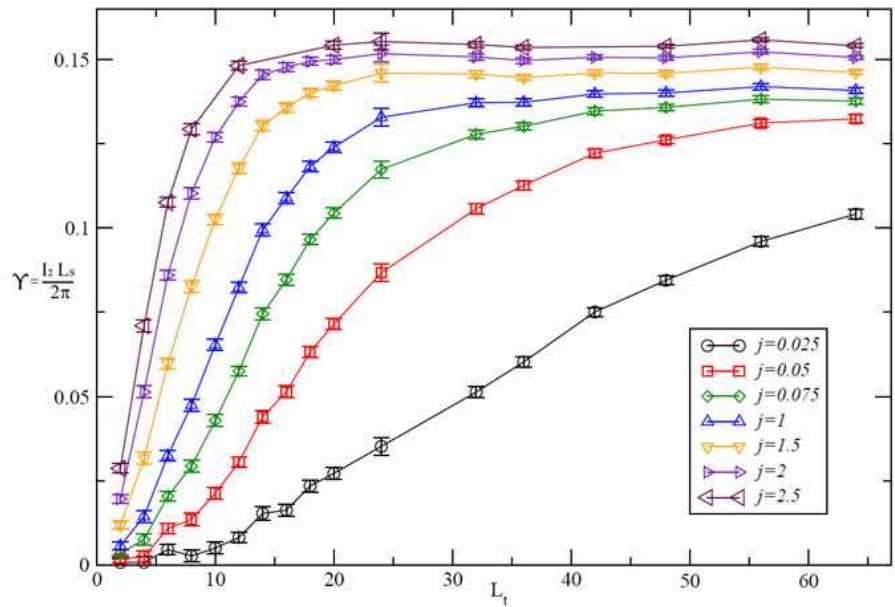}\\
        \caption{$\Upsilon$ as a function of $L_t$ for various $j$ ($\mu=0.8$ and $L_s=32$).}
        \label{y-v-lt-ls=32--j=all}
    \end{center}
\end{figure}
\begin{figure}
    \begin{center}
        \includegraphics[width=0.9\textwidth]{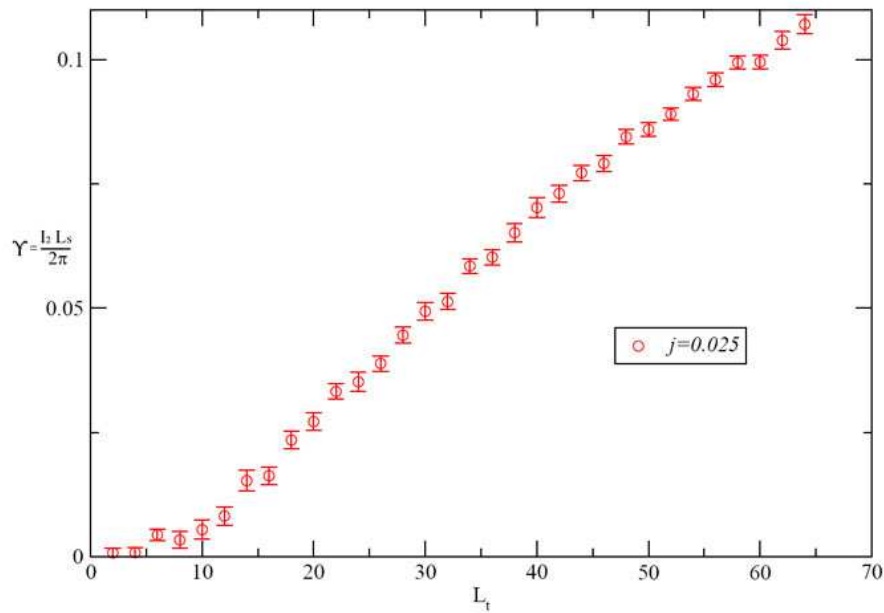}\\
        \caption{A detailed plot of $\Upsilon$ as a function of $L_t$ for $j=0.025$ ($\mu=0.8$ and $L_s=32$).}
        \label{y-v-lt-ls=32--j=0.025}
    \end{center}
\end{figure}
From Fig.~\ref{y-v-lt-ls=32--j=all} we can see that $\Upsilon$
decreases steadily from $L_t=64$ to $L_t=24$, for $j>0.025$.
However, below $L_t=24$ it drops drastically to $\Upsilon=0$.  For
$j=0.025$, in Fig.~\ref{y-v-lt-ls=32--j=0.025}, it can be seen that
$\Upsilon$ follows almost a straight line.  In
Fig.~\ref{y-v-lt-ls=32--j=0025-extrap} we have conducted a straight
line extrapolation to $\Upsilon=0$, from which we found
$L_t=5.54\pm1.66$.
\begin{figure}
    \begin{center}
        \includegraphics[width=0.8\textwidth]{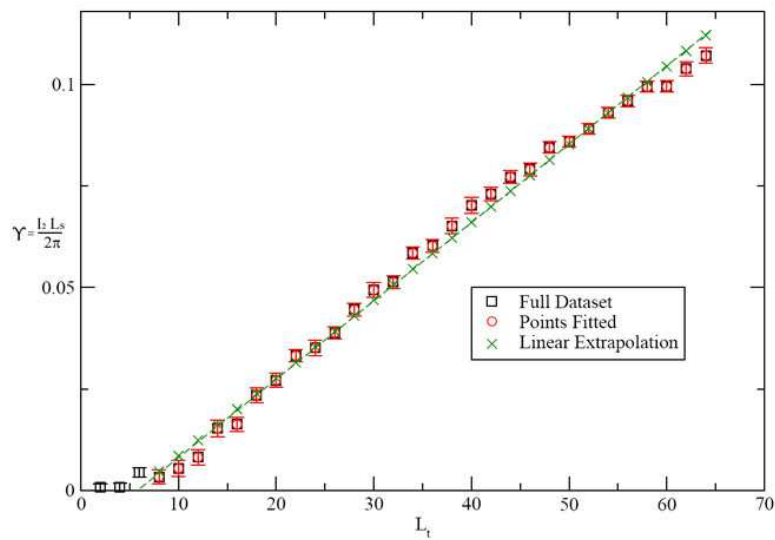}\\
        \caption{$\Upsilon$ as a function of $L_t$ for $j=0.025$ with extrapolation line $y=a_0 x +a_1$
        ($a_0=0.00192\pm0.0005$ and $a_1=-0.01062\pm0.0015$); ($\mu=0.8$ and $L_s=32$).}
        \label{y-v-lt-ls=32--j=0025-extrap}
    \end{center}
\end{figure}
Also, in Fig.~\ref{y-v-lt-ls=32--j=0025-extrap-alternate} we have
conducted another extrapolation, of the same data, but now with only
the first 18 points; from which we find: $L_t=7.24\pm0.9$.  From
these two fits we have the average: $L_t=6.4\pm1.9$.  This value of
$L_t$ is comparable to the one derived in Eqn.~(\ref{kt-derive}).
Therefore, we could conjecture that the critical KT temperature
could be described by the point where $\Upsilon$ intersects the
$L_t$ axis for a fixed small $j$.  In order to confirm this we would
need to conduct further simulations using smaller $j$ (e.g.
$j=0.01$) and maybe larger volumes (in order to reduce finite volume
effects).
\begin{figure}
    \begin{center}
        \includegraphics[width=0.8\textwidth]{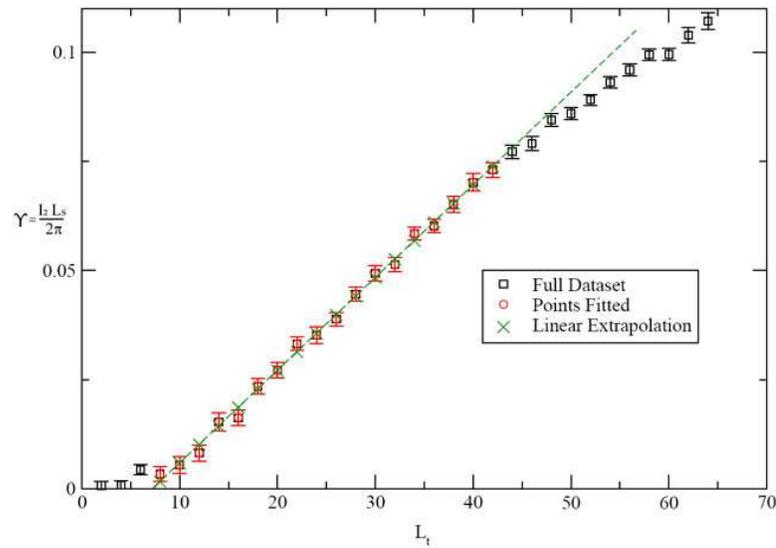}\\
        \caption{$\Upsilon$ as a function of $L_t$ for $j=0.025$ with extrapolation line $y=a_0 x
        +a_1$ using only 18 points
        ($a_0=0.002124\pm0.00025$ and $a_1=-0.015367\pm0.0009$); ($\mu=0.8$ and $L_s=32$).}
        \label{y-v-lt-ls=32--j=0025-extrap-alternate}
    \end{center}
\end{figure}
\subsection{Finite Volume Effects at Low Temperature}
Even though we have located a possible candidate for the critical
temperature, it still does not provide any obvious extrapolation
criterion for the range $L_t=6-24$.

Our next task will be to find a definite lower limit for the
superfluid state, i.e. the smallest $L_t$ point for which infinite
volume extrapolations eliminate the dip in $\Upsilon$ at low $j$. As
for $L_t=64$, Fig.~\ref{y-v-j-ls-lt=64}, we conducted similar
extrapolations for $\Upsilon$ down to $L_t=48$. It was found that
the dip at low $j$ disappears for extrapolations down to $L_t=56$,
Fig.~\ref{y-v-j-lt=56}; these simulations were only done for $4$
points, at low values of $j$, in order to reduce time.
\begin{figure}
    \begin{center}
        \includegraphics[width=0.8\textwidth]{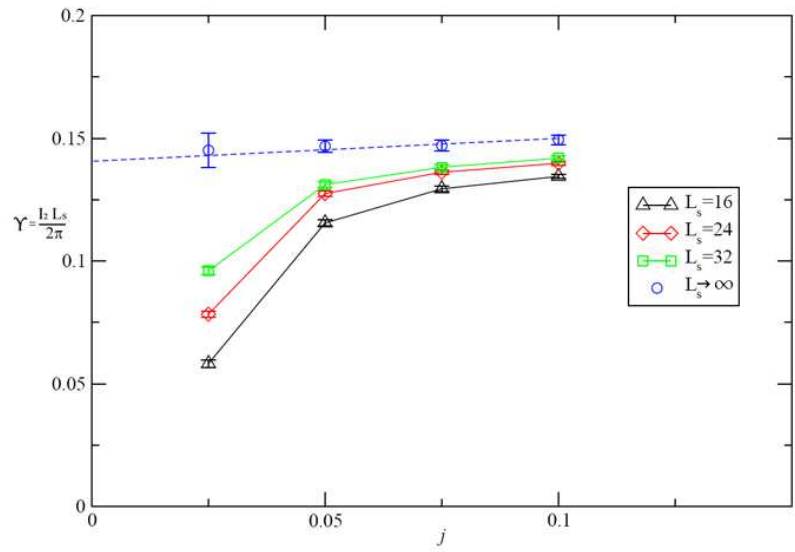}\\
        \caption{The first $4$ points of $\Upsilon$ as a function of $j$ for $L_t=56$ ($\mu=0.8$ and $L_s=32$).}
        \label{y-v-j-lt=56}
    \end{center}
\end{figure}
\begin{figure}
    \begin{center}
        \includegraphics[width=0.8\textwidth]{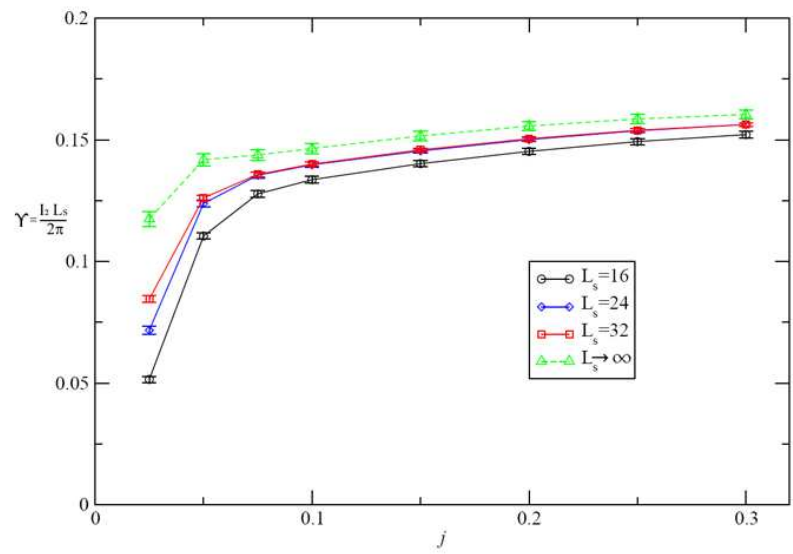}\\
        \caption{$\Upsilon$ as a function of $j$ for $L_t=48$ ($\mu=0.8$ and $L_s=32$).}
        \label{y-v-j-ls=32-lt=48}
    \end{center}
\end{figure}
So, down to and including $L_t=56$ the dip at low $j$ vanishes in
the infinite volume limit.  However, below this point the dip
becomes more and more pronounced, even in the infinite volume limit,
Fig.~\ref{y-v-j-ls=32-lt=48}.  As can be seen for $L_t=24$,
Fig.~\ref{y-v-j-ls-lt=24-extrapolated}, the dip eventually becomes
something that you cannot just ignore.

As in the previous chapter, the infinite volume limit extrapolations
for $L_t=48$ and $L_t=56$, Figs.~\ref{y-v-j-lt=56} and
\ref{y-v-j-ls=32-lt=48}, were done by a straight line fit of
$\Upsilon$ against $1/L_s$, Figs.~\ref{y-v-j-ls-56} and
\ref{y-v-j-ls-48}.
\begin{figure}
    \begin{center}
        \includegraphics[width=0.8\textwidth]{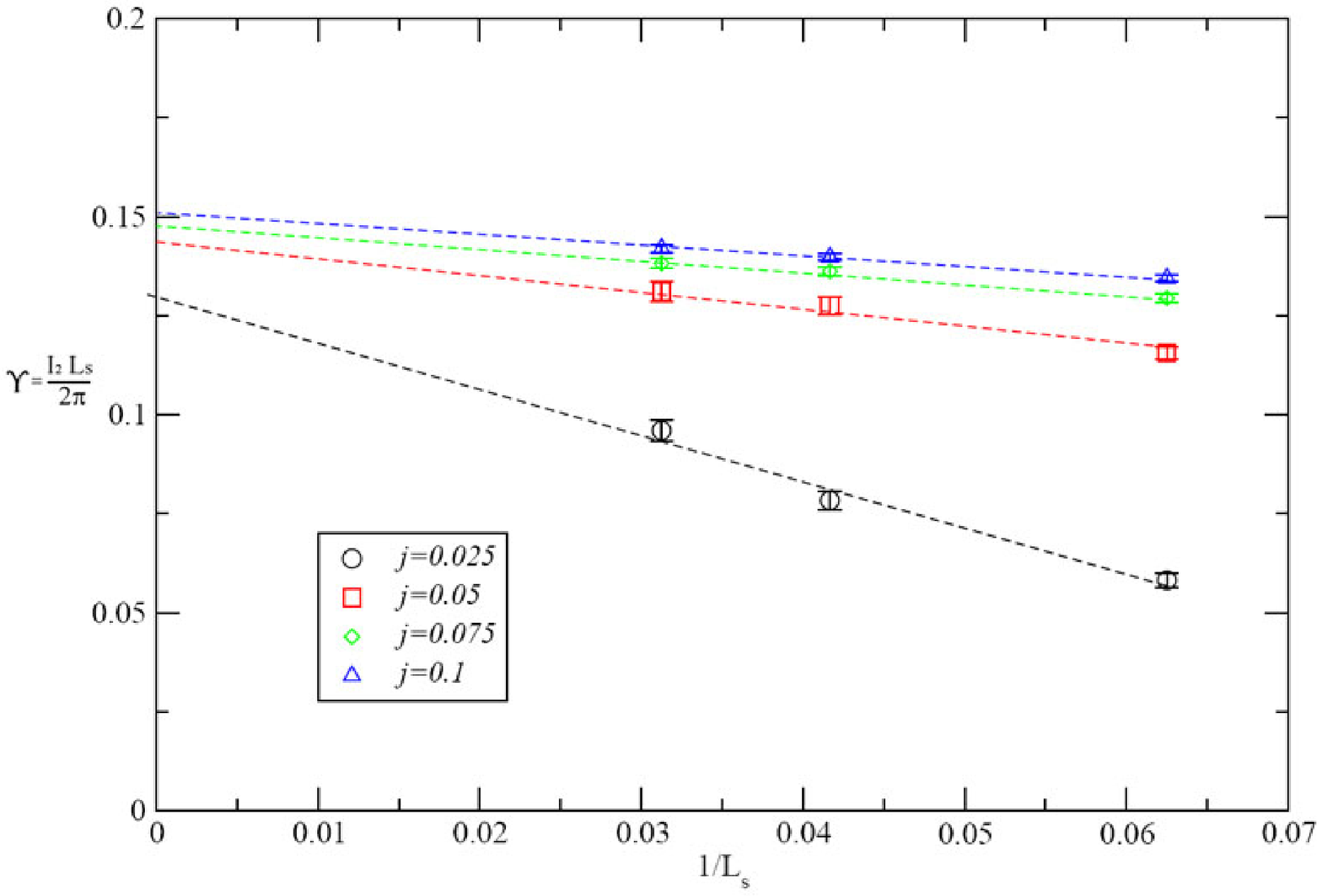}\\
        \caption{$\Upsilon$ as a function of $1/L_s$ for $L_t=56$ ($\mu=0.8$ and $L_s=32$).}
        \label{y-v-j-ls-56}
    \end{center}
\end{figure}
\begin{figure}
    \begin{center}
        \includegraphics[width=0.8\textwidth]{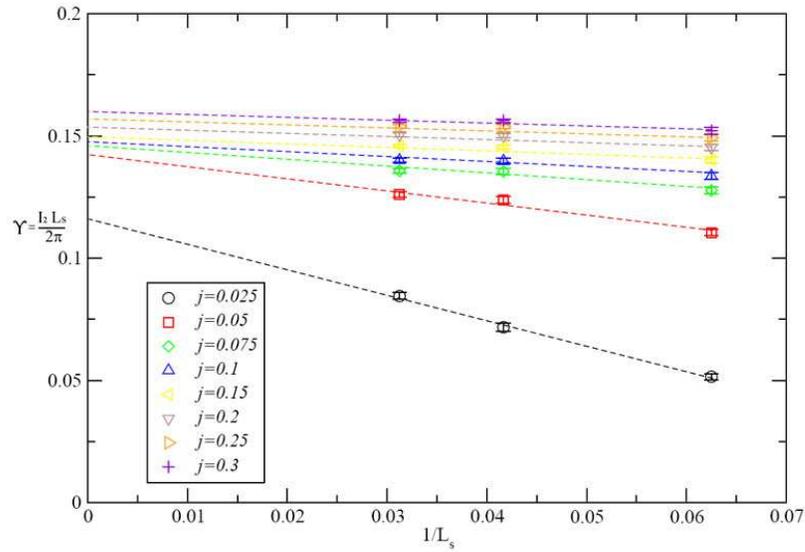}\\
        \caption{$\Upsilon$ as a function of $1/L_s$ for $L_t=48$ ($\mu=0.8$ and $L_s=32$).}
        \label{y-v-j-ls-48}
    \end{center}
\end{figure}
\subsection{Critical Temperature Range}
In a conservative conclusion, the only phase range we can be sure
about is that above $L_t=56$ and the one below $L_t=4$.  For
$L_t\geq56$ infinite volume extrapolations lead to a perfect
straight line, and a non-zero value of $\Upsilon$, implying
superfluidity.  For $L_t\leq4$, even without extrapolation, the
curves are parabolic, Fig.\ref{y-v-j-all-lt}, thus, they could be
associated with the non-superfluid state ($\Upsilon=0$). This then
leaves us with an uncertain range from $L_t=6-54$, which is huge!
However, as I stated before this is a conservative conclusion.  As
we deduced through the plot of $\Upsilon$ against $L_t$ for
$j=0.025$ (Fig.~\ref{y-v-lt-ls=32--j=0.025}) the critical point
could plausibly occur at $L_t\sim6$, in which case all
extrapolations above this point would be of the type
$\Upsilon(j)\propto \Upsilon(0)+Aj$.  This could very well be the
case for $1/T$ above $L_t=24$, even though we can't explain the dips
at low $j$. This then leaves us with the range $L_t=6-22$ as the
uncertain domain, due to the fact that these curves follow a "s"
shape which could be either a linear or parabolic extrapolation.
}

{\typeout{Summary}
\chapter{Summary}
In Chapter~\ref{forced baryon current flow} the calculation of
baryon charge density ($n_B$) was extended to the conserved baryon
3-current $I_{\mu}$, where $n_B$ was the timelike component, $I_0$,
of the current. Then, to force a current flow we implemented a
twisted diquark source i.e. a periodically varying source, spanning
the whole system.

We then investigated the current's behaviour with a  change of
density $\mu$, for two temperatures\footnote{Given as the ratio $L_t
\Upsilon=\frac{\Upsilon_0 a}{T a}=\frac{\Upsilon_0}{T}$.}:
$\frac{\Upsilon_0}{T}=0.565(1)$ and $\frac{\Upsilon_0}{T}=9.043(1)$,
corresponding to lattice lengths $L_t=4$ and $L_t=64$ respectively.
We found that a superfluid phase does exist at high density
($\mu=0.8$) and low temperature $\frac{\Upsilon_0}{T}=9.043(1)$; but
for high temperature, $\frac{\Upsilon_0}{T}=0.565(1)$, the current
is always zero, Figs.~\ref{i2-v-j2-mu-t=64} and
\ref{i2-v-j2-mu-t=4}. For the low temperature case we found that the
current density has a phase transition in the region
$0.65<\mu_c<0.68$, Fig.~\ref{i2-v-j2-mu-t=64}. This transition
region corresponds to that of chiral symmetry restoration in
\cite{hands1}.

These findings show that the low temperature, high density, phase of
the 2+1d NJL model is a thin film superfluid, and the transition
from this superfluid phase at high density to a non-superfluid phase
at low density is of 1st order.

Initially results were presented in a spatially dependent form, so
according to the size of lattice used different results were
obtained for the current. We later showed that it is better to work
with the spatially independent helicity modulus, $\Upsilon$, defined
as the constant of proportionality $I_2=\Upsilon 2\pi/L_2$. The
non-zero value of the helicity modulus was then determined in the
thermodynamic limit for low temperature
($\frac{\Upsilon_0}{T}=9.043(1)$) and high chemical potential
($\mu=0.8$) as $\Upsilon=0.1413(14)$ .  We also showed that this
falls to $\sim 0$ at higher temperatures
($\frac{\Upsilon_0}{T}=0.565(1))$. Therefore, the ratio
$\Upsilon/\Sigma$ that represents the superfluid phase of the $2+1d$
NJL model at high density is given as
$\frac{\Upsilon}{0.71}=0.199(1)\sim 0.2$

Once we had found the presence of a superfluid phase, next, we
wanted to test whether this superfluid behaviour was due to the
binding and unbinding of vortices, as described by the KT theory for
2-dimensional critical systems.  In Chapter~\ref{thin film} we
started by calculating the critical temperature predicted by the KT
theory. Using our value of $\Upsilon$ this critical temperature was
found to be $\frac{\Upsilon_0}{T}=0.637\sim \frac{2}{\pi}$, which
implies a lattice length $L_t=4.51(44)$.

We then showed that by plotting $\Upsilon$ against $L_t$ for low $j$
($j=0.025$ in our case), we obtained a very good straight line fit
that intersects the $L_t$ axis at a point near the KT critical
length prediction, Fig.~\ref{y-v-lt-ls=32--j=0025-extrap} and
\ref{y-v-lt-ls=32--j=0025-extrap-alternate}. Through a straight line
extrapolation ($\Upsilon\propto A L_t+ B$)  this critical point was
found to be at $L_t\sim 6(2)$.  This point corresponds to the KT
prediction within error.  Also, this critical transition point seems
to be of $2nd$ order. However, in order to make any conclusive
statements about the critical point, further study is still needed,
maybe using larger volumes, $L_s>32$, and smaller diquark source
values, $j<0.025$.

We also showed that at low temperature,
$\frac{\Upsilon_0}{T}\geq7.9$ (or in terms of lattice size
$L_t\geq56$), the dip of $\Upsilon$ at low $j$ can be put down to
finite volume effects, which disappears when we extrapolate to the
infinite volume limit. However, even though the plots between
$L_t=24-54$ do follow straight lines for high $j$, similar infinite
volume extrapolations do not eliminate the dips at low $j$. The
domain between $L_t=6-22$ is even more puzzling.  In this area we
cannot easily apply a straight line extrapolation, as the curves of
$\Upsilon$ against $j$ are parabolic at low $j$.

So, in order to construct a conclusive phase diagram for $\Upsilon$
against $L_t$, we not only need to confirm the location of the
critical point, but we also need to investigate and explain the
strange behaviour of $\Upsilon$ in the intermediate temperature
range.
}

{\typeout{Components of the Riemann Tensor in Schwarzschild
Spacetime}
 \appendix{}
\chapter{Schwarzschild Spacetime}
\label{app schwart} The 'Schwarzschild geometry' describes a
spherically symmetric spacetime outside a star, and its properties
are determined by one parameter, the mass M.  The Schwarzschild
metric, in spherical polar coordinates, takes the form:
\begin{equation}
        g_{\mu\nu}=\left(%
        \begin{array}{cccc}
        F & 0 & 0 & 0 \\
        0 & -F^{-1} & 0 & 0 \\
        0 & 0 & -r^2 & 0 \\
        0 & 0 & 0 & -r^2\sin(\theta) \\
        \end{array}%
        \right)
    \end{equation}
This type of spacetime geometry is said to be static, due to the
fact that (i) all metric components are independent of $t$, and (ii)
the geometry is unchanged by time reversal, $t\rightarrow
-t$\footnote{A space time with property (i) but not necessarily (ii)
is said to be stationary i.e. a rotating star/black hole}.
\section{Geodesic Equations}
The equations of motion in Schwarzschild spacetime can be derived by
using the covariant form of Newton's second law of motion,
\begin{eqnarray}
F^{\mu}&=&\frac{D p^{\mu} }{D\tau}\nonumber\\
&=&\frac{dp^{\mu}}{d\tau}+\Gamma^{\mu}_{\nu\rho}\frac{dx^{\nu}}{d\tau}p^{\rho}
\end{eqnarray}
where we have $p^{\mu}=m\frac{dx^{\mu}}{d\tau}$, and for free fall
we have $F^{\mu}=0$, then the equations of motion are:
\begin{equation}
0=\frac{d^2x^{\mu}}{d\tau^2}+\Gamma^{\mu}_{\nu\rho}\frac{dx^{\nu}}{d\tau}\frac{dx^{\rho}}{d\tau}
\end{equation}
Now, using this and the appropriate Christoffels we are able to
determine the equations of motion.  Alternatively, by using the
Schwarzschild Lagrangian
    \begin{equation}
        L=\frac{1}{2}[ (1-\frac{2M}{r}) \dot{dt} ^{2} - (1-\frac{2M}{r}) ^{-1} \dot{dr}^{2}-r^{2} \dot{d\theta}^{2}-(r^{2}
        \sin^{2} \theta) \dot{d\phi}^{2}]
        \label{schwarzschild lagrangian}
    \end{equation}
we find
\begin{eqnarray}
        \frac{\partial L}{\partial t}&=&0 \qquad \frac{\partial L}{\partial \dot{t}}=(1-\frac{2M}{r})\dot{t}\nonumber\\
        \frac{d}{d \lambda}\frac{\partial L}{\partial \dot{t}}&=&(1-\frac{2M}{r})\ddot{t}+\frac{2M}{r^{2}}
        \dot{r}\dot{t}\nonumber\\
\end{eqnarray}
\begin{eqnarray}
        \frac{\partial L}{\partial r}&=&\frac{M}{r^2}\dot{t}^2+\frac{M}{r}^{2}(1-\frac{2M}{r})^{-2}\dot{r}^{2} -r
        \dot{\theta}^{2}-r \dot{\phi}^{2} \sin^{2}\theta\nonumber\\
        \frac{\partial
        L}{\partial\dot{r}}&=&-(1-\frac{2M}{r})^{-1}\dot{r}\nonumber\\
        \frac{d}{d \lambda} \frac{\partial L}{\partial \dot{r}}&=&-(1-\frac{2M}{r})^{-1}\ddot{r}+\frac{2M}{r^{2}}(1-
        \frac{2M}{r})^{-2}\dot{r}\dot{r}\nonumber\\
\end{eqnarray}
\begin{eqnarray}
        \frac{\partial L}{\partial \theta}&=&-r^{2} \sin\theta \cos\theta \dot\phi^{2} \qquad
        \frac{\partial L}{\partial
        \dot{\theta}}=-r^{2}\dot{\theta}\nonumber\\
        \frac{d}{d \lambda}\frac{\partial L}{\partial
        \dot{\theta}}&=&-2r\dot{r}\dot{\theta}-r^{2}\ddot{\theta}\nonumber\\
\end{eqnarray}
\begin{eqnarray}
        \frac{\partial L}{\partial \phi}&=&0 \qquad
        \frac{\partial L}{\partial \dot{\phi}}=-r^{2}\sin^{2}\theta
        \dot{\phi}^{2}\nonumber\\
        \frac{d}{d\lambda}\frac{\partial L}{\partial \dot{\phi}}&=&-2r\dot{r}\sin^2\theta \dot{\phi}-2r^2\sin\theta \cos\theta
                 \dot{\phi}-r^2\sin^2\theta \ddot{\phi}\nonumber\\
        \end{eqnarray}
and substituting into the Euler equation
    \begin{equation}
        \frac{\partial L}{\partial x^{\mu}}-\frac{d}{d\lambda}\frac{\partial L}{\partial \dot{x}^{\mu}}=0
        \label{euler equation}
    \end{equation}
 we have the equations of motion for Schwarzschild spacetime
    \begin{eqnarray}
  \label{equation of motion t}
        0&=&(1-\frac{2M}{r})\ddot{t}+\frac{2M}{r^2}\dot{r}\dot{t}=0\\
  \label{equation of motion r}
        0&=&\frac{M}{r^2}\dot{t}^2-r\dot{\theta}^{2}-r \dot{\phi}^{2}
        \sin^{2}\theta+(1-\frac{2M}{r})^{-1}\ddot{r}\nonumber\\
        &-&\frac{M}{r^{2}}(1-\frac{2M}{r})^{-2}\dot{r}^{2}\\
\label{equation of motion theta}
       0&=& -r^{2} \sin\theta \cos\theta
       \dot\phi^{2}+2r\dot{r}\dot{\theta}+r^{2}\ddot{\theta}\\
 \label{equation of motion phi}
        0&=&2r\dot{r}\sin^2\theta \dot{\phi}+2r^2\sin\theta \cos\theta \dot{\phi}+r^2\sin^2\theta
        \ddot{\phi}\\
        \nonumber
    \end{eqnarray}
\section{Riemann Components in the Orthonormal Frame}
\label{orthonormal} The spacetime line interval,
Eqn.~(\ref{schwarzschild line interval}), can be rewritten using
tetrad transformations (discussed in the next section):
    \begin{equation}
        \omega^t=(1-\frac{2M}{r})^{\frac{1}{2}}dt \qquad
        \omega^r=(1-\frac{2M}{r})^{-\frac{1}{2}}dr \qquad
        \omega^\theta=r d\theta \qquad
        \omega^\phi=r \sin(\theta)d\phi
    \label{omega transforms}
    \end{equation}
as
    \begin{equation}
         ds^{2}=(\omega^{t})^{2}-(\omega^{r})^{2}-(\omega^{\theta})^{2}-(\omega^{\phi})^{2}
        \label{transformed omega interval}
    \end{equation}
This now has the Minkowski metric
    \begin{equation}
    g_{\mu\nu}=\left(%
\begin{array}{cccc}
  1 & 0 & 0 & 0 \\
  0 & -1 & 0 & 0 \\
  0 & 0 & -1 & 0 \\
  0 & 0 & 0 & -1 \\
\end{array}%
\right)
        \label{minkoski metric appedix}
    \end{equation}
Noting that the metric can be written as:
    \begin{equation}
        dg_{\mu\nu}=\omega_{\mu\nu}+\omega_{\nu\mu}
    \end{equation}
and using the fact that:
    \begin{equation}
        dg_{\mu\nu}=\frac{\partial g_{\mu\nu}}{\partial
        x^{\alpha}}dx^\alpha=0
    \end{equation}
we then have\cite{chandrasekhar}:
    \begin{equation}
        \omega_{\mu\nu}=-\omega_{\nu\mu}
    \end{equation}
which implies $\omega$ is antisymmetric, i.e. it has only six unique
components and $\omega_{\mu\nu}=0$ for $\mu=\nu$; and we have the
conditions:
    \begin{equation}
        \omega^0_i=\omega^i_0 \qquad \omega^i_j=-\omega^j_i
    \end{equation}
We can now write the exterior derivatives\footnote{The exterior
derivative: $d=dx^{\mu}\frac{\partial}{\partial x^{\mu}}$ acting on
a 1-form $A^{\nu}=A(x)dx^{\nu}$ gives $dA^{\nu}=dx^{\mu}\wedge
dx^{\nu}\frac{\partial A(x)}{\partial x^{\mu}}$, where
$dx^{\mu}\wedge dx^{\nu}=0$ for $\mu=\nu$ } of $\omega^{\mu}$
    \begin{eqnarray}
  \label{exterior t}
        d\omega^t&=&\frac{1}{2}(1-\frac{2M}{r})^{- \frac{1}{2}}(\frac{2M}{r^2})dr\wedge
        dt\nonumber\\
        &=&\frac{M}{r^2}(1-\frac{2M}{r})^{-\frac{1}{2}}\omega^r\wedge\omega^t\\
     \label{exterior r}
        d\omega^r&=&0\\
  \label{exterior theta}
        d\omega^{\theta}&=&dr\wedge d\theta=\frac{1}{r}(1-\frac{2M}{r})^{\frac{1}{2}}\omega^r \wedge
        \omega^{\theta}\\
 \label{exterior phi}
        d\omega^{\phi}&=&\sin(\theta) dr\wedge
        d\phi+r\cos(\phi)d\theta \wedge
        d\phi\nonumber\\
        &=&\frac{1}{r}(1-\frac{2M}{r})^{\frac{1}{2}}\omega^r \wedge
        \omega^{\phi}+\frac{\cot(\theta)}{r} \omega^\theta \wedge
        \omega^{\phi}\\
        \nonumber
    \end{eqnarray}
Now, using Cartan's equation
        \begin{equation}
            d\omega^\mu=\omega^\alpha \wedge
            \omega^\mu_\alpha+\Omega^\mu
        \end{equation}
    with zero torsion $(\Omega^\mu=0)$, we can write
\begin{eqnarray}
   \label{domega t}
        d\omega^t&=&\omega^r\wedge \omega^t_r+\omega^\theta \wedge
        \omega^t_\theta+\omega^\phi \wedge \omega^t_\phi\\
      \label{domega_r}
        d\omega^r&=&\omega^t\wedge
        \omega^r_t+\omega^\theta\wedge\omega^r_\theta+\omega^\phi\wedge\omega^r_\phi\\
   \label{domega theta}
        d\omega^\theta&=&\omega^t\wedge\omega^\theta_t+\omega^r\wedge\omega^\theta_r+\omega^\phi\wedge\omega^\theta_\phi\\
     \label{domega phi}
        d\omega^\phi&=&\omega^t\wedge\omega^\phi_t+\omega^r\wedge\omega^\phi_r+\omega^\theta\wedge\omega^\phi_\theta\\
\nonumber
    \end{eqnarray}
Comparing Eqns.~(\ref{exterior t})-(\ref{exterior phi})
    with Eqns.~(\ref{domega t})-(\ref{domega phi}) we find the
    six unique components of $\omega^\mu_\nu $ as
    \begin{eqnarray}
        \omega^t_r&=&\frac{M}{r^2}(1-\frac{2M}{r})^{-\frac{1}{2}}\omega^t=\frac{M}{r^2}dt=\omega^r_t\\
        \omega^\theta_r&=&\frac{1}{r}(1-\frac{2M}{r})^{\frac{1}{2}}\omega^\theta=(1-\frac{2M}{r})^{\frac{1}{2}}d\theta=-\omega^r_\theta\\
        \omega^\phi_r&=&\frac{1}{r}(1-\frac{2M}{r})^{\frac{1}{2}}\omega^\phi=\sin(\theta)(1-\frac{2M}{r})^{\frac{1}{2}}d\phi=-\omega^r_\phi\\
        \omega^\phi_\theta&=&\frac{\cot(\theta)}{r}\omega^\phi=\cos(\theta)d\phi=-\omega^\theta_\phi\\
        \omega^\theta_t&=&\omega^t_\theta=0\\
        \omega^\phi_t&=&\omega^t_\phi=0\\
        \nonumber
    \end{eqnarray}
Now, using:
    \begin{equation}
        R^\mu_\nu=d\omega^\mu_\nu+\omega^\mu_\alpha\wedge\omega^\alpha_\nu
    \end{equation}
we can write the six independent components of the Riemann tensor:
\begin{eqnarray}
        R^t_r&=&d\omega^t_r+\omega^t_\alpha\wedge\omega^\alpha_r=d\omega^t_r+\omega^t_\theta\wedge\omega^\theta_r+\omega^t_\phi\wedge\omega^\phi_r\nonumber\\
        &=&-\frac{2M}{r^3}dr\wedge
        dt=-\frac{2M}{r^3}\omega^r\wedge\omega^t\nonumber\\
         &\Rightarrow& R^{t}_{rrt}=R_{trrt}=-\frac{2M}{r^3}\nonumber\\
 \label{R t r r t}
\end{eqnarray}
\begin{eqnarray}
         R^{\theta}_r&=&d \omega^{\theta}_r+ \omega^{\theta}_\alpha \wedge \omega^{\alpha}_r
         =d \omega^{\theta}_r+\omega^{\theta}_t \wedge \omega^t_r
         +\omega^{\theta}_\phi\wedge\omega^{\phi}_r\nonumber\\
         &=&\frac{1}{2}(1-\frac{2M}{r})^{-\frac{1}{2}}\frac{2M}{r^2}dr \wedge d \theta=
         \frac{M}{r^3} \omega^r \wedge \omega^\theta\nonumber\\
        &\Rightarrow& R^{\theta}_{rr\theta}=-R_{\theta
        rr\theta}=\frac{M}{r^3}\nonumber\\
\label{R theta r r theta}
        \end{eqnarray}
\begin{eqnarray}
       R^{\phi}_r&=&d \omega^{\phi}_r+ \omega^{\phi}_\alpha \wedge \omega^{\alpha}_r=d \omega^{\phi}_r+\omega^{\phi}_t \wedge \omega^t_r
         +\omega^{\phi}_\theta\wedge\omega^{\theta}_r\nonumber\\
         &=&\cos(\theta)(1-\frac{2M}{r})^{\frac{1}{2}}d\theta\wedge
         d\phi+\sin(\theta)\frac{M}{r^2}(1-\frac{2M}{r})^{-\frac{1}{2}}dr\wedge
         d\phi\nonumber\\
         &+&\cos(\theta)(1-\frac{2M}{r})^{1}{2}d\phi\wedge
         d\theta\nonumber\\
         &=&\frac{\cot(\theta)}{r^2}(1-\frac{2M}{r})^{\frac{1}{2}}
         \omega^\theta\wedge \omega^\phi +
         \frac{M}{r^3}\omega^r \wedge
         \omega^\phi\nonumber\\
         &+&\frac{\cot(\theta)}{r^2}(1-\frac{2M}{r})^{\frac{1}{2}}\omega^\phi \wedge
         \omega^\theta=\frac{M}{r^3}\omega^r\wedge \omega^\phi\nonumber\\
        &\Rightarrow& R^{\phi}_{rr\phi}=-R_{\phi
        rr\phi}=\frac{M}{r^3} \nonumber\\
   \label{R phi r r phi}
        \end{eqnarray}
\begin{eqnarray}
         R^{\phi}_\theta&=&d \omega^{\phi}_\theta+ \omega^{\phi}_\alpha \wedge \omega^{\alpha}_\theta=d \omega^{\phi}_\theta+\omega^{\phi}_t \wedge
         \omega^t_\theta+\omega^{\phi}_r\wedge\omega^r_\theta\nonumber\\
         &=&-\sin(\theta)d\theta\wedge
         d\phi-\sin(\theta)(1-\frac{2M}{r})d\phi\wedge
         d\theta\nonumber\\
         &=&-\frac{1}{r^2}\omega^\theta\wedge \omega^\phi +
         \frac{1}{r^2}(1-\frac{2M}{r})\omega^\theta \wedge
         \omega^\phi=-\frac{2M}{r^3}\omega^\theta\wedge\omega^\phi\nonumber\\
         &\Rightarrow&
         R^\phi_{\theta\theta\phi}=-R_{\phi\theta\theta\phi}=-\frac{2M}{r^3} \nonumber\\
 \label{R phi theta theta phi}
        \end{eqnarray}
\begin{eqnarray}
         R^{\theta}_t&=&d \omega^{\theta}_t+ \omega^{\theta}_\alpha \wedge \omega^{\alpha}_t=
         d \omega^{\theta}_t+\omega^{\theta}_r \wedge \omega^r_t
         +\omega^{\theta}_\phi\wedge\omega^{\phi}_t\nonumber\\
         &=&\frac{M}{r^2}(1-\frac{2M}{r})^{\frac{1}{2}}d\theta \wedge dt
         =\frac{M}{r^3} \omega^\theta \wedge \omega^t\nonumber\\
         & \Rightarrow &R^{\theta}_{t\theta t}=-R_{\theta t\theta
         t}=\frac{M}{r^3} \nonumber\\
\label{R theta t theta t}
         \end{eqnarray}
\begin{eqnarray}
         R^{\phi}_t&=&d \omega^{\phi}_t+ \omega^{\phi}_\alpha \wedge \omega^{\alpha}_t=
         d \omega^{\phi}_t+\omega^{\phi}_r \wedge \omega^r_t
         +\omega^{\phi}_\theta\wedge\omega^{\theta}_t\nonumber\\
         &=&\frac{M}{r^2}(1-\frac{2M}{r})^{\frac{1}{2}}\sin(\theta)d\theta \wedge dt
         =\frac{M}{r^3} \omega^\phi \wedge \omega^t\nonumber\\
        &\Rightarrow& R^{\phi}_{t\phi t}=-R_{\phi t\phi
        t}=\frac{M}{r^3}\nonumber\\
  \label{R phi t phi t}
        \end{eqnarray}
\section{Tetrad Transformation}
\label{tetrad} In Eqn.~(\ref{omega transforms}) we transformed from
the coordinate frame to an orthonormal frame. This is achieved by
using the tetrad of the form:
       \begin{equation}
            e^{a}_{\mu}=\left(%
            \begin{array}{cccc}
            F^{\frac{1}{2}} & 0 & 0 & 0 \\
            0 & F^{-\frac{1}{2}} & 0 & 0 \\
            0 & 0 & r & 0 \\
            0 & 0 & 0 & r\sin(\theta) \\
            \end{array}%
            \right)
            \label{tetrad}
    \end{equation}
This takes a 4-vector, given in the coordinate frame, and maps it to
the equivalent in the orthonormal frame.  We can then determine the
inverse tetrad, i.e. a tetrad that takes vectors in the orthonormal
frame and maps them to the coordinate frame.  We find this by using
the inverse of (\ref{tetrad}),
    \begin{equation}
        (e^{-1})^{\mu}_{a}=\left(%
        \begin{array}{cccc}
        F^{-\frac{1}{2}} & 0 & 0 & 0 \\
        0 & F^{\frac{1}{2}} & 0 & 0 \\
        0 & 0 & \frac{1}{r} & 0 \\
        0 & 0 & 0 & \frac{1}{r\sin(\theta)} \\
        \end{array}%
        \right),
        \label{inverse tetrad}
    \end{equation}
which has the property:
    \begin{equation}
        e^{a}_{\mu}(e^{-1})^{\mu}_{b}=\delta^a_b
    \end{equation}
Therefore, (\ref{inverse tetrad}) takes vectors in the orthonormal
frame and maps them to the coordinate frame.
\section{Riemann Components in the Coordinate Frame}
In Sec.~\ref{orthonormal} the unique components of the Riemann
tensor were calculated in the orthonormal frame.  In order to
determine the components in the coordinate basis we will use the
transformation tetrad given in Eqn.~\ref{inverse tetrad}.  Before we
can us the tetrad, (\ref{inverse tetrad}), we need to adjust it so
it is able to take covectors in the orthonormal frame, rather than
vectors:
    \begin{equation}
        (e^{-1})^{\mu}_ag_{\mu\nu}\eta^{ab}=(e^{-1})_\nu^{b}=\left(%
        \begin{array}{cccc}
          F^{\frac{1}{2}} & 0 & 0 & 0 \\
          0 & F^{-\frac{1}{2}} & 0 & 0 \\
          0 & 0 & r & 0 \\
          0 & 0 & 0 & r \\
          \end{array}%
            \right),
            \label{modified inverse tetrad}
    \end{equation}
which is just Eqn.~(\ref{tetrad}).  Then using this we have:
    \begin{displaymath}
        \Rightarrow \qquad A_\mu(\textrm{coordinate
        frame})=(e^{-1})_{\mu}^aA_a(\textrm{orthonormal frame})
    \end{displaymath}
Now using this tetrad we can transform the orthonormal Riemann
tensor components as:
        \begin{displaymath}
            R_{\mu\nu\alpha\beta}=(e^{-1})_{\mu}^{a}
            (e^{-1})_{\nu}^{b}(e^{-1})_{\alpha}^{c}(e^{-1})_{\beta}^{d}R_{abcd}
        \end{displaymath}
Now using the components given in Sec.~\ref{orthonormal} we have:
    \begin{eqnarray}
        R'_{trrt}&=&(e^{-1})_{t}^{t}
            (e^{-1})_{r}^{r}(e^{-1})_{r}^{r}(e^{-1})_{t}^{t}R_{trrt}=F^{\frac{1}{2}}
            F^{-\frac{1}{2}}F^{-\frac{1}{2}}F^{\frac{1}{2}}(-\frac{2M}{r^3})=-\frac{2M}{r^3}\nonumber\\
        R'_{\theta r r \theta}&=&(e^{-1})_{\theta}^{\theta}
            (e^{-1})_{r}^{r}(e^{-1})_{r}^{r}(e^{-1})_{\theta}^{\theta}R_{\theta r r \theta}=r F^{-\frac{1}{2}}
            F^{-\frac{1}{2}}r(-\frac{M}{r^3})=-\frac{MF^{-1}}{r}\nonumber\\
        R'_{\phi rr \phi}&=&(e^{-1})_{\phi}^{\phi}
            (e^{-1})_{r}^{r}(e^{-1})_{r}^{r}(e^{-1})_{\phi}^{\phi}R_{\phi r r
            \phi}=r
            F^{-\frac{1}{2}}F^{-\frac{1}{2}}r(-\frac{M}{r^3})=-\frac{MF^{-1}}{r}\nonumber\\
        R'_{\phi \theta \theta \phi}&=&(e^{-1})_{\phi}^{\phi}
            (e^{-1})_{\theta}^{\theta}(e^{-1})_{\theta}^{\theta}(e^{-1})_{\phi}^{\phi}R_{\phi \theta \theta
            \phi}=r r r r(\frac{2M}{r^3})=2Mr\nonumber\\
        R'_{\theta t \theta t}&=&(e^{-1})_{\theta}^{\theta}
            (e^{-1})_{t}^{t} (e^{-1})_{\theta}^{\theta} (e^{-1})_{t}^{t}  R_{\theta t \theta t}=r F^{\frac{1}{2}}
             r F^{\frac{1}{2}}(-\frac{M}{r^3})=-\frac{M
             F}{r}\nonumber\\
        R'_{\phi t \phi t}&=&(e^{-1})_{\phi}^{\phi}
            (e^{-1})_{t}^{t}(e^{-1})_{\phi}^{\phi} (e^{-1})_{t}^{t} R_{\phi
            t \phi} t=r  F^{\frac{1}{2}} r F^{\frac{1}{2}}(-\frac{M}{r^3})=-\frac{M
            F}{r}\nonumber\\
    \end{eqnarray}
Therefore, the Riemann tensor components in the coordinate frame are
given as:
    \begin{eqnarray}
   \label{coordinate frame Riemann components}
        R'_{trrt}&=&-\frac{2M}{r^3} \qquad  R'_{\theta r r
        \theta}=-\frac{MF^{-1}}{r} \qquad R'_{\phi rr
        \phi}=-\frac{MF^{-1}}{r}\nonumber\\
         R'_{\phi \theta \theta
        \phi}&=&2Mr \qquad  R'_{\theta t \theta t}=-\frac{M F}{r}
        \qquad R'_{\phi t \phi t}=-\frac{M F}{r}\nonumber\\
    \end{eqnarray}
}

{\typeout{Appendix A}
 \chapter{XY Model and the \textit{KT} Transition}
\label{condensed matter}
\section{Spin Systems}
\subsection{The Ising Model}
The most basic spin system is the Ising model, which is generally
used to describe magnetisation.  It consists of $N$ spin sites, each
taking a value of $+1$ or $-1$ (i.e. up and down respectively). In
its simplest form it describes a 1-dimensional lattice, usually with
the $N$ sites in a chain formation and periodic boundaries.  Other
simplifications include a uniform external field at each site,
$H_i=H$, so that the only interaction between spins is that between
the nearest neighbours, symbolised by $J$.  This system is then
described by the Hamiltonian:
\begin{equation}
-\mathcal{H}=\sum_{i=1}^{N}H_i S_{i}+J\sum_{<i,j>}S_i S_j
\label{ising model}
\end{equation}
The partition function of this model is then written as:
\begin{equation} Z=\sum_{[S_i]}\exp(-\beta
\mathcal{H}(S_i)) \label{ising partition}
\end{equation}
where $\beta=1/(k_B T)$, and the the sum is taken over all possible
states i.e. a total of $2^N$ possible spin states.  Then, to extract
thermodynamic information from the partition function we define the
free energy of the Ising model as,
\begin{displaymath}
F=-k_B T \ln Z \label{free energy}
\end{displaymath}
From this we can determine the magnetisation of the
system\footnote{For $H_i=H$
 we have $<S>=\sum_i<S_i>=\frac{\partial F}{\partial H}$},
\begin{displaymath}
M=\frac{1}{N}<S> \qquad
<S>=\sum_{i=1}^{N}<S_i>=-\sum_{i=1}^{N}\frac{\partial F}{\partial
 H_i}
\end{displaymath}
where $<S_i>$ is the average spin of the $i^{th}$ site.  The
correlation function, a measure of the influence exerted by a given
spin $S_i$ on other spin sites $S_j$, is given as:
\begin{displaymath}
G_{ij}=<S_i S_j>-<S_i><S_j>;
\end{displaymath}
Since the interaction between spins favours alignment, a nearby spin
$S_j$ will tend to assume the same orientation as $S_i$; however
thermal fluctuations counteract this tendency and exert a
de-correlating effect.  Thus we expect some correlation that weakens
as the distance between $S_i$ and $S_j$ increases, also at a fixed
distance apart the correlation will be stronger when the temperature
is lower, see Fig.~\ref{corr-fig} \cite{bellac}.
\begin{figure}
    \begin{center}
        \includegraphics[width=0.6\textwidth]{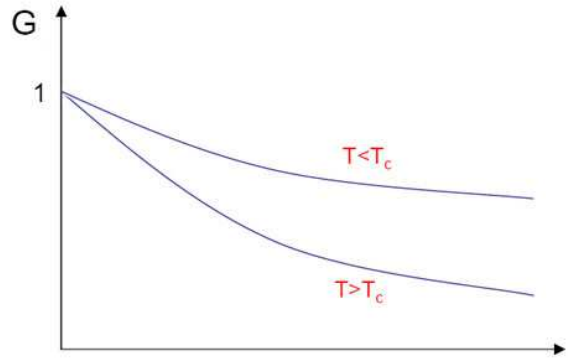}\\
        \caption{Ratio of the correlation function $G=G(r_{ij})/G(0)$, for two different temperatures
        $T$ (not to scale).}
        \label{corr-fig}
    \end{center}
\end{figure}
So, when the temperature is low we can assume that the majority of
spin states are correlated, which in the absence of an external
field leads to the phenomenon of spontaneous magnetization, $M_s$.
For the 1-dimensional Ising model this does not occur for $T\neq0$,
but there is a trivial phase change at $T=0$, i.e. when the thermal
energy is zero all the spins are aligned. However, in 2-dimensions
there is a second order phase transition at the Curie temperature
$T=T_c$ \cite{bellac}\cite{goldenfield}, Fig.~\ref{spon-mag}.
\begin{figure}
    \begin{center}
        \includegraphics[width=0.54\textwidth]{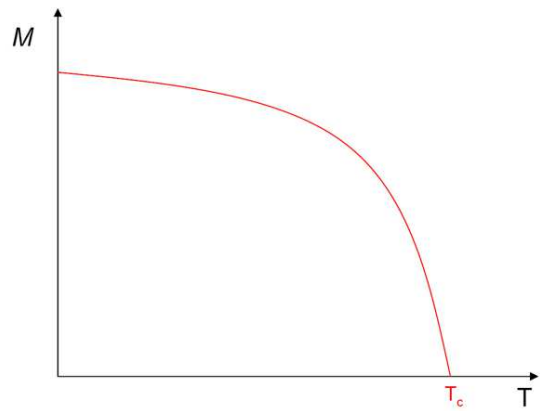}\\
        \caption{Spontaneous magnetization in 2-dimensions against temperature.}
        \label{spon-mag}
    \end{center}
\end{figure}
\subsection{The XY Model}
The restriction of the Ising model is that the spin vector can only
assume one of 2 discrete values. Even though for some cases this is
fine, a more realistic model is the XY model. In this case spin is a
continuous vector in a plane (for 2 dimensions). This model is
characterized by the Hamiltonian:
\begin{equation}
-\mathcal{H}=\sum_{i=1}^{N}H_i
S^z_{i}+J\sum_{<i,j>}\vec{S}_i\cdot\vec{S}_j
\end{equation}
where the spins are classical vectors of unit length aligned in the
plane.  For zero external field we can write:
\begin{equation}
-\mathcal{H}=J\sum_{<i,j>}\vec{S}_i\cdot\vec{S}_j=J\sum_{<i,j>}\cos(\Phi_i-\Phi_j),
\label{xy model}
\end{equation}
where $\Phi$ is the angle between the spin direction and an
arbitrarily chosen axis, and is a continuous variable.

In this 2-dimensional model, spin excitations (spin waves) are
easier to excite thermally than for the Ising model, they are so
strong that they can destroy long-range order at any finite T; so,
in the case of 2-dimensions, symmetry breaking for $T\geq0$ does not
occur (Mermin-Wagner theorem \cite{mermin}). On the other hand, in
1973, Kosterlitz and Thouless showed that a kind of phase transition
can exist for the $2d$ XY model, or more accurately there is a
critical point at non-zero temperature.  It arises not from the
long-range ordering like spontaneous magnetization but from the
excitations of vortex-antivortex pairs
\cite{kosterlitz}\cite{kosterlitz2}.
\section{Phase Transitions in the XY Model}
\subsection{The Critical Phase in the XY Model}
\subsubsection{Vortices}
If we consider small spin variations,$\Phi_i-\Phi_j\ll1$, then we
can expand \ref{xy model} as:
\begin{equation}
\mathcal{H}\simeq
E_0+\frac{1}{2}J\sum_{<i,j>}(\Phi_i-\Phi_j)^2=\frac{J
a^2}{2}\sum_i|\vec{\nabla}\Phi(i)|^2 \label{xy model expanded}
\end{equation}
where $a$ is the lattice spacing.  This can now be taken to the
continuum limit, so that $\Phi(r)$ becomes a scalar field, and the
sum becomes an integral over the $2d$ area:
\begin{equation}
\mathcal{H}-E_0=\frac{J}{2}\int d^2 r |\vec{\nabla}\Phi(r)|^2
\label{energy of continuous xy model expanded}
\end{equation}
In this $2d$ state of continuous spin states a possible excitation
above the (perfectly ordered) ground state is a vortex\footnote{Spin
waves are another form of excitation, and are responsible for
destroying long-range order at any finite T.}, Fig.~\ref{vortex}.
Any line integral along a closed path around the center of the
vortex will give:
\begin{equation}
\oint \vec{\nabla}\Phi(\vec{r})\cdot d\vec{r}= 2 \pi q \qquad
\Rightarrow \qquad
\begin{array}{c}
  \Phi=q\phi \\
 |\vec{\nabla}\Phi(r)|=\frac{q}{r}
\end{array}
\label{contour vortex}
\end{equation}
where $q$ is the vorticity and $\phi$ is the angle to
$\vec{r}$.\footnote{ $\vec{\nabla}q
\phi=\frac{1}{r}\frac{\partial}{\partial \phi}  q \phi=\frac{q}{r}$}
\begin{figure}
    \begin{center}
        \includegraphics[width=0.8\textwidth]{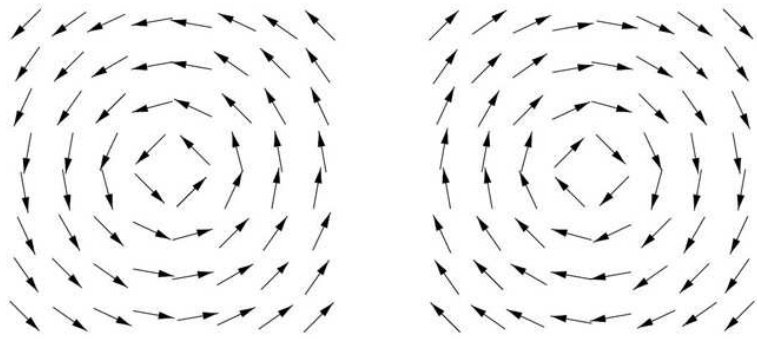}\\
        \caption{Vortex with vorticity $|q|=1$}
        \label{vortex}
    \end{center}
\end{figure}
Now using (\ref{energy of continuous xy model expanded}) and
(\ref{contour vortex}) we can find the energy of this vortex above
the ground state energy:
\begin{equation}
E=\int_a^R 2 \pi r dr \frac{J}{2} (\frac{q}{r})^2=\pi J\ln
(\frac{R}{a})q^2 \label{vortex energy}
\end{equation}
here, $R$ is the size of the system \cite{vilfan}\cite{bellac}, and
$a$ is the lower cut off (as the theory diverges as
$r\rightarrow0$). Now it can be seen that the energy of the vortex
increases logarithmically with the size of the system.
\subsubsection{Thermodynamics of Single Vortices}
A vortex can be put onto any lattice site, thus the number of
configurations is $(R/a)^2$, and then we can write the entropy of
the vortex as
\begin{equation}
S=2 k_B \ln(\frac{R}{a})
\end{equation}
Therefore, the free energy, $F=E-TS$, is positive and large at low
temperatures.  Thus, it is not very easy to thermally excite single
vortices at low temperature.  But, as temperature increases, the
free energy decreases, and at a critical temperature $T_{KT}$ the
free energy vanishes:
\begin{equation}
k_B T_{KT} =\frac{\pi J}{2}\label{single critical temperature}
\end{equation}
Then, above $T_{KT}$ a large number of vortices are thermally
excited.
\subsubsection{Vortex-Antivortex "Dipole" Pairs}
Even though it is difficult to excite single vortices at low
temperature, it does not mean none exist.  In the low temperature
phase it becomes more favorable to excite pairs of vortices, i.e. a
vortex-antivortex pair.  To see this we start by writing the energy
of a system of two vortices, with vorticity $q_1$ and $q_2$:
\begin{eqnarray}
\Phi&=&\Phi_1(\vec{r}_1)\Phi_2(\vec{r}_2)\nonumber\\
E(q_1,q_2)&=&-\pi J (q_1+q_2)^2\ln(\frac{R}{a})\nonumber\\
&-&2\pi J q_1 q_2\ln(\frac{|\vec{r}_1-\vec{r}_2|}{a}) \label{pair
vortex energy}
\end{eqnarray}
where for a vortex and antivortex pair $q_1=-q_2$, thus we have:
\begin{equation}
E(q_1,q_2)=2\pi J q^2\ln(\frac{|\vec{r}_1-\vec{r}_2|}{a})
\label{vortex anti-vortex energy}
\end{equation}
From this you can see that the self-energy of a vortex-antivortex
pair is small when the vortices are close together. Thus, as the
energy of a "dipole" pair is smaller than a single vortex, it is
easier to excite a pair of opposite vortices than a single vortex.
\subsubsection{Critical Phase of the XY Model}
We can now see that at low temperature, $0<T<T_{KT}$, there will be
a finite density of low energy vortex-antivortex pairs, of zero
total vorticity.  Then, at high temperatures, $T > T_{KT}$, there
will be a large concentration of unbound vortices.

In this picture the critical temperature is then associated with the
unbinding of vortex-antivortex pairs; the pairs are bound strongly
to each other at low temperature, and as the temperature increases,
the number of pairs increases.  Then, at the critical temperature
$T_{KT}$, vortex-antivortex pairs start to separate and become free;
this is called the \textit{KT} (Kosterlitz and Thouless) phase
transition.
\subsubsection{Correlation Length of Vortex Pairs}
Due to the $KT$ transition the correlation function of the XY model
changes behaviour as it crosses the critical point $T_{KT}$.  Above
the critical point, when the vortices are unbound and free to
interact with each other, the correlation function behaves in a
classic way:
\begin{equation}
<S_i S_j>=e^{-\frac{r}{\xi}},
\end{equation}
with the correlation length:
\begin{equation}
\xi=\frac{a}{\ln T/J}
\end{equation}
However, below the critical temperature the correlation function
shifts to a power-law:
\begin{equation}
<S_i S_j>=(\frac{a}{r})^{\frac{1}{\xi}}
\end{equation}
and the critical exponent becomes:
\begin{equation}
\frac{1}{\xi}=\frac{T}{2\pi J}
\end{equation}
This can be understood in the following way: at low temperature
vortices bind into dipole pairs, whose influence on the system is
confined to small distances, thus the only important large distance
interactions are spin waves.  This phase is then described by the
power-law behaviour of the correlation function.  As the temperature
rises the size of the dipole pairs increases, and diverges at
$T=T_{KT}$ when free vortices appear. These free vortices then
interact, and in the process, the low temperature dominance of spin
waves is destroyed.  This phase is then described by an exponential
behaviour of the correlation function.
\section{The \textit{KT} Transition and Superfluidity in $2d$}
\label{kt transition and supefluid} One application of the
\textit{KT} transition, i.e the binding and unbinding of vortices at
the critical temperature $T_{KT}$, is in understanding superfluidity
in thin films. For $d>2$ Bose-Einstein condensation occurs in a Bose
gas at low temperature, and this condensed phase is defined by the
breaking of a $U(1)$ global gauge symmetry. However, for systems
with $d\leq2$ the breaking of a $U(1)$ global gauge symmetry is
prohibited by the Mermin-Wagner theorem, i.e. no phase transition to
a state with long-range order can occur at finite temperatures. On
the other hand, topological phase transitions are possible; and the
binding and unbinding of vortices in two dimensions leads to a type
of phase transition.
\subsection{Superfluids and the XY model} Our previous
calculations, relating to vortices in the XY model can be directly
applied to the study of superfluids.  For single vortices, the XY
spin angle $\Phi$ (\ref{xy model expanded}) can be related to the
phase of a particle's wave function in the superfluid condensate
state:
\begin{equation}
\Psi=\sqrt{n_0}\exp(-i\Phi(t,\vec{r})) \label{condensate state}
\end{equation}
where $n_0$ is the number of condensate particles (superfluid
density). Then the current density of the superfluid is given by:
\begin{eqnarray}
I_{cond}&=&\frac{i\hbar}{2m}(\Psi\vec{\nabla}\Psi^*-\Psi^*\vec{\nabla}\Psi)\nonumber\\
&=&\frac{\hbar}{m}n_0\vec{\nabla}\Phi\nonumber\\
&=&\Upsilon\vec{\nabla}\Phi \qquad \rightarrow \qquad
\Upsilon=\frac{\hbar}{m}n_0 \label{condenstae current}
\end{eqnarray}
where $m$ is the mass of the superfluid particle; and $\Upsilon$ is
the constant of proportionality between the current density and the
phase angle, and is known as the helicity modulus ($\Upsilon$). The
phase angle $\Phi$ can be associated with the spin angle as in the
XY-model or with a particles phase as in the superfluid condensate.

If we divide both sides of (\ref{condenstae current}) by the density
$n_0$, we have the velocity of the condensate particles:
\begin{equation}
\vec{v}_{cond}(\vec{r})=\frac{\hbar}{m}\vec{\nabla} \Phi
\end{equation}
This can now be inserted into the energy of a superfluid velocity
fluctuation,
\begin{equation}
\mathcal{H}=\frac{1}{2}mn_0 \int d^2r |\vec{v}_{cond}(\vec{r})|^2
\end{equation}
where $mn_0 $ is the mass density of the superfluid.  This then
gives us:
\begin{equation}
\mathcal=\frac{\hbar\Upsilon}{2} \int d^2 r (\vec{\nabla\Phi})^2
\qquad \rightarrow \qquad \Upsilon=\frac{\hbar}{m}n_0
\label{superfluid energy}
\end{equation}
It can now be seen that this is similar to (\ref{energy of
continuous xy model expanded}) if, as stated before, we identify the
phase of the superfluid condensate with the angle of spin in the XY
model and $\hbar\Upsilon=J$ .  Therefore, due to this link between
the XY model and superfluids, the transition from a superfluid state
to a normal state for a $2d$ superfluid can be associated with the
binding and unbinding of vortex pair excitations with opposite
circulation. Evidence of this $KT$ behaviour is seen in superfluid
Helium 4. In $3d$ there is a continuous decrease of the superfluid
density with temperature, but in the 2 dimension case, for example
in thin films of $^4He$, a jump in the superfluid density occurs at
the critical temperature $T_{KT}$ \cite{nelson}.
\subsubsection{$T_{KT}$ for Superfluids} By combining
Eqns.~(\ref{single critical temperature}) and (\ref{superfluid
energy}) we are able to determine the critical temperature,
$T_{KT}$, for superfluids:
\begin{equation}
\frac{1}{T_{KT}}=\frac{2}{\pi}\frac{K_B}{J}=\frac{2}{\pi}\frac{K_B}{\hbar\Upsilon}
\label{tkt-superfluid}
\end{equation}
where the final form, in terms of the helicity modulus, is the most
useful form.  This can be written in lattice field theory (using
lattice units $K_B=1$, $\hbar=1$ and $L_t=1/T$) as:
\begin{equation}
L_{t}=\frac{1}{T_{KT}}=\frac{2}{\pi}\frac{1}{\Upsilon},
\label{tkt-superfluid}
\end{equation}
}

{\typeout{Appendix B}
 \chapter{Changes to QCD}
\label{lattices}
\section{Lattice QCD at $\mu\neq0$}
\label{lattice qcd mu} Lattice calculations cannot be extended to
$\mu\neq0$ due to technical problems.  A quantum system, recast in
Euclidean space-time, can be represented statistically in terms of
the partition function:
\begin{equation}
    \mathcal{Z}=\int dU d\bar{\psi} d\psi e^{-S[U,\bar{\psi},\psi]},
    \label{calculate z}
\end{equation}
where $S=\sum_x\mathcal{L}_{QCD}$ is the QCD action.  Then, in order
to calculate this, or related expectation values,
\begin{equation}
<\mathcal{O}>\equiv \frac{1}{\mathcal{Z}}\int dU d\bar{\psi}d\psi
\mathcal{O} e^{-S[U,\bar{\psi},\psi]},
\end{equation}
one can explicitly integrate out the bilinear fermionic dependence
leaving
\begin{equation}
    \mathcal{Z}=\int dU \det M e^{-S_g[U]}
    \label{calculate z}
\end{equation}
where $S_g$ is the gauge contribution to the action and
$M=(\Dslash+m_0)$ is the fermion kinetic matrix.  Now, from the
direct correspondence between Euclidean quantum field theory and
statistical mechanics, it is possible to carry out the integral over
gauge field configurations using Monte Carlo methods, in which the
highly peaked nature of $e^{-S}$ is exploited by using it as a
sampling weight.  This is possible because $\Dslash$ is an
anti-hermitian operator which obeys chiral symmetry, so we have
$[\Dslash,\gamma_5]=0$ and its eigenvalues come in complex conjugate
pairs, $\pm i\lambda$. Therefore, the determinant of $\Dslash$ can
be written as a product of the eigenvalues, which in this case is
real.  Hence, we can use: $\det \Dslash e^{-S_g}$ as the full
functional weight.  This is still a valid argument even with the
inclusion of the chiral symmetry breaking mass term, as the
eigenvalues of $\Dslash +m_0$ are $m_0\pm i\lambda$, which remain
complex conjugate pairs

However, when a chemical potential is introduced we have,
$M\rightarrow\Dslash+m_0+\mu\gamma_0$.  As the chemical potential
term is hermitian, $M$ then has complex eigenvalues\footnote{The
determinant of $Hermitian+AntiHermitian$ matrices is complex.},
making its determinant complex; and so the importance sampling
weight becomes $|\det M|e^{i\theta}e^{-S_g}$.  This can be shown as
follows\footnote{Using $\gamma_{\nu}=\gamma_{\nu}^{\dag}$ and
$\partial_{\nu}=-\partial_{\nu}^{\dag}$}:
\begin{eqnarray}
\bar{\psi}M(\mu)\psi&=&\bar{\psi}(\gamma_{\nu}\partial_{\nu}+\mu\gamma_0+m)\psi\nonumber\\
&=&\bar{\psi}\gamma_5\gamma_5(\gamma_{\nu}\partial_{\nu}+\mu\gamma_0+m)\psi\nonumber\\
&=&\bar{\psi}\gamma_5(-\gamma_{\nu}\partial_{\nu}-\mu\gamma_0+m)\gamma_5\psi\nonumber\\
&=&\bar{\psi}\gamma_5M^{\dag}(-\mu)\gamma_5\psi
\end{eqnarray}
Therefore,
\begin{eqnarray}
M(\mu)&=&\gamma_5M^{\dag}(-\mu)\gamma_5\nonumber\\
\Rightarrow \qquad\det M(\mu)&=&\det
\gamma_5M^{\dag}(-\mu)\gamma_5\nonumber\\
&=& \det M^{\dag}(-\mu)\nonumber\\
& =& \det M^{*}(-\mu)
\end{eqnarray}
Then, for $\mu=0$, $\det M$=$\det M^*$  implies $\det M$ is real,
but if $\mu\neq 0$, we have $\det M(\mu)$=$\det M^*(-\mu)$, implying
$\det M$ is complex.

Therefore, the case for $\mu\neq 0$ leads to a complex phase, which
in turn leads to configurations with large $e^{-S_g}$ cancelling and
the effectiveness of the importance sampling weight becoming
suppressed\cite{walters}.
\subsection{$\mu\neq0$ in the NJL Model}
In the NJL model, it can be shown that even with the introduction of
a chemical potential term, the importance sampling weight remains
real.  Using the NJL Lagrangian, Eqn.~(\ref{bosonized path integral
njl simple}), we have:
\begin{equation}
M\rightarrow\gamma_{\nu}\partial_{\nu}+\mu\gamma_0+m+(\sigma+i\gamma_5\vec{\tau}\cdot\vec{\pi})
\end{equation}
Now, we have\footnote{Using $C\gamma_{\nu}C^{-1}=-\gamma_{\nu}^*$
where the charge conjugation operator satisfies: $C^{-1}=-C$, and
$\tau_2\vec{\tau}\tau_2=-\vec{\tau}^*$}:
\begin{eqnarray}
\bar{\psi}M(\mu)\psi=&\bar{\psi}&(C\gamma_5 \otimes
\tau_2)(C^{-1}\gamma_5 \otimes \tau_2)
(\gamma_{\nu}\partial_{\nu}+\mu\gamma_0+m\nonumber\\
&+&\sigma+i\gamma_5\vec{\tau}\cdot\vec{\pi})\psi\nonumber\\
=&\bar{\psi}&(C\gamma_5 \otimes \tau_2)C^{-1}\gamma_5
(\gamma_{\nu}\partial_{\nu}+\mu\gamma_0+m\nonumber\\
&+&\sigma-i\gamma_5\vec{\tau}^*\cdot\vec{\pi})\tau_2\psi\nonumber\\
=&\bar{\psi}&(C\gamma_5 \otimes \tau_2)C^{-1}
(-\gamma_{\nu}\partial_{\nu}-\mu\gamma_0+m\nonumber\\
&+&\sigma-i\gamma_5\vec{\tau}^*\cdot\vec{\pi})\gamma_5\otimes\tau_2\psi\nonumber\\
=&\bar{\psi}&(C\gamma_5 \otimes \tau_2)
(\gamma_{\nu}^*\partial_{\nu}+\mu\gamma_0^*+m\nonumber\\
&+&\sigma-i\gamma_5^*\vec{\tau}^*\cdot\vec{\pi})(C^{-1}\gamma_5\otimes\tau_2)\psi\nonumber\\
=&\bar{\psi}&(C\gamma_5 \otimes \tau_2)M^*(\mu)(C^{-1}\gamma_5
\otimes \tau_2)\psi
\end{eqnarray}
Here we have $\det M=\det M^*$, implying $\det M$ is real.

\section{Staggered Fermions}
The Euclidean action for a free fermion field $\psi$, in $2+1d$, is
given by
\begin{equation}
S=\int d^3x\bar{\psi}(\Dslash+m_0)\psi,
\end{equation}
which, when discretised becomes
\begin{equation}
S=a^3\sum_x \left[
    \bar{\psi}_x\sum_{\nu}\left\{
\gamma_{\nu}\frac{(\psi_{x+\hat{\nu}}-
\psi_{x-\hat{\nu}})}{2a}\right\}+m_0\bar{\psi}_x\psi_x \right]
\label{staggered action prem}
\end{equation}
Fourier transforming this to momentum space this becomes
\begin{equation}
S=a^3\int^{\frac{\pi}{a}}_{-\frac{\pi}{a}}\frac{d^3p}{(2\pi)^3}\bar{\psi}
\left[
     \frac{i}{a}\sum_{\nu} \gamma_{\nu} \sin(ap_{\nu})\psi +m_0\psi
 \right]
\end{equation}
The inverse of $M$, where $S=\int \bar{\psi}M\psi$, is found to be
\begin{equation}
S_F(p)=\frac{-\frac{i}{a}\sum_{\nu} \gamma_{\nu} \sin(ap_{\nu})
+m_0}{\frac{1}{a^2}\sum_{\nu}\sin ^2(ap_{\nu})+(m_0)^2}
\end{equation}
We can see that in the case of small momentum (long wavelength)
limit, the small angle approximation implies $\sin ap_{\nu} \approx
ap_{\nu}$ and we recover the continuum Euclidean propagator.
However, in general this represents $8$ species of fermion.

This "doubling problem" can be resolved by the staggered
formulation.  In this way we can reduce the number of species by a
factor of $4$, which can therefore by interpreted as the physical
flavours \cite{kogut}.  By making the transformation:
\begin{equation}
\psi_x\rightarrow\gamma^{x_1}_1\gamma^{x_2}_2\gamma^{x_3}_3 \chi_x
\qquad
\bar{\psi}_x\rightarrow\bar{\chi}_x\gamma^{x_3}_3\gamma^{x_2}_2\gamma^{x_1}_1
\end{equation}
so that
$\bar{\psi}_x\gamma_{\nu}\psi_{x\pm\hat{\nu}}\rightarrow\eta_{\nu}(x)\bar{\chi}_x\chi_{x\pm\hat{\nu}}$,
where $\eta_{\nu}(x)=(-1)^{x_1+\ldots + x_{\nu+1}}$, then the
action, Eqn.~(\ref{staggered action prem}), becomes:
\begin{equation}
S=a^3\sum_x \left[
    \bar{\chi}_x\sum_{\nu}\left\{
\eta_{\nu}(x)\frac{(\chi_{x+\hat{\nu}}-
\chi_{x-\hat{\nu}})}{2a}\right\}+m_0\bar{\chi}_x\chi_x \right]
\label{staggered action}
\end{equation}
Now, $M$ is in a reducible diagonal form, so disregarding all but
one component of $\nu$, we are left with the reduced number of $2$
quark species.
}

 {\typeout{HMD}
 \chapter{Hybrid Molecular Dynamics Code}
\label{hmd}

For the  2+1d Euclidean NJL model (in the Gor'kov basis),
expectation values of an observable $O$ as a function of the bosonic
field configuration $[\Phi]$ are given by:
\begin{equation}
\langle O\rangle \equiv \mathcal{Z}^{-1}\int \mathcal{D}\Phi
\mathcal{D}\Phi^{\dag}O[\Phi]\sqrt{\det 2
\mathcal{A}}[\Phi]\exp{-S_{bos}[\Phi]}
\end{equation}
where $\mathcal{Z}$ is the path integral (\ref{partition function
for the njl model in bosonic form}). As in statistical physics, such
expectation values can be calculated using Monte Carlo or molecular
dynamic techniques\footnote{Where one evolves trajectories according
to a stochastic process, the other evolves them along classic
trajectories represented by Newton's laws}.  In the most basic form
we can consider that a statistical sample \{${\Phi_n;
n=1,2,\ldots,N}$\} is used to approximate the full ensemble
${\Phi}$; then the average of $O[\Phi]$ is taken with respect to the
randomly selected configuration of fields. However, as $\sqrt{\det 2
\mathcal{A}}[\Phi]e^{-S_{bos}[\Phi]}$ is highly peaked about certain
configurations, it is then used as an importance sampling weight.

In this work the expectation values were calculated using a hybrid
molecular dynamics code \cite{gottlieb}, where the field
configurations were evolved along classic trajectories with a time
step of $dt=0.04$.  Then the expectation values, $\langle O
\rangle$, were conducted using complex Gaussian vectors (noisy
estimators). This process was carried out for 300 trajectories.
Then the final measurement was obtained by taking the arithmetic
average of the various instantaneous values assumed by the HMD run.

As an example we can look at the chiral condensate expectation
value, Eqn.~(\ref{chiral expectation}).  Once a trajectory is
complete, and the end point is selected as the field configuration,
the program calls the measure subroutine.  In this part of the
program a set of Gaussian vectors are defined as: $\eta^p_q$, where
$p=1,\ldots,4$ are the isospin components and $q=1,\ldots, vol$ are
the lattice sites; and if we average over the noise we obtain:
\begin{equation}
\overline{\eta^{p\dagger}_q \eta^{p'}_{q'}}=\delta^{pp'}\delta_{qq'}
\end{equation}
Using this we can estimate a value for the chiral condensate.
Starting with the matrix multiplication:
\begin{equation}
\eta^{\dag} \mathcal{A}^{-1}[\Phi] \left(
                                          \begin{array}{cc}
                                            0 & \frac{1}{2}\delta \\
                                            -\frac{1}{2}\delta & 0   \\
                                          \end{array}
                                        \right) \eta=\eta^{\dag} \mathcal{X} \eta
\end{equation}
We can then average over the noise, as before, and we obtain:
\begin{equation}
\overline{\eta^{\dag} \mathcal{X}
\eta}=\mathcal{X}^{pp'}_{qq'}\delta^{pp'}\delta_{qq'}=\textrm{tr}\left[
\mathcal{A}^{-1}[\Phi] \left(
                                          \begin{array}{cc}
                                            0 & \frac{1}{2}\delta \\
                                            -\frac{1}{2}\delta & 0   \\
                                          \end{array}
                                        \right)\right]
\end{equation}
This will then be the instantaneous value assumed by the HMD run.
Once this is carried out for the $300$ trajectories, we then find
their arithmetic average and error.
\\
\\
In this work we used an HMD code for the 3D four-fermi model with
$SU(2)\times SU(2)$ symmetry, which was based on the algorithm in
\cite{duane}.  
}

%%%%%%%%%%%%%%%%%%%%%%%%%%%%%%%%%%%%%%%%%%%%%%%%%%%%%%%%%%%%%%%%%%%%%%%%%%%%%%%%%%%%%
%%%%%%%%%%%%%%%%%%%%%%%%%%%%%%%%%%%%%%%%%%%%%%%%%%%%%%%%%%%%%%%%%%%%%%%%%%%%%%%%%%%%%
%%%%%%%%%%%%%%%%%%%%%%%%%%%%%%%%%%%%%%%%%%%%%%%%%%%%%%%%%%%%%%%%%%%%%%%%%%%%%%%%%%%%%
%%%%%%%%%%%%%%%%%%%%%%%%%%%%%%%%%%%%%%%%%%%%%%%%%%%%%%%%%%%%%%%%%%%%%%%%%%%%%%%%%%%%%
\backmatter

 \setlinespacing{1.50}

\listoffigures

\end{document}